\documentclass{jfm}
\usepackage{graphicx}
\usepackage{epstopdf, epsfig}
\usepackage{color,soul}
\usepackage{amsmath,amssymb}
\usepackage{mathtools}
\usepackage{mathrsfs}
\usepackage{xfrac}
\setul{0.5ex}{0.2ex}
\setulcolor{red}

\usepackage[colorlinks=true,allcolors=blue]{hyperref}
\usepackage{xcolor}

\newcommand{\blangle}{\langle}
\newcommand{\kflow}{k_f}
\newcommand{\brangle}{\rangle}

\newcommand{\mcP}{\mathcal{P}}

\newcommand{\dd}{\mathrm{d}}
\newcommand{\br}{\boldsymbol{r}}
\newcommand{\bx}{\boldsymbol{x}}
\newcommand{\bs}{\boldsymbol{s}}

\newcommand{\bxp}{\boldsymbol{x}'}
\newcommand{\bu}{\boldsymbol{u}}
\newcommand{\bv}{\boldsymbol{v}}
\newcommand{\bF}{\boldsymbol{f}}
\newcommand{\bP}{\boldsymbol{P}}

\newcommand{\bB}{\boldsymbol{B}}

\newcommand{\bA}{\boldsymbol{A}}

\newcommand{\bw}{\boldsymbol{w}}

\newcommand{\bk}{\boldsymbol{k}}

\newcommand{\intpikp}{\int \frac{\dd^3 \bk'}{(2\pi)^3}}

\newcommand{\intpikpp}{\int \frac{\dd^3 \bk''}{(2\pi)^3}}

\newcommand{\intkpp}{\int \dd^3 \bk''}

\newcommand{\mcE}{\mathcal{E}}
\newcommand{\const}{\mathrm{const}}

\newcommand{\uu}{\blangle  \bu \bcdot \bu' \brangle}

\newcommand{\uf}{\blangle  \bu \bcdot \bF' \brangle}

\newcommand{\tnl}{t_{\mathrm{nl}}}

\newcommand{\ol}[1]{\overline{#1}}
\newcommand{\wt}[1]{\widetilde{#1}}

\shorttitle{Emergence of long-range correlations and thermal spectra in forced turbulence}
\shortauthor{D. N. Hosking and A. A. Schekochihin}

\title{Emergence of long-range correlations and thermal spectra in forced turbulence}

\author{D. N. Hosking\aff{1,}\aff{2,}\aff{4,}\aff{5}
  \corresp{\email{dhosking@princeton.edu}} \and A. A. Schekochihin\aff{3,}\aff{4}}

\affiliation{
\aff{1}Oxford Astrophysics, Denys Wilkinson Building, Keble Road, Oxford, OX1 3RH, UK
\aff{2}Princeton Center for Theoretical Science, Princeton University, Princeton, NJ 08544, USA
\aff{3}Rudolf Peierls Centre for Theoretical Physics, Clarendon Laboratory, Parks Road, Oxford, OX1 3PU, UK
\aff{4}Merton College, Merton Street, Oxford, OX1 4JD, UK
\aff{5}Gonville \& Caius College, Trinity Street, Cambridge, CB2 1TA, UK}

\begin{document}

\maketitle

\begin{abstract} 
Recent numerical studies have shown that forced, statistically isotropic turbulence develops a `thermal equilibrium' spectrum, $\mcE(k) \propto k^2$, at large scales. This behaviour presents a puzzle, as it appears to imply the growth of a non-zero Saffman integral, which would require the longitudinal velocity correlation function, $\chi(r)$, to satisfy $\chi(r\to \infty)\propto r^{-3}$. As is well known, the Saffman integral is an invariant of decaying turbulence, precisely because non-local interactions (i.e., interactions via exchange of pressure waves) are too weak to generate such correlations. Subject to certain restrictions on the nature of the forcing, we argue that the same should be true for forced turbulence. We show that long-range correlations and a $k^2$ spectrum arise as a result of the turbulent diffusion of linear momentum, and extend only up to a maximum scale that grows slowly with time. This picture has a number of interesting consequences. First, if the forcing generates eddies with significant linear momentum (as in so-called Saffman turbulence), a thermal spectrum is not reached---instead, a shallower spectrum develops. Secondly, the energy of turbulence that is forced for a while and then allowed to decay obeys Saffman's decay laws for a period that is much longer than the duration of the forcing stage.
\end{abstract}

\section{Introduction\label{sec:introduction}}

Probably the best-known result in the theory of turbulence is Kolmogorov's law for the spectral energy density in the inertial range, $\mcE(k)\propto k^{-5/3}$. This law follows from the conjecture of a constant flux of energy in $k$-space, from the large scales at which it is injected, to the small scales at which it is dissipated by molecular viscosity \citep{Kolmogorov41a}. However, a power-law spectrum can also be found at scales larger than the outer scale of the turbulence, if that scale is small compared to the system's size. Unlike the inertial-range spectrum, this small-$k$ spectral tail does not correspond to large eddies with size $k^{-1}$---instead, it is controlled by statistical properties of the eddies at the outer scale~\citep{Davidson15}. Provided the two-point velocity correlation function, ${\blangle \bu (\bx) \bcdot \bu (\bx + \br) \brangle \equiv \uu}$, decays sufficiently quickly with distance, a purely kinematic calculation shows that the energy spectrum of statistically isotropic and homogeneous turbulence satisfies
\begin{equation}
    \mcE(k\to 0) = \frac{L k^2}{4 \pi ^2} + \frac{I k^4}{24 \pi^2} + O(k^5), \label{Eexpansion}
\end{equation}
where
\begin{equation}
    L = \int \dd^3 \br \uu, \label{saffman_integral}
\end{equation}
and
\begin{equation}
    I = - \int \dd^3 \br r^2 \uu,\label{Loitsyansky}
\end{equation}are known as the Saffman\footnote{The Saffman integral is sometimes known as the Saffman-Birkhoff integral, in recognition of the work by~\cite{Birkhoff54}---for convenience, we shall use the more economical standard name in this work.} and Loitsyansky integrals, respectively \citep{Saffman67, Loitsyansky39}. These integrals encode information about the distribution of linear and angular momentum in real space \citep{LandauLifshitzFluids, Saffman67, Davidson09}. Owing to the conservation of these momenta, it turns out that~$L$ and~$I$ are invariants of unforced, decaying turbulence\footnote{More precisely, $I$ is related to a weighted integral of angular momentum density, and is invariant only if correlations decay sufficiently rapidly with distance \citep{Davidson09}.}, leading to a phenomenon often called the \textit{`permanence of large-scale structure'}---as turbulence decays, the small-$k$ part of the spectrum remains unchanged. This observation, together with the assumption of self-similarity, allows the decay of kinetic energy to be computed as a function of time~(\citealt{Saffman67, BatchelorProudman56}; see \citealt{Davidson15} for a review).

While these results are well established in the theory of decaying turbulence, the large-scale properties of \emph{forced} turbulence, i.e., one into which energy is continually injected, are usually described in very different terms. In that context, the small-$k$ part of the energy spectrum has received particular attention in recent years, owing to an attractive analogy with statistical mechanics. It has been shown in numerical simulations that there is no net $k$-space energy flux to these scales \citep{Dallas15}, as is to be expected on physical grounds. Accordingly, it has been argued that the largest scales of steady-state forced turbulence might constitute a subsystem in \emph{thermal equilibrium} with the separate, non-equilibrium subsystem represented by the rest of the flow \citep{Dallas15, Cameron17, AlexakisBiferale18,AlexakisBrachet19}. This idea leads immediately to a prediction for the large-scale spectrum: energy should be equipartitioned between Fourier modes\footnote{If the large-scale Fourier modes of Navier-Stokes turbulence are taken to constitute a separate system to their smaller-scale forced and dissipating counterparts, then their thermal-equilibrium spectrum follows formally from the statistical mechanics of the truncated Euler equations,
$\p_t \bu +\mathcal{P}_{K}\left[\bu \bcdot \bnabla \bu + \bnabla p\right] = 0, \label{truncatedeuler}$
where $\bu$ is the incompressible velocity field, $p$ is the pressure, determined by $\bnabla \bcdot \bu = 0$, and $\mathcal{P}_K$ is a truncation operator that sets to zero all Fourier modes with $k>K$. This system satisfies a Liouville's theorem, and has an absolute equilibrium state that satisfies \eqref{k2} in the absence of net kinetic helicity [\citealt{Lee52, Orszag77, Kraichnan73}, see §1 of \citealt{AlexakisBrachet19} for a review].}, so, in 3D, 
\begin{equation}
    \mcE(k) \propto k^2. \label{k2}
\end{equation}

In reality, the large-scale modes do not constitute an isolated system, but, if their nonlinear interaction with the turbulent scales is weak, it may be expected that they should develop a close-to-equilibrium state \citep{AlexakisBrachet19}. Indeed, \eqref{k2} is well supported by a number of numerical studies conducted in recent years \citep{Dallas15, Cameron17, AlexakisBiferale18,AlexakisBrachet19}. Furthermore, the validity of thermal-equilibrium spectra in more general types of turbulence has been demonstrated experimentally for capillary-wave turbulence by \cite{Michel17}. An experiment to study the large scales of hydrodynamic turbulence is also in development by the same group.

However, like any statistical-mechanics argument, the reasoning outlined above does not elucidate the mechanism by which the equilibrium spectrum is attained. Furthermore, it is unclear what relation \eqref{k2} has to the expansion of $\mcE(k)$ in terms of the Saffman and Loitsyansky integrals, \eqref{Eexpansion}. Until now, it has been assumed that the connection between forced turbulence and the concept of decaying `Saffman turbulence' (i.e., that with $L\neq 0$) is superficial, despite both having the same large-scale spectral power law. This is because (\textit{i}) analysis of the former is mostly concerned with the statistical steady state, obtained by taking a long-time average, while decaying turbulence is, by definition, transient; and (\textit{ii}) large scales in the former may interact with the forcing, which is absent from decaying turbulence \citep{AlexakisBrachet19}. Nonetheless, it should be noted that \eqref{Eexpansion} is a purely kinematic result, and must, therefore, apply equally to the forced and decaying cases.

The central goal of the present work is to reconcile the kinematic and statistical-mechanical points of view. This problem turns out to be non-trivial, because of the invariance of the Saffman integral. As we shall show in §\ref{sec:correlations}, this invariance is not restricted to decaying turbulence, but should also apply to forced turbulence, subject to certain reasonable conditions on the nature of the forcing.
In particular, if the forcing is solenoidal and sufficiently local in real space (i.e., if its correlations decay sufficiently quickly), then non-local interactions via pressure waves are too weak to generate the long-range longitudinal velocity correlations, $\chi(r\to\infty)\propto r^{-3}$, required for a non-zero Saffman integral, as is the case in decaying turbulence \citep{Saffman67, BatchelorProudman56, Davidson15}. Thus, the na\"{i}ve conclusion that the equilibrium spectrum \eqref{k2} simply corresponds to $L\neq 0$ cannot be correct. For consistency with~\eqref{Eexpansion}, therefore, it must always be the case that the equilibrium, $\propto k^2$, part of the spectrum terminates at some large cutoff scale, provided it is smaller than the system size. Above the cutoff scale,~\eqref{Eexpansion} demands that $\mcE(k\to 0)\propto k^4$.

In §\ref{sec:momentum}, we shall argue that the physical mechanism by which the equilibrium part of the spectrum develops is the stochasticisation of the distribution of linear momentum, an inevitable consequence of interactions between eddies, even if each of them individually has zero net momentum when it forms. We shall show that this process leads naturally to a split-power-law spectrum at the large scales, with \eqref{k2} satisfied up to a cutoff scale that grows with time, corresponding to the largest scale at which eddies have been able to stochasticise their momentum distribution. The requirement of momentum conservation in these interactions means that different eddies become correlated, which generates the long-range correlations, $\chi(r)\propto r^{-3}$, required for $\mcE(k)\propto k^2$, though only up to the cutoff scale, above which correlations decay rapidly.

In §\ref{sec:passive}, we propose a simple, though non-rigorous, model of this phenomenon, in which the large-scale momentum distribution of the flow evolves due to turbulent diffusion caused by flow-scale structures. Under this model, we find that the development of a~$k^2$ spectrum is recovered for local, solenoidal forcing, with the cutoff scale growing like~$t^{1/2}$. This prediction, along with a number of others, is borne out well in the numerical simulations that we present. Under the same model, we also consider forcing that is local in real space, but not solenoidal---arguably, a more generic situation. Making use of a theorem due to \cite{Saffman67}, we show that such turbulence need not equilibrate at large scales, on account of the long-range real-space correlations present in the solenoidal part of the forcing. Instead, the turbulent diffusion of injected momentum leads to a shallower spectrum than \eqref{k2}.  

Finally, in §\ref{sec:decaying_connections}, we investigate the implications of the equilibration phenomenon for decaying turbulence. We find \textit{inter alia} that the energy $E$ of turbulence forced solenoidally without long-range correlations and then allowed to decay satisfies Saffman's law ${E\propto t^{-6/5}}$~\citep{Saffman67} for a time period that is much larger than the period of forcing if the latter is large compared to the turnover time of the largest eddies.

§\ref{sec:conclusion} contains a short summary of our findings, followed by a discussion of their possible applications, implications and extensions in both hydrodynamical contexts and beyond---viz., in astrophysical MHD turbulence.

\section{Long-range correlations and the invariance of Saffman's integral \label{sec:correlations}}

Let us begin by reviewing an important kinematic result: turbulence with an energy spectrum satisfying $\mcE(k\to 0) \propto k^2$ necessarily has strong long-range correlations in real space \citep{BatchelorProudman56, Saffman67, Davidson15}. 

\subsection{A \texorpdfstring{$k^2$}{thermal} spectrum requires strong long-range correlations \label{sec:kinematic_correlations}}
The energy spectrum is the Fourier transform of the two-point velocity correlation function, $\blangle \bu (\bx) \bcdot \bu (\bx + \br) \brangle \equiv \uu$, where angle brackets indicate an ensemble average. For statistically homogeneous and isotropic turbulence, $\uu$ is a function of $r= |\br|$ only, and then the energy spectrum is
\begin{equation}
    \mcE(k) = \frac{k^2}{4\pi^2} \int \dd^3 \br\, \uu e^{-i\bk \bcdot \br} = \frac{1}{\pi} \int^\infty_0 \dd r \uu kr \sin (kr).\label{E(k)exact}
\end{equation}If correlations between points separated by distances much greater than the energy-containing scale of the turbulence, $l$, decay sufficiently quickly, then \eqref{E(k)exact} may be Taylor-expanded for small $k$. Namely, if $\uu = o(r^{-5})$ as $r \to \infty$, then \eqref{Eexpansion} holds.

From \eqref{Eexpansion}, it would appear that the `thermal' $k^2$ spectrum corresponds to $L\neq 0$. However, this conclusion is problematic, because $L$ is an invariant. This fact is well known in the context of decaying turbulence, for which the conservation of $L$ implies a meaningful distinction between turbulence with finite $L$, called `Saffman turbulence', and that with $L=0$, called `Batchelor turbulence'. These two canonical types of turbulence have a number of differences, chief among them their laws for the decay of energy with time (see \citealt{Davidson15} for a review). As we shall show in §\ref{sec:batchelor_argument}, conservation of $L$ should also be expected in \emph{forced} turbulence, provided that long-range correlations in the forcing function are sufficiently weak to prohibit injection of $L$. As a result, if $L=0$ at $t=0$, $L=0$ at all subsequent times.

The relevance of correlations in the forcing function is that sufficiently strong long-range correlations in the velocity field are required for $L$ to be non-zero. Statistical isotropy and homogeneity, together with incompressibility, restrict the allowed form of the two-point velocity correlation tensor $\langle u_i u_j'\rangle$ to
\begin{equation}
    \langle u_i u_j'\rangle = \frac{u^2}{2r}\left[(r^2 \chi)' \delta_{ij} - \chi'(r) r_i r_j\right], \label{uiuj(f)}
\end{equation}where $\chi(r) = \langle u_r (\bx) u_r (\bx+\br) \rangle/u^2$ is the longitudinal correlation function \citep[see, e.g.,][]{Davidson15, LandauLifshitzFluids}, and we follow the convention ${u^2 \equiv \langle u_x^2 \rangle = \langle |\bu| ^2\rangle}$/3. Equation \eqref{uiuj(f)} implies
\begin{equation}
     \uu = \frac{1}{r^2}\frac{\p}{\p r} (r^3 u^2 \chi).\label{uu(f)} 
\end{equation}Integrating \eqref{uu(f)} over all space, we find that the Saffman integral, \eqref{saffman_integral}, is
\begin{equation}
    L = 4\pi u^2 \lim_{r\to \infty} r^3 \chi (r). \label{L=limr3f}
\end{equation}Thus, $L$ is finite if and only if 
\begin{equation}
    \chi(r\to\infty) = O(r^{-3}).\label{fsimr-3}
\end{equation}

Note that, somewhat counter-intuitively, \eqref{fsimr-3} need not mean that $\uu$ decays slowly with $r$, as may be shown by substituting \eqref{fsimr-3} in \eqref{uu(f)}. As a consequence, the long-range correlations implied by \eqref{fsimr-3} do not necessarily invalidate the expansion \eqref{Eexpansion}, which required $\uu = o(r^{-5})$. An extreme example is a white-noise velocity field,
\begin{equation}
	\langle \bu (\bx)\bcdot \bu(\bx +\br) \rangle \propto \delta^3 (\br) \implies \uu (r) \propto \frac{\delta (r)}{r^2}. \label{whitenoise}
\end{equation}It follows immediately from \eqref{saffman_integral} and \eqref{whitenoise} that $L\neq 0$ for such a field (in this case, the $k^2$ spectrum extends to all scales). However, we see from~\eqref{uu(f)} that $\chi(r)=1/r^3$, and so, from~\eqref{uiuj(f)}, $\langle u_i u_j'\rangle = 3u^2 r_i r_j/2r^5$ for $i\neq j$. This means that even a white-noise velocity field, if incompressible, must have long-range correlations hidden in the off-diagonal components of its spectral tensor.

\subsection{Non-local fluid processes are insufficient to generate long-range correlations \label{sec:batchelor_argument}}

Intuitively, no local (in real space) forcing mechanism can set up correlations between infinitely separated points, at least in the absence of non-local fluid processes. Of course, this need not be an obstacle to the development of a non-zero Saffman integral, and hence a thermal-equilibrium $k^2$ spectrum, because incompressible turbulence \emph{is} subject to non-local interactions: physically, incompressibility is enforced via the action of pressure waves, which propagate at infinite speed through the fluid. Let us estimate the strength of correlations that can develop as a consequence of the pressure-mediated interactions.

Taking the divergence of the Navier-Stokes equation,
\begin{align}
    \frac{\p \bu}{\p t} + \bu \bcdot \bnabla \bu = -\bnabla p + \nu \nabla^2 \bu + \bF,\quad 
    \bnabla \bcdot \bu = 0,\label{NavierStokes}
\end{align}and assuming that $\bF$ is solenoidal (i.e., that $\bnabla \bcdot \bF = 0$---we shall return to the case of $\bnabla \bcdot \bF \neq 0$ in §\ref{sec:nonsolenoidal}\footnote{The reader may wonder why this distinction is necessary. After all, only the solenoidal part of~$\bF$ is dynamically significant; the compressive part is negated by the pressure in an incompressible fluid. The problem is that when $\bnabla \bcdot \bF \neq 0$, the solenoidal part of $\bF$ is not necessarily local in real space, even if $\bF$ is. Remarkably, we shall find in §\ref{sec:nonsolenoidal} that when $\bnabla \bcdot \bF\neq 0$, locally forced turbulence does not generically tend to equilibrate towards $\mcE(k) \propto k^2$ at large scales.}), we have
\begin{align}
    \nabla^2 p = - \bnabla \bcdot (\bu \bcdot \bnabla \bu) , \label{p=2}
\end{align}so the pressure is always exactly what is required to negate the non-solenoidal part of the inertial force. Inverting~\eqref{p=2}, we find that the far-field pressure generated by an eddy localised at $\bx=0$ in an otherwise quiescent fluid is
\begin{align}
    p(\bx) & = \frac{1}{4\pi} \int \frac{\dd^3 \bx'}{|\bx' -\bx|} \frac{\p}{\p x_i'}\frac{\p}{ \p x_j'}u_i(\bx') u_j(\bx') \label{p_integral}\\ & = \frac{1}{4\pi} \frac{\p}{\p x_i}\frac{\p}{ \p x_j}\frac{1}{x}\int \dd^3 \bx' u_i(\bx')u_j(\bx') + O(x^{-4}) \\ & = O(x^{-3}), \label{p=3}
\end{align}where we have Taylor-expanded the Green's function $|\bx'-\bx|^{-1}$ in \eqref{p=3} in small~$|\bx'|$. Thus, a localised eddy generates a pressure field that extends to arbitrarily large distances, falling off as $x^{-3}$, with the corresponding $\bnabla p$ force decaying as $x^{-4}$. 

Informally, we can imagine constructing a homogeneous and isotropic turbulence as a random assembly of many such eddies. Each would exert a force on distant ones that scales with their separation $r$ as $r^{-4}$. The strength of correlations that would develop due to these pressure-mediated interactions may be estimated using the von K\'{a}rm\'{a}n-Howarth equation \citep{KarmanHowarth38}, which follows from \eqref{NavierStokes} under the assumptions of statistical isotropy and homogeneity:
\begin{equation}
    \frac{\p}{\p t} \uu = \frac{1}{r^2} \frac{\p}{\p r} \frac{1}{r} \frac{\p}{\p r} (r^4 u^3 K) + 2 \nu \nabla^2 \uu + 2 \uf, \label{KH}
\end{equation} where ${K(r) = \langle u_r(\bx) u_r(\bx) u_r(\bx+\br)\rangle / u^{3/2}}$ is the longitudinal triple-correlation function. Pressure does not appear in \eqref{KH}---this is a consequence of statistical homogeneity, which demands that $\p\langle u_i p'\rangle/\p r_j = 0$. Instead, pressure enters implicitly via the coupling of~\eqref{KH} to higher-order correlators, i.e., the term containing $K(r)$. The analogue of~\eqref{KH} for triple correlations is
\begin{multline}
    \frac{\p }{\p t} \langle u_i u_j u'_k\rangle=  \frac{\p}{\p r_l}\langle u_i u_j u'_k u_l\rangle - \frac{\p}{\p r_l}\langle u_i u_j u'_k u'_l\rangle - \bigg\langle u_i u_j \frac{\p p'}{\p x'_k}\bigg\rangle - \bigg\langle u'_k\left( u_i \frac{\p p}{\p x_j}+u_j\frac{\p p}{\p x_i}\right)\bigg\rangle \\ + \mathrm{viscous\,\, terms}  +\langle u_i u_j f'_k \rangle+\langle (u_i f_j+f_i u_j) u'_k \rangle, \label{KH_for_K}
\end{multline}where the terms involving $p$ do not vanish. In particular, for our ensemble of randomly distributed eddies, the correlator $\langle u_i u_j \p_k' p'\rangle $ is $O(r^{-4})$ as $r\to\infty$, because the part of $p'=p(\bx+\br)$ that is correlated with the velocity field at position $\bx$ decays like $r^{-3}$. This suggests that
\begin{equation}
    K(r\to \infty) = O(r^{-4}). \label{Ksim-4}
\end{equation}

The above argument for the scaling~\eqref{Ksim-4} is informal: there is an obvious inconsistency in evaluating $\langle u_i u_j \p_k' p'\rangle $ by assuming that different patches of the turbulence are uncorrelated with the conclusion that correlations $K(r\to\infty) = O(r^{-4})$ must develop. However, the argument can be formalised---we prove in Appendix~\ref{BatchelorProof} that \eqref{Ksim-4} holds for real turbulence provided spatial correlations in the forcing decay sufficiently quickly (viz., exponentially) under suitable assumptions [this proof is a generalisation to forced turbulence of arguments presented by~\cite{BatchelorProudman56} and~\cite{Saffman67} for decaying turbulence]. Importantly, \eqref{Ksim-4} is too weak a correlation to permit $L\neq 0$: integrating~\eqref{KH} over all $\br$, we find
\begin{equation}
     \frac{\dd L}{\dd t} = 4\pi \lim_{r\to\infty}\left[\frac{1}{r} \frac{\p}{\p r} (r^4 u^3 K)\right] +2 \int \dd^3 \br \uf, \label{dL/dt}
 \end{equation}where the term involving $K$ vanishes for $K(r\to\infty) = O(r^{-4})$.
 
 \subsection{Correlations generated directly by the forcing\label{sec:correlations3}}
 
The argument in §\ref{sec:batchelor_argument} indicates that non-local interactions between fluid elements are too weak to allow the Saffman integral to change with time. Another concern is that correlations in the forcing function itself might decay sufficiently slowly with distance to permit development of $L\neq 0$; this effect is encoded in the second term on the right-hand side of \eqref{dL/dt}. We can estimate how slowly these correlations need to decay for $\dd L/\dd t$ to be non-zero by examining the formal solution for $\uf$ that is obtained by integrating the  Navier-Stokes equation in time:
\begin{align}
    \uf = \int^t_0 \dd s\, \bigg[ & \frac{\p}{\p r_j}\langle u_i(s) u_j(s) f'_i(t) \rangle - \frac{\p}{\p r_i} \langle p(s) f_i'(t) \rangle \nonumber \\ & + \nu \nabla^2 \langle \bu(s) \bcdot \bF'(t)\rangle + \langle \bF(s) \bcdot \bF'(t) \rangle\bigg].\label{uf_soln}
\end{align}Of the four terms in the right-hand side of \eqref{uf_soln}, the first and third give rise to surface terms in \eqref{dL/dt}, which vanish provided the relevant correlators fall off faster than $r^{-2}$ and $r^{-1}$, respectively.\footnote{A proof that they do, under the assumption that correlations in the forcing decay exponentially with distance, is presented in Appendix~\ref{BatchelorProof}.} The second term is identically zero by the solenoidality of $\bF$. The fourth term is a two-point, two-time correlation function of~$\bF$, which, because $\bF$ is a solenoidal, statistically isotropic vector field, satisfies [cf.~\eqref{uu(f)}]
\begin{equation}
    \int^t_0 \dd s\, \langle \bF(s) \bcdot \bF'(t)\rangle = \frac{1}{r^2}\frac{\p}{\p r} \big[r^3 H(t,r)\big]\label{ff'},
\end{equation}where $H(t,r)$ is the time-integrated longitudinal correlation function of $\bF$. From \eqref{dL/dt}, we find that the contribution of this term to the rate of change of the Saffman integral vanishes if 
\begin{equation}
    H(t,r)=o(r^{-3}),\label{h=O(r-3)}
\end{equation}which is unsurprising, given \eqref{fsimr-3}. 

The arguments presented in Sections~\ref{sec:batchelor_argument} and~\ref{sec:correlations3} indicate that, if the forcing function is solenoidal and sufficiently localised, then correlations between infinitely separated points that are strong enough to change the Saffman integral cannot arise in finite time, even accounting for the non-local nature of the pressure force.\footnote{The reader used to thinking of forcing whose properties are specified in spectral, rather than real, space, might wonder whether the condition of ``sufficient localisation'' is satisfied for the common choice of forcing in a finite spectral band. In Appendix~\ref{app:finite_band_forcing}, we show that a finite-band forcing with a smooth spectrum has very weak correlations at the largest scales (it decays faster with $r$ than any power law), as is intuitive, given the absence of energy in large-scale modes. If the spectrum of $\bF$ is not smooth, but instead has sharp discontinuities at the edges of the band, correlations are induced in $\uu$ that oscillate in $r$ at the wavenumbers of the edges, and decay in amplitude as $r^{-3}$. While these correlations may, in principle, propagate into all other correlators, we show in Appendix~\ref{app:finite_band_forcing} that any oscillatory component of $\uu$ always has a negligible effect on $\mcE(k)$ at small $k$, so these oscillatory correlations are of little dynamical significance.} Because the Saffman integral was zero at $t=0$, when $\bu = 0$, it remains zero at all times, and therefore it might appear that the system is forbidden from developing a $k^2$ spectrum at $k \to 0$.

\subsection{Long-range correlations as a cumulative effect of short-range interactions \label{sec:cumulative_correlations}}

How, then, does one explain the numerical evidence for a thermal-equilibrium $k^2$ spectrum in forced turbulence~\citep{Dallas15, Cameron17, AlexakisBiferale18, AlexakisBrachet19}? The answer is that the $k^2$ spectrum is established not by non-local processes (in real space), but by the cumulative effect of \emph{local} ones. Then, while infinitely separated points can never be strongly correlated enough to induce a $k^2$ spectrum, points separated by a large but finite distance can be (as long as one is prepared to wait long enough), leading to a $k^2$ spectrum that spans a finite, time-dependent range of wavenumbers. 

\begin{figure}
    \centering
    \includegraphics[width=0.9\textwidth]{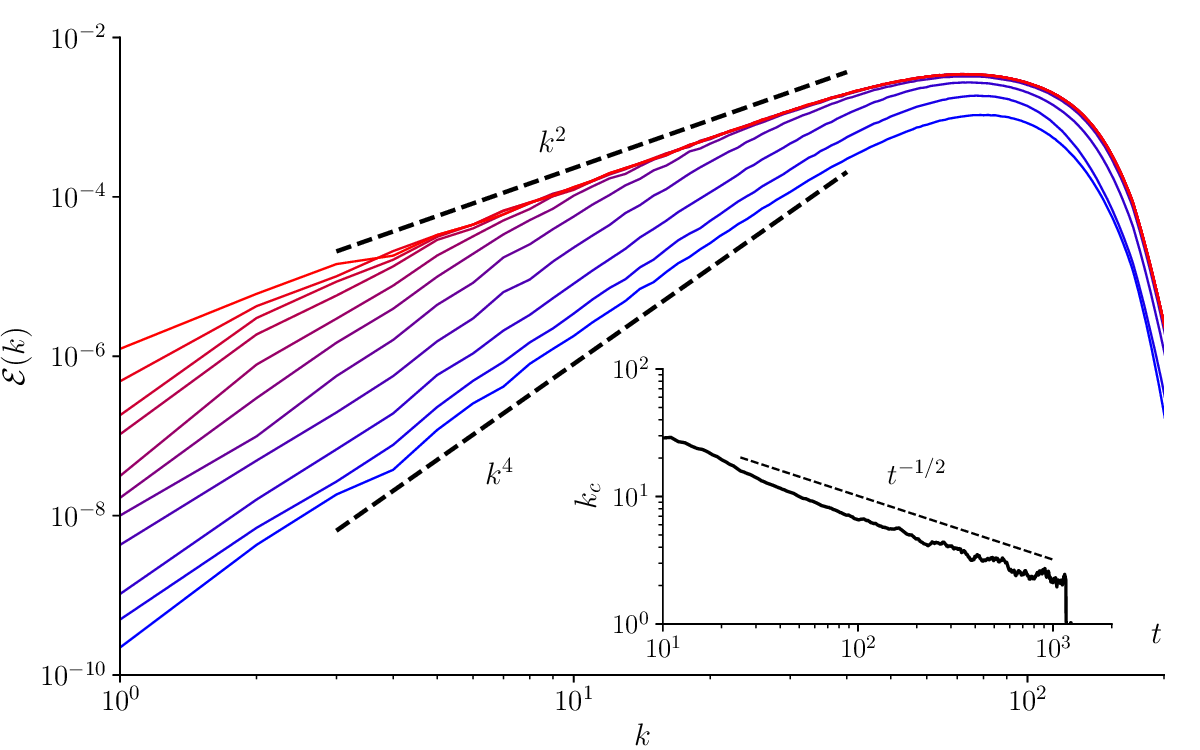}
    \caption{Saturation of the large scales in simulated Navier-Stokes turbulence forced by a delta-correlated, Gaussian random field with weak long-range spatial correlations [so that $F(k\to0\propto k^4)$, as explained in the text]. Displayed spectra are logarithmically spaced in time, with blue~$\to$~red indicating earlier~$\to$~later times. The inset shows the evolution of the knee wavenumber, $k_c(t)$, that separates the $\propto k^4$ and $\propto k^2$ parts of the spectrum. In the chosen units, the energy-injection rate is $0.7$, and the r.m.s. velocity is $\simeq 0.5$.}
    \label{fig:k2schematic}
\end{figure}

Let us now demonstrate that this is indeed the typical behaviour of forced turbulence, by means of a numerical simulation. We take the forcing function $\bF$ to be a solenoidal, Gaussian random field (as is a common choice for forced-turbulence studies), and to be delta-correlated in time, so the spectrum of energy injection is
\begin{equation}
    F(k) \equiv \frac{k^2}{2\pi^2} \int_0^t \dd s\int \dd^3 \br\, \langle \bF(t) \bcdot \bF'(s)\rangle e^{-i\bk \bcdot \br} \propto k^4 \exp(-k^2/k_p^2),
\end{equation}where the peak wavenumber $k_p=80$. Because the power injected into the $k=0$ mode is always zero, the average of the velocity (momentum) over the periodic box is zero for all~$t$. The large-scale $k^4$ tail of $F(k)$ is consistent with the generic spectral tail of an isotropic field with short spatial correlations\footnote{\label{artificial_forcing_footnote}An expansion of $F(k\to0)$ analogous to \eqref{Eexpansion} yields $F(k\to0)\propto k^4$ if $\langle \bF \bcdot \bF'\rangle$ decays rapidly with $r$. A faster decay of $F(k\to 0)$ would require the equivalent of the Loitsyansky integral~\eqref{Loitsyansky} for $\bF$, $I_{\bF}\equiv-\int^t_0\dd s \int \dd^3 \br r^2 \langle \bF(t) \bcdot \bF'(s)\rangle$, to be zero, which is an artificial situation.}, although this choice is not essential to observe the development of a $k^2$ band---other studies have used finite-band forcing~\citep{Dallas15, Cameron17, AlexakisBrachet19}. The algorithm that we employ to generate $\bF$ is described in Appendix~D of \cite{HoskingSchekochihin20decay}. With this choice, we solve the Navier-Stokes equations~\eqref{NavierStokes} in a periodic domain of size~$2\pi$ using the pseudo-spectral code Snoopy \citep{Lesur15} with $512^3$ resolution. We employ de-aliasing according to the $2/3$-rule, and use eighth-order hyper-dissipation, i.e., $\nu \nabla^2$ is replaced by $\nu_8 \nabla^8$ in \eqref{NavierStokes}, where $\nu_8 = 10^{-16}$. The use of hyper-dissipation ensures that the effect of viscosity on the development of the large-scale structure is negligible. The simulation time step $\Delta t$ is chosen automatically by the code so as to be sufficiently small to maintain the stability of the simulation according to the CFL criterion. The spectral scheme allows the viscous term to be solved exactly at each time step.

The results of this simulation are shown in figure~\ref{fig:k2schematic}. We observe that $\bu$ gradually develops a $k^2$ spectrum at large scales, with a spectral knee at a time-dependent wavenumber $k_c(t)$ separating the $\propto k^4$ and $\propto k^2$ parts, as anticipated. By fitting the numerical spectrum to a trial function of the form $k^2[1-\exp(-k^2/k_c(t)^2)]$, we find that $k_c(t)\propto t^{-1/2}$ (see inset to figure~\ref{fig:k2schematic}). At large enough times, the $k^2$ part of the spectrum extends all the way to the box size, which is the steady state (close to the box scale, i.e., say, at $k<4$, we observe some deviation from the $\propto k^2$ scaling at late times, which presumably is due to the absence of statistical isotropy at these scales). The ability of the system to reach a steady state hinges on the finite size of the simulation box---in an infinite system, $\mcE(k)\propto k^2$ would only ever be satisfied in an ever-broadening but finite band of wavenumbers.

We note that the turbulence in this simulation is not fully developed---we sacrifice the resolution of the $k^{-5/3}$ inertial range to facilitate resolving many forced structures, so that we may invoke statistical isotropy and homogeneity in our analysis, and also to allow a wide-band forcing function, so as to eliminate spurious effects that occur when forcing is restricted to a narrow spectral band (see Section 4.4). We do not expect that the development of the $k^2$ band is a consequence of our failure to resolve the inertial range, as the small-$k$ spectral asymptotic is determined by the statistical properties of outer-scale structures [via~\eqref{Eexpansion}]. This view is supported by the study of \cite{AlexakisBrachet19}, which presents simulations of a turbulence that appears closer to being fully developed than ours (achieved by forcing in a narrow spectral band) but still develops the thermal spectrum. We do not expect that the use or order of hyperdiffusion affects the process of thermalisation, for the same reason.

To summarise our progress so far, we have seen that the law of conservation of the Saffman integral, ported from the theory of decaying turbulence, also holds for forced turbulence, and that this law prohibits the thermal equilibration of arbitrarily large scales in finite time. Nonetheless, thermal equilibration up to a large but finite scale is not prohibited, and indeed this is the behaviour that turbulence whose forcing is spatially decorrelated tends to adopt (as is shown by figure~\ref{fig:k2schematic}). However, we still lack a physical mechanism for the development of the thermal spectrum. In the next section, and the one that follows it, we shall argue that this mechanism is turbulent diffusion of linear momentum.

\section{The large-scale spectrum and linear momentum \label{sec:momentum}}

Assuming the equivalence of volume and ensemble averages, the definition of the Saffman integral, \eqref{saffman_integral}, is equivalent to
\begin{equation}
    L = \lim_{V\to \infty}\frac{1}{V}\bigg\langle \left[\int_V \dd^3 \br \,\bu\right]^2 \bigg\rangle \equiv \lim_{V\to \infty} \frac{\blangle \bP_V^2 \brangle}{V}.\label{Saffman_momentum}
\end{equation}The Saffman integral, therefore, is a measure of how much linear momentum $\bP_V$ is contained in a large control volume $V$ \citep{Saffman67, Davidson15}. For instance, in Saffman turbulence, where each eddy in~$V$ has non-zero, but random, linear momentum, $\blangle \bP_V^2 \brangle \propto V$ (accumulating as a random walk), so $L$ is finite. If, instead, each eddy has vanishing total momentum, as in Batchelor turbulence, then~\eqref{Saffman_momentum} is dominated by the contributions of eddies at the surface of $V$. In that case, $\blangle \bP_V^2 \brangle \propto V^{2/3}$, and so $L=0$.

This idea immediately provides a physical explanation for why long-range correlations~\eqref{fsimr-3} are required for a finite $L$. Consider an isolated turbulent eddy in an otherwise quiescent fluid. The linear momentum contained in a large sphere $V$ of radius~$R$, centred on the eddy, is
\begin{equation}
    \bP_{\mathrm{eddy}}=\int_V \dd^3 \br \, \bu = \int_{\p V} \dd \boldsymbol{S} \times \bA,
\end{equation}where we have represented the solenoidal velocity field as $\bu = \bnabla \times \bA$. Clearly, $\bP_{\mathrm{eddy}}$ vanishes unless the average of $\bA$ over $\p V$ scales as $R^{-2}$ as $R\to\infty$, implying that the mean velocity on $\p V$ scales as $R^{-3}$. One can imagine building a synthetic $L\neq 0$ turbulence by superimposing such eddies with random positions and orientations; this velocity field will necessarily have long-range correlations, owing to the long tails of the component eddies.

In fact, there is a deep connection between the linear-momentum content of the turbulence and the large-scale spectrum, that goes beyond the finiteness of the Saffman integral and the Taylor expansion \eqref{Eexpansion}. \cite{Davidson15} has shown that, in incompressible, homogenenous and isotropic turbulence, $\langle \bP_V^2 \rangle$ is a functional of $\chi(r)$: if $V$ is a sphere of radius $R$,
\begin{equation}
    \langle \bP_V^2 \rangle= 4\pi^2 R^2 u^2 \int^{2R}_0 \dd r \,r^3 \chi(r) \left[1-\left(\frac{r}{2R}\right)^2 \right].
\end{equation}
It is convenient to integrate this formula by parts, which gives
\begin{equation}
    \langle \bP_V^2 \rangle = 2\pi^2 u^2 \int^{2R}_0 \dd r r \, \int_0^r \dd r' r'^3 \chi(r'),\label{P2_bp}
\end{equation}boundary terms having vanished exactly. From \eqref{P2_bp}, it is clear that $\langle \bP_V^2 \rangle \propto R^3$ only if $\chi(r'\to\infty)\propto r'^{-3}$. If, instead, $\chi(r')$ decays quickly, viz., $\chi(r'\to \infty)=o(r'^{-4})$, the $r'$ integral in \eqref{P2_bp} is dominated by small $r'$, and hence $\langle \bP_V^2 \rangle \propto R^2$, which is the scaling $\langle \bP_V^2 \rangle \propto V^{2/3}$ obtained above. Equation \eqref{P2_bp} is also readily inverted, to yield
\begin{equation}
    u^2 \chi(2R) = \frac{1}{128\,\pi^2} \frac{1}{R^3} \frac{\p}{\p R} \frac{1}{R} \frac{\p \langle \bP_V^2 \rangle}{\p R}.
\end{equation}Therefore, full knowledge of $\langle \bP_V^2 \rangle$ as a function of $R$ is equivalent to full knowledge of~$\chi(r)$, and hence, via \eqref{E(k)exact} and \eqref{uu(f)}, of the energy spectrum. This observation suggests that one might seek the explanation of the growth of the thermal spectrum in figure \ref{fig:k2schematic} as a consequence of the evolution of $\langle \bP_V^2 \rangle$.

\subsection{Broken-power-law spectra and their momentum content \label{sec:localk2}}

\begin{figure}
    \centering
    \includegraphics[width=0.8\textwidth]{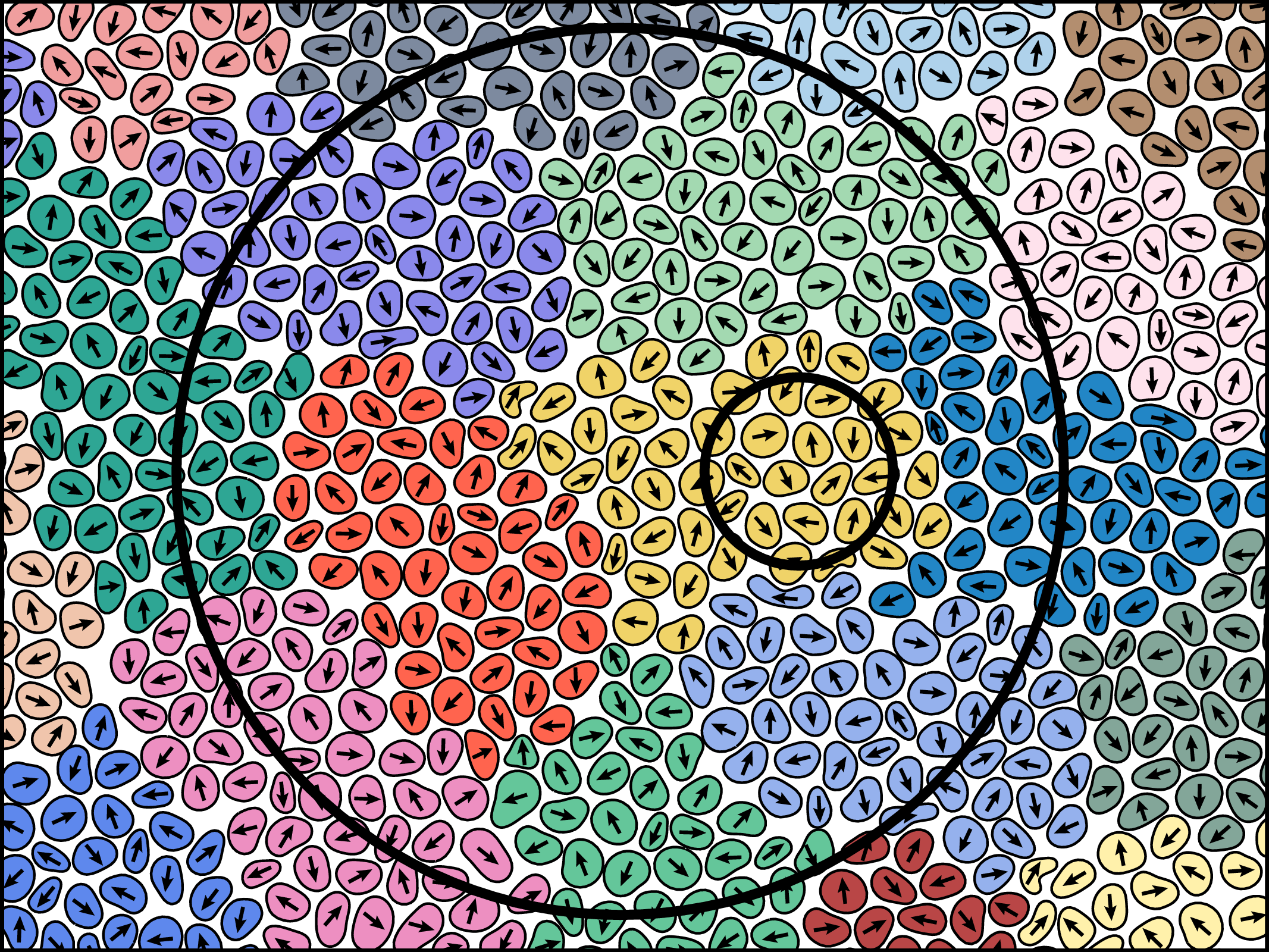}
    \caption{Schematic of a `quasi-random' distribution of linear momentum, i.e., one that would result in a broken-power-law spectrum, as in figure~\ref{fig:k2schematic}. Sufficiently large patches of turbulence have vanishing total momentum---a number of such patches (identified in a non-unique manner) are shown in different colours in the figure. For a control volume $V$ that is larger than the outer scale of the turbulence but smaller than the characteristic scale of the net-zero-momentum patches (e.g., the smaller circle in the figure), $\langle\bP^2\rangle \propto R^3$ because the eddies contained by $V$ (represented by individual blobs) have uncorrelated, random momenta (represented by arrows). On the other hand, $\langle\bP^2\rangle \propto R^2$ for $V$ much larger than the zero-net-momentum patches, because then only patches at the surface of $V$ contribute to the sum---in the figure, the central orange and yellow patches do not contribute to the total momentum contained within the larger circle.}
    \label{fig:quasirandom}
\end{figure}

The above discussion suggests that we might interpret the growth of a $k^2$ spectrum over a finite range of wavenumbers as indicating the development of random fluctuations in momentum that satisfy $\blangle \bP_V^2 \brangle \propto R^3$ over the corresponding range of scales. More precisely, these fluctuations would be quasi-random, in that the momenta of the eddies contained within $V$ would cancel more precisely when $R$ was large enough, so that $\blangle \bP_V^2 \brangle$ would be dominated by eddies at the surface of $V$, so that $\blangle \bP_V^2 \brangle\propto R^2$. Then, ${\mcE(k\to 0)\propto k^4}$. A schematic representation of the distribution of momentum of this ``quasi-Saffman turbulence'', similar to those presented by \cite{Davidson15} for the true Saffman and Batchelor turbulence, is shown in figure \ref{fig:quasirandom}.

Let us now check that these intuitive expectations hold up mathematically, i.e., that broken-power-law spectra do correspond to broken power laws in the dependence of $\langle\bP_V^2\rangle$ on $R$. From \eqref{uu(f)} and
\begin{equation}
    \uu = 2 \int^{\infty}_0 \dd k \mcE(k) \frac{\sin (kr)}{kr},\label{E(k)exact_inv}
\end{equation}which is the inverse of \eqref{E(k)exact}, it follows that
\begin{equation}
    u^2 \chi(r) = 2 \int^{\infty}_0\dd k\,  \mcE(k)\, \frac{\sin(kr)-kr\cos(kr)}{(kr)^3}.\label{f(E)_maintext}
\end{equation}In Appendix~\ref{app:asymptotics}, we present a formal asymptotic expansion of \eqref{f(E)_maintext}, assuming that ${\mcE(k)\propto k^a}$ for ${k_1 \leq k \leq k_2}$ with ${k_2 \gg k_1}$ [we remain agnostic about $\mcE(k)$ outside of this range]. This choice for ${\mcE(k)}$ models the broken-power-law spectrum shown in figure~\ref{fig:k2schematic}. We show that, for ${1/k_2 \ll r\ll 1/k_1}$,
\begin{equation}
    u^2 \chi(r) \simeq \begin{dcases*}
            \mathrm{constant} \sim \int^{k_1}_{0} \dd k \mcE(k) & if  $a<-1$; \\
                    \frac{1}{3} \frac{\ln(k_1 r)}{\ln (k_2/k_1)} \int^{k_2}_{k_1} \dd k \mcE(k) & if  $a=-1$; \\
                    \mathrm{undetermined,\,\,}\, \lesssim (k_2 r)^{-1-a} \int^{k_2}_{k_1} \dd k \mcE(k) & if  $a = 4,6,8\dots$;\\
                    -\Gamma(a-2)(a^2-1)\sin\left(\frac{a\pi}{2}\right) (k_2 r)^{-1-a}\int^{k_2}_{k_1} \dd k \mcE(k) & otherwise.
                 \end{dcases*}\label{asymptotic_result}
\end{equation} 

Let us explain qualitatively each case in turn. 

If $a<-1$, $\chi(r)\sim \const$, which is intuitive: the energy contained in the band $\{k_1,\,k_2\}$ is dominated by the largest structures, while we are looking at correlations on scales much smaller than them ($r\ll k_1^{-1}$). 

If $a=-1$, then every scale in the band $\{k_1,\,k_2\}$ contributes equally to the energy contained within it---this energy diverges in the limit $k_2/k_1\to \infty$, explaining the factor of $\ln (k_2/k_1)$ in the denominator of \eqref{asymptotic_result}. It turns out that the $r$ dependence of $\chi(r)$ is logarithmic in this case.

If $a = 4,6,8\dots$, then although $\chi(r)$ must decay faster than $r^{-1-a}$ in the range ${1/k_2 \ll r\ll 1/k_1}$, its behaviour is not uniquely determined by our assumption of a power-law scaling for $\mcE(k)$. This was to be expected, because even-power spectra are precisely the ones generated in the expansion~\eqref{Eexpansion}, and no specific strength of long-range correlations in the velocity field is required for the coefficients of $k^a$ with $a = 4,6,8\dots$ in this expansion to be non-zero (unlike for $a=2$).

Finally, for all other cases, including that of $a=2$, $\chi(r)$ decays like $r^{-1-a}$ for ${1/k_2 \ll r\ll 1/k_1}$. In particular, note that~\eqref{fsimr-3} may be recovered from~\eqref{asymptotic_result} for $a=2$, as ${\lim_{a\to2}\Gamma(a-2)\sin(a\pi/2)=-\pi/2}$. 

Our motivation in deriving \eqref{asymptotic_result} was to obtain the dependence of $\langle \bP^2_V\rangle$ on $R$ that is associated with a finite-extent large-scale power law in $\mcE(k)$. Let us consider scales $1/k_2 \ll R \ll 1/k_1$, where $k_2$ is now identified with the outer scale of the turbulence, i.e., $k_2\sim 1/l$, and $k_1$ is identified with the scale of the spectral knee $k_c$ in figure \eqref{fig:k2schematic}. Then, from~\eqref{P2_bp},
\begin{align}
    \langle \bP_V^2 \rangle & = 2\pi^2 u^2 \int^{2R}_0 \dd r r \,\left[\int_0^{X/k_2} \dd r' r'^3 \chi(r')+ \int_{X/k_2}^r \dd r' r'^3 \chi(r')\right],\label{P2_asymptotic}
\end{align}where $X$ is chosen so that $1 \ll X \ll k_2/k_1$, in which case \eqref{asymptotic_result} is applicable in the second integral appearing inside the brackets in \eqref{P2_asymptotic}. This integral dominates over the first one for $r\gg X/k_2$ as long as $r^3 \chi(r)\geq O(1/r)$, which, according to \eqref{asymptotic_result}, it will do if the spectrum follows a local power law with exponent $a \leq 3$. Otherwise, the first integral, which is independent of $r$, dominates. Thus, we have
\begin{align}
    \langle \bP_V^2 \rangle \propto \begin{dcases*}
            R^2 & if  $a>3$, \\
            R^2 \ln R & if  $a=3$, \\
            R^{5-a} & if  $-1<a<3$, \\
            R^{6}\ln R & if  $a=-1$, \\
            R^{6} & if  $a<-1$. \\
                 \end{dcases*}\label{P2_asymptotic2}
\end{align}We note that the classical scalings (see \citealt{Davidson15}) are readily recoverable from~\eqref{P2_asymptotic2}: the intuitive ``surface-term-dominated'' $\langle \bP_V^2 \rangle \propto R^2$ is recovered for steep spectral slopes, ${a>3}$, corresponding to weak long-range correlations, while the Saffman scaling ${\langle \bP_V^2 \rangle \propto R^3}$ is recovered for $a=2$. The scaling $\langle \bP_V^2 \rangle \propto R^6$ for $a<-1$ is also an intuitive one: such a spectrum is energetically dominated by structures with characteristic scale much larger than $R$, therefore control volumes $V\propto R^3$ will contain a total amount of momentum that is proportional to $V$. Though these results are familiar, \eqref{P2_asymptotic2} has the important new feature that it does not require the spectral power law to extend all the way to $k=0$---it is sufficient for $\mcE(k)\propto k^a$ only for $k_1 \leq k\leq k_2$, as long as we restrict attention to volumes with $1/k_2 \ll R\ll 1/k_1$.\footnote{\label{footnote_Loitsyansky}Aside from the generalisation of previous results to a finite-band power law in $\mcE(k)$, the other qualitatively new feature in \eqref{P2_asymptotic2} is that we have allowed for non-integer $a$. In this respect, \eqref{asymptotic_result} and \eqref{P2_asymptotic2} can be viewed as an extension of the results for $\chi(r\to\infty)\propto r^{-m}$ for integer $m$ derived by \cite{Davidson11}. While non-integer large-scale spectral power laws may be of limited applicability to real turbulence (though they can, of course, be manufactured numerically), they nonetheless have pedagogical value, particularly for $3<a<4$, when $\langle \bP_V^2 \rangle \propto R^2$, implying that arguments for the invariance of the large-scale spectrum in decaying turbulence that are based on momentum conservation (\citealt{Saffman67}, \citealt{Davidson11}; also see~§\ref{sec:decaying_connections}) do not apply. If it is true that the large-scale asymptotic of the energy spectrum is indeed invariant in decaying turbulence with $3<a<4$, then this must be a result of the conservation of some other quantity. The arguments presented in \cite{Davidson09, Davidson11} suggest that angular-momentum conservation, if it holds, would result in the invariance of this asymptotic; direct numerical simulations of turbulence with $3<a<4$ might therefore shed some light on the unsolved problem of angular-momentum conservation in turbulence in open domains. Interestingly, large-scale spectra with $3<a<4$ are not invariant under the EDQNM closure model, whereas those with $a<3$ are~\citep{EyinkThomson00, Lesieur05, Lesieur08}.}

\subsection{The development of ``quasi-random'' momentum fluctuations \label{sec:randomisation_mechanism}}

Having confirmed that broken-power-law spectra, of the form shown in figure \ref{fig:k2schematic}, do correspond to $\blangle \bP_V^2 \brangle \propto R^3$ over a finite range of scales, we now turn to the question of the physical mechanism that is responsible for the development of such a scaling. 

Intuitively, $\blangle \bP_V^2 \brangle \propto R^3$ can arise as a simple consequence of momentum transport by the flow. Consider a localised fluid motion that develops at $t=0$ as a result of the forcing. While the total linear momentum associated with this structure will be zero, its momentum density will be transported under the action of the flow (i.e., the sum of the eddy's own motion and that of the rest of the flow), and therefore will become distributed over an ever-increasing volume as time advances. When this volume becomes large compared to the control volume $V$ for which we are interested in computing the total square momentum, this structure will contribute to $\bP_V$ as a ``volume term'', rather than as a surface one. The occurence of this process at all points in space will then lead to a ``quasi-random'' momentum distribution, of the form depicted in figure~\ref{fig:quasirandom}.

\begin{figure}
    \centering
    \includegraphics[width=1.0\textwidth]{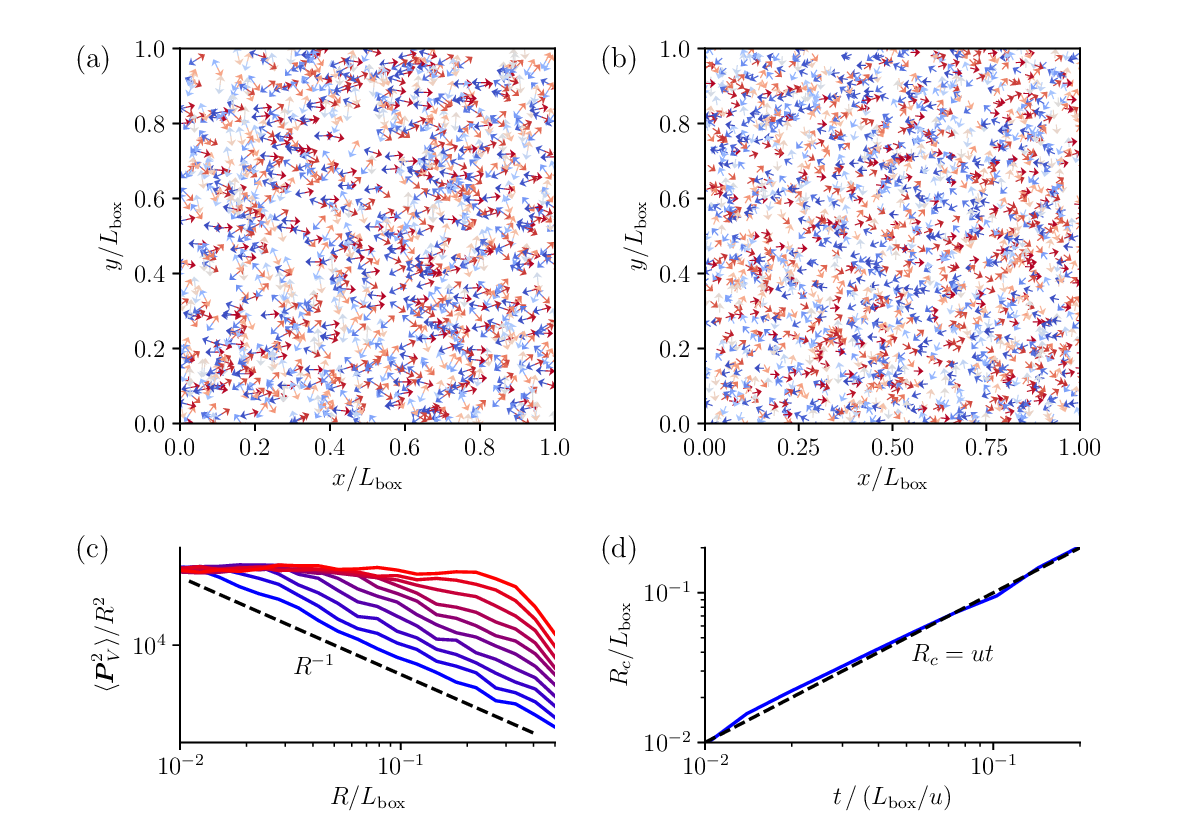}
    \caption{A toy model to illustrate quasi-randomisation of eddy momentum. Eddies are represented by pairs of particles that are initialised with equal and opposite momenta, but at the same position in space, shown as red and blue arrows in panel (a). Panel (b) shows the state of the system at $t = l_{\mathrm{box}}/5u$. Panel (c) shows the evolution of $\blangle \bP_V^2 \brangle$ with time (blue~=~early, red~=~late), demonstrating the development of the ``stochastic'' momentum scaling, $\blangle \bP_V^2 \brangle \propto R^2$, as explained in the text. Panel (d) shows that the position $R_c(t)$ of the ``knee'' in the scaling behaviour of $\blangle \bP_V^2 \brangle$, between $\propto R$ and $\propto R^2$, grows linearly with time, $R_c = ut$.}
    \label{fig:ballisticarrows}
\end{figure}

In figure~\ref{fig:ballisticarrows}, we present a simple toy model to illustrate this idea. In this model, turbulent eddies initialised by the forcing at $t=0$ are represented by pairs of particles. Each particle in the pair is initialised with the same random position in 2D space, though they have opposite momenta -- see panel (a). This means that at $t=0$, $\blangle \bP_V^2 \brangle \propto R$, because only ``eddies'' at the boundary of $V$ contribute, much as in a real forced turbulence that has just reached saturation at the forcing scale (note that, naturally, the surface-dominated and volume-dominated scalings are different in 2D). Subsequently, the particles move ballistically, i.e., without interacting, all at the same speed, $u$, but in random directions. At later times, the distribution of their momenta becomes quasi-random, in the sense described above. For $R \ll ut\equiv R_c(t)$, $\blangle \bP_V^2 \brangle \propto R^2$, because the control volume $V$ will only contain one particle from each pair. For $R\gtrsim R_c(t)$, $\blangle \bP_V^2 \brangle \propto R$, as the volume will contain both particles, whose contributions will cancel, unless they straddle the boundary.

While ballistic streaming is likely a poor model of the real motions of turbulent fluid structures, and while this toy model also neglects the effect of continuous forcing and energy dissipation, it captures the essential idea: transport of linear momentum means that highly ordered states where the total momentum of each flow structure vanishes cannot be maintained. Instead, it seems inevitable that the distrubution of momentum will become quasi-random, i.e., that $\blangle \bP_V^2 \brangle \propto R^3$ (in 3D) up to a finite $R=R_c(t)$, which will grow with time.

Intuitively, ballistic streaming represents the upper bound permitted by causality on the rate at which the distribution of momentum can become stochasticised in the absence of significant non-local interactions. In Appendix~\ref{momentum_evolution_appendix}, we show how this causal bound, $R_c \sim ut$, can be recovered from the Navier-Stokes equation directly. In real turbulence, however, momentum is transported chaotically, rather than ballistically, so we should expect that $R_c \ll ut$. Owing to the long non-linear timescale associated with interactions between structures on the largest scales, it is reasonable to suppose that ``momentum density'' is transported \emph{passively} by the turbulent diffusivity of the flow, at least insofar as the large scales are concerned. In that case, we expect a diffusive scaling $R_c \sim 1/k_c \propto t^{1/2}$, rather than the ballistic one, $R_c \propto t$. As we shall see in~§\ref{sec:passive}, the diffusive scaling is indeed in excellent agreement with direct numerical simulations.

\subsection{\texorpdfstring{$R_c \propto t^{1/2}$}{Diffusive scalings} due to linear growth of the Loitsyansky integral \label{sec:dodgyderiv}}

Before we explore this topic further, however, let us pause to consider a tempting, if dangerous, argument that would appear to guarantee the scaling $R_c \sim 1/k_c \propto t^{1/2}$ without any further assumptions.

Let us accept, on the basis of the intuitive momentum-stochasticisation argument of~§\ref{sec:randomisation_mechanism}, that a $k^2$ spectrum will develop in a limited range of $k$, as depicted in figure \ref{fig:k2schematic}.

It follows from integrating \eqref{KH} over $r$ that the Loitsyansky integral \eqref{Loitsyansky} grows according to
\begin{equation}
    \frac{\dd I}{\dd t} = 8\pi u^3 \lim_{r\to\infty} \left[ r^4 K(r)  \right] - 12 \nu L - 2 \int \dd ^3 \br\, r^2 \uf. \label{dI/dt}
\end{equation}Because the system always has a vanishing Saffman integral, the second term on the right-hand side of \eqref{dI/dt} is zero. For simplicity, let us assume that the forcing has a short correlation time, in which case~\eqref{dI/dt} becomes, after substitution of~\eqref{uf_soln} and~\eqref{ff'},
\begin{equation}
    \frac{\dd I}{\dd t} = 8\pi u^3 \lim_{r\to\infty} \left[ r^4 K(r)  \right] + 8\pi \int \dd r\, r^4 H(r)  \label{dI/dt_2}.
\end{equation}

It is often conjectured that $I$ is conserved by an isotropic turbulence decaying from an initial state with a $k^4$ spectrum \citep{Kolmogorov41c, LandauLifshitzFluids, Davidson15, Ishida06}. While this point is not universally accepted (e.g., $I$ is not conserved under the popular EDQNM closure; see \citealt{Lesieur08} and references therein), the evidence from direct numerical simulations appears to support the conservation of $I$, at least after an initial transient period \citep{Ishida06}. As may be seen from \eqref{dI/dt_2}, the invariance of $I$ in the absence of forcing requires that $K(r\to\infty)=o(r^{-4})$, i.e., long-range triple correlations must decay faster than the $K (r\to\infty) = O(r^{-4})$ that follows from considering long-range pressure-mediated interactions (\citealt{BatchelorProudman56,Davidson15}; see our §\ref{sec:batchelor_argument}). Supposing that a state with $K(r\to\infty)=o(r^{-4})$ can also arise in forced turbulence, \eqref{dI/dt_2} implies linear growth of $I$, whence, by  \eqref{Eexpansion},
\begin{equation}
    \mcE (k \to 0) \propto k^4 t. \label{I_propto_t}
\end{equation}From our expectation that the system will saturate with a spectrum $\mcE (k) \propto k^2$, the wavenumber $k_c$ that has just saturated at time $t$ satisfies
\begin{equation}
    k_c^4 t \propto k_c^2 \implies k_c \propto t^{-1/2},\label{dodgyderiv}
\end{equation}which is precisely the diffusive scaling suggested above.

However, this argument should be treated with caution, because it seems unlikely that $K(r\to\infty)=o(r^{-4})$ could be realised in forced turbulence. Numerical evidence suggests that this condition is only satisfied in decaying turbulence after an initial transient period \citep{Ishida06}, during which the system loses memory of the initial conditions. Prior to this, growth of $I$ is observed, which requires $K(r\to\infty)=O(r^{-4})$. Forced turbulence, of course, is essentially always in this `transient' regime, as the system never loses memory of the statistical properties of the forcing. Indeed, it is clear that $K(r\to\infty)=O(r^{-4})$ from the fact that $\mcE(k)\propto k^4$ does develop at the largest scales in turbulence forced with $I_{\bF}=0$, which is the case, e.g., for forcing in a finite spectral band (see §\ref{sec:simulations}). 

On the other hand, the diffusive scaling \eqref{dodgyderiv} may still be obtained if ${\lim_{r\to\infty} r^4 K(r)}$ is constant in time. Admittedly, it is not \emph{a priori} clear that this should be the case, because the value of this limit can change as a result of the long-range, pressure-mediated interactions between eddies, whose statistical properties do, after all, change with time as a result of the stochasticisation of linear momentum. Nonetheless, in the next section, we show that a passive model of the large-scale dynamics reproduces the linear growth~\eqref{I_propto_t} of $I$, indicating that ${\lim_{r\to\infty} r^4 K(r)}=\const$ may be a reasonable approximation in real forced turbulence.

\section{A solvable model of passive momentum diffusion \label{sec:passive}}

In §\ref{sec:momentum}, we argued that the development of a thermal $k^2$ spectrum over a finite, but growing, large-scale band is a consequence of the quasi-randomisation of the linear momentum distribution. In this section, we consider a model of this process in which the momentum density is a \emph{passive} quantity, in which case its randomisation can be understood as a consequence of turbulent diffusion. 

To motivate the model, let us consider the evolution of a velocity field $\bw$ under the Navier-Stokes equations. Let $\bw = \ol{\bw} + \wt{\bw}$, where $\ol{\bw}$ is the large-scale part of $\bw$, formally defined as the result of applying a Fourier-space filter to $\bw$ to isolate only those modes with $k<K$, for some $K$ much smaller than the characteristic wavenumber of the forcing, while $\wt{\bw}$ is the remaining smaller-scale part, consisting of modes with $k>K$. Then the evolution of $\bw$ proceeds according to
\begin{equation}
    \frac{\p \bw}{\p t} + \mathcal{P}\big[\ol{\bw}\bcdot \bnabla \ol{\bw} + \ol{\bw}\bcdot \bnabla \wt{\bw} + \wt{\bw}\bcdot \bnabla \ol{\bw} + \wt{\bw}\bcdot \bnabla \wt{\bw}\big] = \nu \nabla^2 \bw + \bF,
\end{equation}where $\mathcal{P}$ is the Fourier-space projection operator that returns the solenoidal part of the field on which it operates: $[\mathcal{P}\bw]_i \equiv (\delta_{ij}-k_ik_j/k^2)w_j$. Let us assume that, because the large-scale modes are energetically subdominant to the rest of the flow, advection by them is unimportant. Then we are left with
\begin{equation}
    \frac{\p \bw}{\p t} + \mathcal{P}\big[\wt{\bw}\bcdot \bnabla \bw\big] = \nu \nabla^2 \bw + \bF,\label{p2}
\end{equation}so the only important non-linearity is advection by the small-scale part of $\bw$. The small-scale part of~\eqref{p2} is
\begin{equation}
    \frac{\p \wt{\bw}}{\p t} + \mathcal{P}\big[\wt{\wt{\bw}\bcdot \bnabla \wt{\bw}} + \wt{\wt{\bw}\bcdot \bnabla \ol{\bw}}\big] = \nu \nabla^2 \wt{\bw} + \wt{\bF}.
\end{equation}Again, owing to the energetic subdominance of $\ol{\bw}$ (and its small gradients), let us assume that the term involving $\ol{\bw}$ is negligible compared to the other term inside the square brackets, so
\begin{equation}
    \frac{\p \wt{\bw}}{\p t} + \mathcal{P}\big[\wt{\wt{\bw}\bcdot \bnabla \wt{\bw}} \big] = \nu \nabla^2 \wt{\bw} + \wt{\bF}.\label{smsceq}
\end{equation}Equation \eqref{smsceq} shows that, under the approximations outlined so far, the evolution of $\wt{\bw}$ is entirely decoupled from that of $\ol{\bw}$. Taking the large-scale part of \eqref{p2}, we find that $\ol{\bw}$ satisfies
\begin{equation}
    \frac{\p \ol{\bw}}{\p t} + \mathcal{P}\big[\,\ol{\wt{\bw}\bcdot \bnabla \wt{\bw}} + \ol{\wt{\bw}\bcdot \bnabla \ol{\bw}} \,\big] = \nu \nabla^2 \ol{\bw} + \ol{\bF}.\label{lasceq}
\end{equation}In a sense, therefore, $\ol{\bw}$ is a passive field: although~\eqref{lasceq} shows that its evolution is affected by $\wt{\bw}$, $\wt{\bw}$ is not affected by $\ol{\bw}$, according to~\eqref{smsceq}. 

Motivated by this property, we propose to replace $\wt{\bw}$ with an artificial field, $\bu$, wherever the former appears as an advecting field in \eqref{smsceq} and \eqref{lasceq}. This model can be summarised by
\begin{equation}
    \frac{\p \bw}{\p t} + \mathcal{P}\big[\bu\bcdot \bnabla \bw\big] = \nu \nabla^2 \bw + \bF.\label{passiveNS}
\end{equation}This equation is sometimes called the ``linear pressure model'' of the Navier-Stokes equation; a number of its properties have been studied by~\cite{Benzi01}, \cite{Adzhemyan01b, Adzhemyan01a}, \cite{Antonov03}, and~\cite{Arponen09}. Physically, \eqref{passiveNS} describes ``eddies'' of the field $\bw$ interacting nonlinearly with eddies of the field $\bu$, rather than other $\bw$-eddies. The $\bw$-eddies can receive momentum from their interaction with the $\bu$-eddies, whereas the latter do not get anything back, as their motion is externally prescribed. Nonetheless, the receipt of momentum by $\bw$-eddies still occurs in a way that locally conserves momentum, because $\int \dd^3\bx\,\bw$ is an invariant of \eqref{passiveNS}. Ultimately, then, the $\bw$-eddies do have local interactions that satisfy net-momentum conservation, which is the key ingredient for the stochasticisation of their momentum distribution. If the field $\bu$ is chosen so that its statistical properties are close to those of real turbulence, it may be hoped that the evolution of $\ol{\bw}$ should mimic that of the large-scale part of a real velocity field (the same need not be true of $\wt{\bw}$, though see~\citealt{Benzi01} for some similarities in small-scale properties).

In §\ref{sec:passive_theory}, we shall present an analytic treatment of \eqref{passiveNS}, taking $\bu$ to be the so-called Kraichnan ensemble~\citep{Kraichnan65, Kraichnan94}. First, however, we present numerical simulations to demonstrate the validity of the model~\eqref{passiveNS}.

\subsection{Assessing the passive-velocity model in simulated turbulence\label{sec:simulations}}

\begin{figure}
    \centering
    \includegraphics[width=1.0\textwidth]{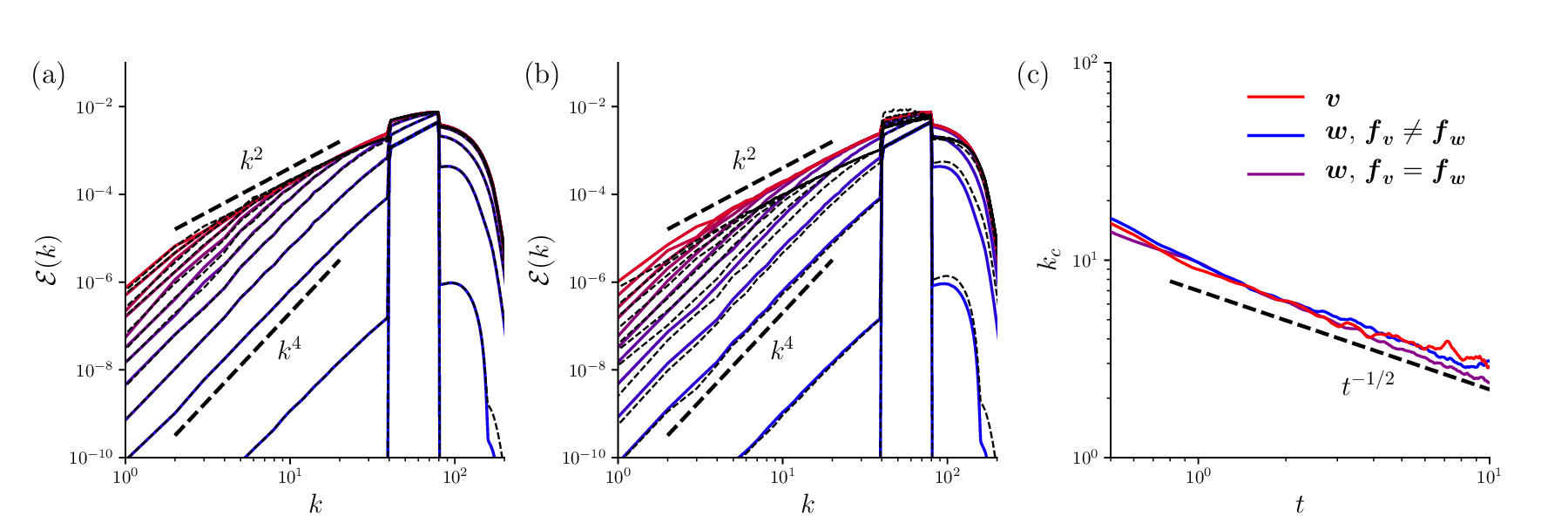}
    \caption{Development of the ``thermal'' $k^2$ spectrum by a Navier-Stokes velocity field, $\bv$, described by~\eqref{forcedNS}, and a ``passive velocity field'', $\bw$,  described by \eqref{passive_velocity_eqn}. Panel (a) shows the case where $\bw$ and $\bv$ are forced by the same function, $\bF_{\bv}=\bF_{\bw}$, while panel (b) shows the case where $\bw$ and $\bv$ are forced independently. Spectra of $\bw$, $\mcE_{\bw}(k)$, are plotted with dashed black lines, while spectra of $\bv$, $\mcE_{\bv}(k)$, are plotted with solid coloured lines: blue~$\to$~red indicates earlier~$\to$~later times.  Panel (c) shows the evolution of the knee wavenumber $k_c(t)$ between the $k^4$ and $k^2$ parts of the spectrum. In the chosen units, the energy injection rate into each of $\bv$ and $\bw$ is $2.5$, and the r.m.s. values of all velocity fields are $\simeq 1.0$. 
    \label{fig:passive_justification}}
\end{figure}

In figure \ref{fig:passive_justification}, we present results of simulations with both ``active'' and ``passive'' velocity fields. Specifically, we plot the evolution of the energy spectra of the fields $\bv$ and $\bw$, denoted $\mcE_{\bv}(k)$ and $\mcE_{\bw}(k)$, respectively, where $\bv$ is determined by the forced Navier-Stokes equation,
\begin{equation}
    \frac{\p \bv}{\p t} + \bv\bcdot \bnabla \bv  = - \bnabla p_{\bv} + \nu \nabla^2 \bv + \bF_{\bv},\label{forcedNS}
\end{equation}while $\bw$ is governed by the passive-velocity equation~\eqref{passiveNS}, viz.,
\begin{equation}
    \frac{\p \bw}{\p t} + \bu\bcdot \bnabla \bw  = - \bnabla p_{\bw} + \nu \nabla^2 \bw + \bF_{\bw}\label{passive_velocity_eqn},
\end{equation}where~$\bu = \mathcal{P}_K \bv$, $\mathcal{P}_K$ is the truncation operator that removes all Fourier modes with $k<K=40$, and $\bF_{\bv}$ and $\bF_{\bw}$ are forcing functions that are delta-correlated in time and inject an equal amount of energy into each Fourier mode in the band $40<k<80$ at every timestep (the box size is $2\pi$). The other details of the simulations are as described in §\ref{sec:correlations3}.

Panel (a) of figure~\ref{fig:passive_justification} shows the results of a simulation where $\bF_{\bv}=\bF_{\bw}$. The only difference between $\bv$ and $\bw$ in this case is that $\bw$ evolves without being advected by the modes in the large-scale tail of the spectrum of $\bv$. We see that both $\bv$ and $\bw$ develop $k^2$ spectra at large scales gradually, with a spectral knee separating the $\propto k^4$ and $\propto k^2$ parts, as we anticipated in figure \ref{fig:k2schematic}. The spectra $\mcE_{\bw}(k)$ and $\mcE_{\bv}(k)$ are almost the same at all scales and at all times. This finding demonstrates that the role of the large-scale structure of the turbulence in advecting itself and the small-scale flow is of negligible importance to the development of the thermal spectrum. 

In panel (b), we plot the same spectra for a simulation where $\bF_{\bv}$ and $\bF_{\bw}$ are independent random variables. In this case, the small-scale field that advects $\bw$ resembles $\bw$'s small-scale part only in a statistical sense. Nonetheless, $\bw$ develops a $k^2$ band at roughly the same rate as $\bv$. We view this finding as numerical justification of the passive-velocity model.

Finally, panel (c) shows the wavenumbers of the spectral knees in $\mcE_{\bv}(k)$ and $\mcE_{\bw}(k)$ as functions of time, computed by fitting a trial function of the form $k^2[1-\exp({-k^2/k_c^2})]$ to the large-scale tail of the spectra. In all three cases, $k_c \propto t^{1/2}$, which is the diffusive scaling anticipated in §\ref{sec:randomisation_mechanism}. In the next section, we shall derive this result (including the functional form of the knee) analytically from the passive-velocity equation~\eqref{passiveNS}.

\subsection{Advection by a Kraichnan flow \label{sec:passive_theory}}

In this section, we shall compute $\mcE_{\bw}(k, t)$ from \eqref{passiveNS} analytically. The price we pay to do so is the need to make modelling assumptions about the advecting velocity field~$\bu$ --- specifically, we take both $\bu$ and $\bF_{\bw}$ to be delta-correlated in time. Under this assumption, the $\bk$-space correlation function of $\bu$ is
\begin{align}
    \blangle u_i(t, \bk) u_j (t', \bk')\brangle & = (2\pi)^3  \delta (t-t') \delta (\bk + \bk') \kappa_{ij}(\bk), \label{uu_kraichnan_mt}
\end{align}where the appearance of~$\delta (\bk + \bk')$ in this expression is a consequence of statistical homogeneity. A similar expression is adopted for $\bF_{\bw}$ (see Appendix~\ref{app:passive}). Together, incompressibility and isotropy further imply that
\begin{align}
    \kappa_{ij}(\bk) & = \kappa (k) \mcP_{ij}(\bk),
\end{align} where $\mcP_{ij}(\bk) = \delta_{ij} - k_i k_j/k^2$ is the usual $\bk$-space projection operator. A synthetic velocity field satisfying \eqref{uu_kraichnan_mt} is often called the (incompressible) Kraichnan ensemble, after \cite{Kraichnan68, Kraichnan94}, who proposed it as a model for studying the behaviour of a passive scalar advected by a turbulent flow. The same model was used independently by \cite{Kazantsev68} to study the growth of magnetic fields via the turbulent dynamo effect. In both of these applications, the model gave rise to a lively analytical following (see reviews by~\citealt{Falkovich01} and~\citealt{Rincon19}). The inertial-range statistics of the passive-velocity equation~\eqref{passiveNS} with $\bu$ the Kraichnan ensemble have also been studied in detail by~\cite{Benzi01}, \cite{Adzhemyan01b, Adzhemyan01a}, and~\cite{Arponen09}. The short-correlation-time approximation is a natural one for our problem, because the timescale on which large-scale structures diffuse is much longer than the correlation time of the outer-scale turbulence.

Finally, we assume further that
\begin{equation}
    \kappa (k) = \kappa_0 \delta (k-k_f),\label{kappa}
\end{equation}i.e., that the advecting field has a single wavenumber, $k_f$. While this assumption is not strictly required to produce a closed set of equations, it nonetheless greatly simplifies the calculation, and is not particularly limiting considering the simplifications already adopted. It should be noted that~\eqref{kappa} does not restrict the applicability of the model to turbulence that is forced at a single scale, as $\bF_{\bw}$ can still be multi-scale. We also note that there is little to be gained by choosing $\bu$ to have large-scale structure, because in any situation in which the large-scale structure of the advecting flow is important, the passive model of momentum diffusion will not be appropriate anyway.

We show in Appendix~\ref{app:passive} that under these assumptions, the spectrum $\mcE_{\bw}(k)$ of the passive velocity field $\bw$ satisfies the following mode-coupling equation 
\begin{align}
      & \p_t \mcE_{\bw}(k) + 2 \left[\nu + \nu_T(k)\right] k^2 \mcE_{\bw}(k) = \frac{\kappa_0 \kflow}{ (2\pi)^2} k \int^{k+\kflow}_{|k-\kflow|} \frac{\dd k' }{k'}   K(k',k) \mcE_{\bw} ( k') + F_{\bw}(k),\label{mode_coupling_equation_kraichnan_mt}
\end{align}where the spectrum of energy injection is
\begin{equation}
    F_{\bw}(k) = \frac{k^2}{2\pi^2} \int_0^t \dd s\int \dd^3 \br\, \langle \bF_{\bw}(t) \bcdot \bF'_{\bw}(s)\rangle e^{-i\bk \bcdot \br}.
\end{equation}The turbulent viscosity $\nu_T(k)$ and the kernel $K(k', k)$ that appears in the mode-coupling integral are both unwieldy functions whose precise forms are given in Appendix~\ref{app:passive}. However, because our interest is in wavenumbers~$k \ll k_f$, we only need the small-$k$ part of \eqref{mode_coupling_equation_kraichnan_mt}, and thus only the small-$k$ limits of~$\nu_T(k)$ and~$K(k', k)$. These are
\begin{equation}
    \lim_{k\to 0}\nu_T(k)=\frac{\kappa_0 k_f^2}{10\pi^2}, \quad \lim_{k,\,q\,\to 0}\left[\frac{k_f k}{k_f + q}K(k'=k_f+q,k)\right]=\frac{k^4 - q^4 }{2 k}.\label{asy_limits}
\end{equation}
Substituting~\eqref{asy_limits} into  \eqref{mode_coupling_equation_kraichnan_mt} yields, for $k \ll k_f$,
\begin{align}
      & \p_t \mcE_{\bw}(k) + \beta k_f^2 k^2 \mcE_{\bw}(k) = \frac{5}{8} \frac{\beta}{k} \int^{k}_{-k} \dd q (k^4 - q^4) \mcE_{\bw} ( k_f+q) + F_{\bw}(k),\label{penultimate}
\end{align}where we have defined $\beta = \kappa_0 / 5\pi^2$ and assumed that the turbulent viscosity dominates over the molecular one. Finally, if $k$ is small compared to the wavenumber scale on which $\mcE_{\bw}(k)$ varies in the vicinity of $k_f$, then we may take $\mcE_{\bw}(k_f+q)\simeq \mcE_{\bw}(k_f)$ in~\eqref{penultimate} (we shall consider the effect of relaxing this assumption in §\ref{sec:narrow_band}). Then~\eqref{penultimate} becomes
\begin{align}
      \p_t \mcE_{\bw}(k) + \beta \kflow^2 k^2 \mcE_{\bw}(k) = \beta k^4 \mcE_{\bw} (\kflow) + Ck^b, \label{large_scale_equation}
\end{align}where we have replaced $F_{\bw}(k)$ by its small-$k$ asymptotic form, taken to be a power law with exponent $b$ (note that the case of finite-band forcing may be recovered by setting $C=0$ in what follows).

Equation~\eqref{large_scale_equation} is coupled to the forcing-scale modes via the appearance of $\mcE_{\bw}(k_f)$. Therefore, in order to calculate the growth of $\mcE_{\bw}(k\ll k_f)$ from an initial state with $\mcE_{\bw}(k)=0$, we should, strictly speaking, compute the evolution of $\mcE_{\bw}(k_f)$ from \eqref{mode_coupling_equation_kraichnan_mt} and substitute the result into \eqref{large_scale_equation}. However, we expect the spectrum to saturate much more quickly at the forcing scale than at $k\to 0$, so we may, with negligible error, take $\mcE_{\bw}(k_f)$ to be equal to its saturated value at all times. Then, solving~\eqref{large_scale_equation} subject to $\mcE_{\bw}(t=0,k)=0$ gives
\begin{equation}
    \mcE_{\bw}(k) = \left(1-e^{-\beta k^2 k_f^2 t}\right) \frac{C k^b + \beta \mcE_{\bw}(k_f)k^4}{\beta k^2 k_f^2}.\label{general_timedepsoln}
\end{equation}As anticipated, \eqref{general_timedepsoln} exhibits a split power law. For any value of $b$, the critical wavenumber demarcating the two regimes is
\begin{equation}
    k_c \sim \frac{1}{\sqrt{\beta k_f^2 t}} \sim \frac{1}{l} \sqrt{\frac{\tnl}{t}},\label{kc_passive}
\end{equation}where $l\sim k_f^{-1}$ and $\tnl\sim 1/\beta k_f^4$ is the characteristic nonlinear advection time at the injection scale. This is the diffusive scaling for the spectral knee anticipated at the end of~§\ref{sec:randomisation_mechanism}.

For $k\ll k_c$, \eqref{general_timedepsoln} reduces to
\begin{equation}
    \mcE_{\bw}(k) = \left[C k^b + \beta \mcE_{\bw}(k_f)k^4\right]t.\label{general_tll}
\end{equation}Therefore, at small enough $k$ (or early enough times), $\mcE_{\bw}(k)$ has a $k^b$ power law if $b\leq 4$, or a $k^4$ power law if $b>4$. In the case of solenoidal forcing that is local in real space, which has been our focus so far, $b=4$, so $\mcE_{\bw}(k\to 0)\propto k^4$, consistent with the numerical results presented in figures~\ref{fig:k2schematic} and~\ref{fig:passive_justification}. The development of a $k^4$ spectrum in the case of~$b>4$ or~$C=0$ reflects the fact that turbulence with zero Loitsyansky integral is unsustainable: even if the forcing has $I_{\bF}=0$, the flow will develop $I\neq 0$ on a dynamical timescale, owing to interactions between eddies [cf.~\eqref{dI/dt_2}]. We note that the linear dependence of the right-hand side of~\eqref{general_tll} on $t$ indicates that our passive model of momentum diffusion corresponds to real turbulence with ${\lim_{r\to\infty} r^4 K(r)}$ constant in time [see §\ref{sec:dodgyderiv}].

In the opposite limit, $k\gg k_c$, \eqref{general_timedepsoln} becomes
\begin{equation}
    \mcE_{\bw}(k) = \frac{C k^b + \beta \mcE_{\bw}(k_f)k^4}{\beta k^2 k_f^2},\label{general_tgg}
\end{equation}so $\mcE_{\bw}(k)\propto k^2$ if $b \geq 4$. Thus, we recover the expected thermal spectrum for $b\geq 4$, i.e., for real-space correlations in the forcing function that satisfy ${H_{\bw}(r\to\infty)\leq O(r^{-5})}$ (where $H_{\bw}$ is the analogue of $H$ for $\bF_{\bw}$).\footnote{This statement follows from a calculation directly analogous to the one that showed that $\mcE(k\to0)\propto k^a \iff \chi(r\to\infty)\leq O (r^{5})$ for $a\geq 4$; see \eqref{asymptotic_result}.} As explained in §\ref{sec:localk2}, this corresponds to the development of a quasi-random momentum distribution, ${\langle \bP_V^2\rangle\propto R^3}$. If long-range correlations in the forcing are stronger, i.e., $b<4$, then the thermal spectrum is not realised: instead, a shallower $k^{b-2}$ spectrum develops. 

According to \eqref{P2_asymptotic2},~\eqref{general_tll} and~\eqref{general_tgg} correspond to
\begin{equation}
    \langle \bP^2_V \rangle \propto \begin{dcases*}
            R^{7-b} & if  $l\ll R\ll R_c$, \\
            R^{5-b} & if  $R \gg R_c$ \& $b<3$, \\
            R^{2} & if  $R \gg R_c$ \& $3<b<4$, \\
            \end{dcases*}  \label{anomalousP2}
\end{equation}where $R_c \equiv k_c^{-1}$. The $R\gg R_c$ scalings are easily interpreted: for large volumes for which there has not been enough time for momentum diffusion to act, the turbulence inherits the momentum scaling dictated by the forcing. The $\langle \bP^2_V \rangle \propto R^{7-b}$ scaling in the range $l\ll R\ll R_c$ is less intuitive. Interestingly, different values of $b$ in the range $3<b<4$ tend to saturate with different power laws, even though all $\mcE_{\bw}\propto k^a$ spectra with $3<a<4$ have~$\langle \bP^2_V \rangle \propto R^2$, as~\eqref{P2_asymptotic2} shows. To understand the origins of these scalings, it is convenient to consider the momentum-diffusion process as the net result of a series of instances of a decaying passive vector field. While the characteristic scale of the diffusing momentum grows like~$t^{1/2}$ in all cases, the energy of the diffusing field decays at a rate that depends on the exponent~$b$, and therefore the contribution of each instance of forcing to $\langle \bP^2_V \rangle \propto R^2$ depends on~$b$, even when~$3<b<4$. In Appendix~\ref{app:anomalousP2}, we show how the scalings~\eqref{anomalousP2} can be derived directly by thinking about the diffusion of momentum in such terms.

\subsection{Local, non-solenoidal forcing \label{sec:nonsolenoidal}}

While forcing with $b\leq 4$ is easy to implement in numerical simulations, where complete control of the forcing spectrum is possible, such forcing is artificial in the sense that $b<4$ corresponds to long-range correlations that decay with distance in a very particular way [see~\eqref{asymptotic_result}]. 
\begin{figure}
    \centering
    \includegraphics[width=1.0\textwidth]{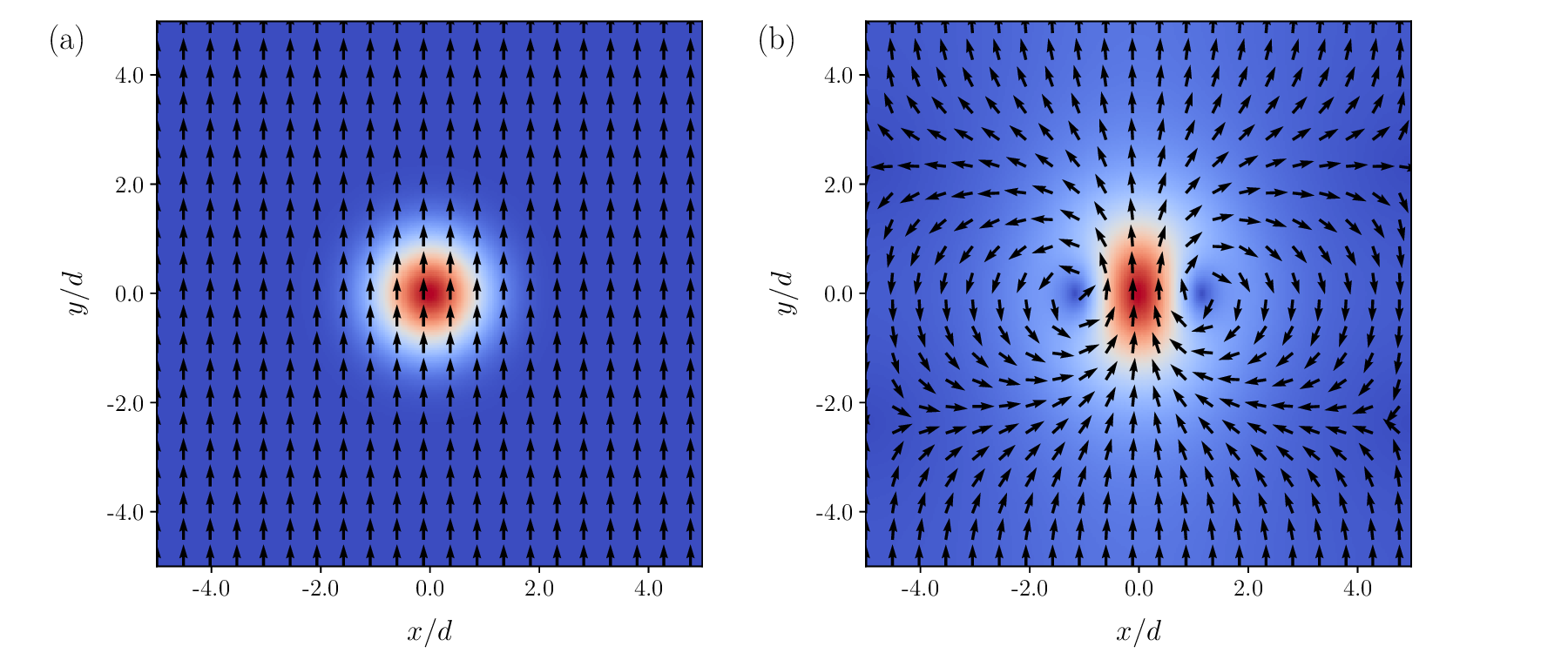}
    \caption{The effect of Fourier-space projection of a non-solenoidal forcing. Panel (a) shows a uniformly directed 2D impulse that decays exponentially with distance from the origin. Panel (b) shows the result of removing the non-solenoidal part of this impulse by application of the Fourier-space operator $\mathcal{P}_{ij}= \delta_{ij}-k_ik_j/k^2$; the impulse now falls off much more slowly with distance from the origin.}
    \label{fig:projection}
\end{figure}
An important exception to this statement is the case of $b=2$. An expansion of ${F_{\bw}(k\to0)}$ analogous to \eqref{Eexpansion} yields
\begin{equation}
    F_{\bw}(k\to0) = \frac{L_{\bF_{\bw}}k^2}{2\pi^2}+\dots\,,
\end{equation}so~$b=2$ corresponds to a finite value of
\begin{equation}
    L_{\bF_{\bw}} \equiv \int_0^t \dd s \int \dd^3 \br \langle \bF_{\bw}(t) \bcdot \bF_{\bw}'(s) \rangle,\label{Lf}
\end{equation}which is the analogue of the Saffman integral for $\bF_{\bw}$. If the forcing is solenoidal, then strong long-range correlations in $\bF_{\bw}$ are required for $L_{\bF_{\bw}}$ to be finite, because then~\eqref{ff'} yields
\begin{equation}
    L_{\bF_{\bw}} = 4\pi \lim_{r\to\infty} r^3 H_{\bw}(r),
\end{equation}so $L_{\bF_{\bw}}\neq 0 $ requires $H_{\bw}(r\to\infty)=O(r^{-3})$. However, these long-range correlations need not be present if the forcing is non-solenoidal, as it turns out that they are generated naturally when the non-solenoidal part of $\bF$ is removed by the action of the projection operator $\mathcal{P}$---this effect is illustrated in figure~\ref{fig:projection}. Physically, the correlations arise because non-solenoidal forcing generates pressure gradients that decay slowly with distance from the point at which an impulse is applied (these gradients are established instantaneously in an incompressible fluid). This result is due to \cite{Saffman67}, who used it to argue that naturally occurring decaying turbulence need not have ${\mcE(k\to0)\propto k^4}$, as had been supposed by \cite{BatchelorProudman56}. A proof (closely following the one presented by \citealt{Saffman67}) and some further comments are given in Appendix~\ref{app:nonsolenoidal}.

In summary, there are two values of $b$ that are relevant to turbulence that is forced locally in real space. If the forcing is solenoidal (or non-solenoidal but with $L_{\bF_{\bw}}=0$), $b=4$, in which case our passive diffusion model predicts saturation with a thermal $k^2$ spectrum. If the forcing is non-solenoidal and has $L_{\bF}\neq0$, then $b=2$, and \eqref{general_timedepsoln} predicts saturation with a flat spectrum: from~\eqref{general_timedepsoln}, with $C= L_{\bF_{\bw}}/2\pi^2$,
\begin{equation}
    \mcE_{\bw}(k) \simeq (1-e^{-\beta k^2 k_f^2 t})\frac{L_{\bF_{\bw}}}{2\pi^2 \beta k_f^2}.
\end{equation}

\begin{figure}
    \centering
    \includegraphics[width=1.0\textwidth]{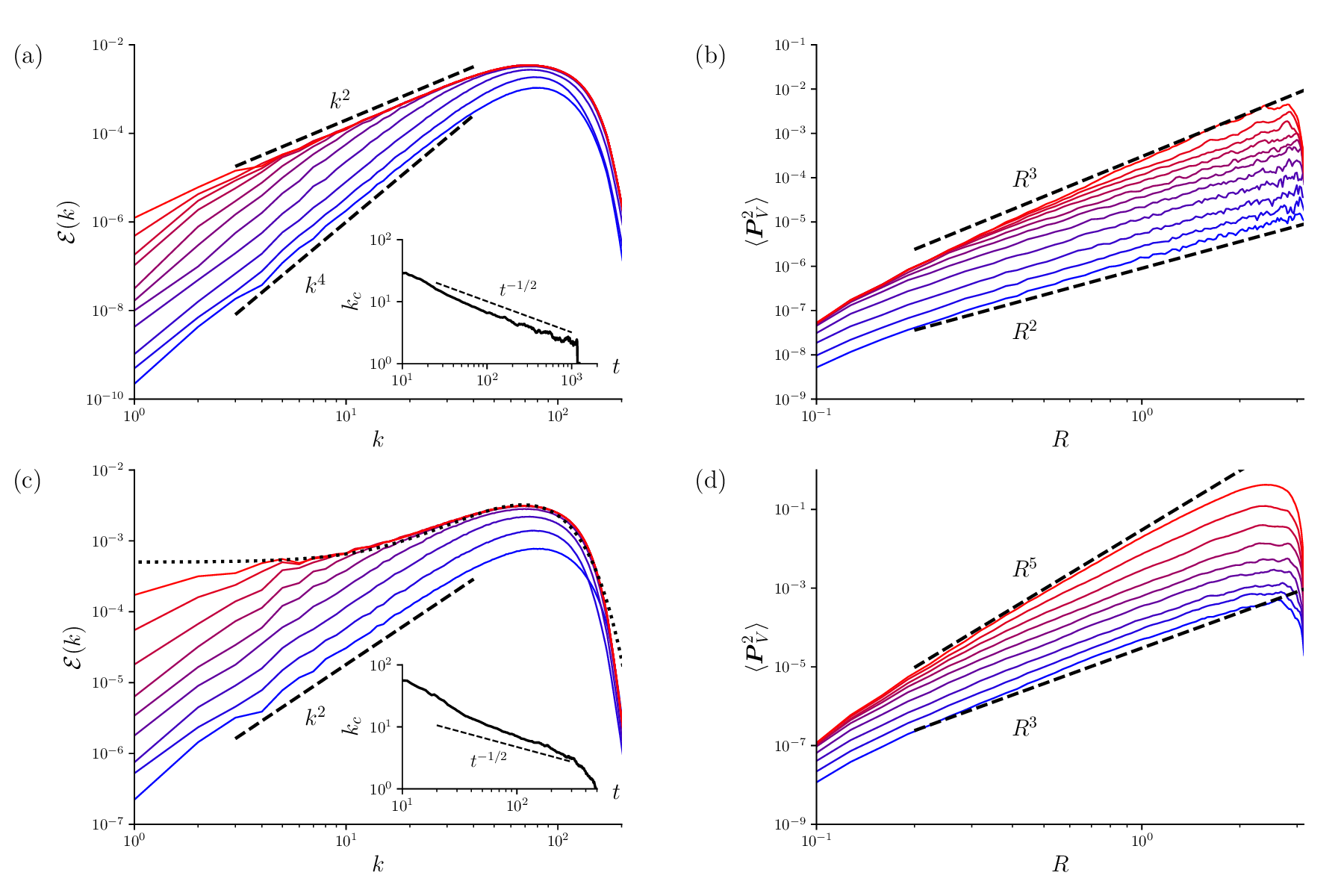}
    \caption{Saturation of the large scales in Navier-Stokes turbulence with a delta-correlated Gaussian random forcing. Panel (a) (the same as figure~\ref{fig:k2schematic}) shows the evolution of the energy spectrum, while (b) shows the evolution of the mean square momentum $\langle \bP^2_V \rangle$, here computed for cubic subvolumes of the box with side length $2R$. Simulations shown in Panels (a) and (b) had the forcing spectrum $F(k)\propto  k^4 \exp(-k^2/k_p^2)$. Panels (c) and (d) show the same quantities for $F(k)\propto k^2 \exp(-k^2/2k_p^2)$. In both cases, the peak of $F(k)$ is at $k_p = 80$. In panel (c), a numerical fit of the data to $(Ak^2/k_p^2+B)\exp (-Ck^2)$, as explained in the text, is plotted as a dotted line. Insets to panels (a) and (c) show the evolution of the spectral knee $k_c(t)$. In the chosen units, the energy injection rate is $0.7$ in both cases, and the r.m.s. velocities are $\simeq 0.5$. The plotted curves are logarithmically spaced in time, with blue~$\to$~red indicating earlier~$\to$~later times. Details of the numerical setup are described in~§\ref{sec:simulations} and~§\ref{sec:cumulative_correlations}.}
    \label{fig:k2k4}
\end{figure}

As we show in figure~\ref{fig:k2k4}, these predictions hold up reasonably well in our numerical simulations of Navier-Stokes turbulence, as do the corresponding scalings for linear momentum. For the $b=4$ case, panel~(a) shows that a finite-band $k^2$ spectrum develops over a wavenumber interval that widens over time, in close agreement with~\eqref{kc_passive} and~\eqref{general_tgg} [panel~(a) is the same as figure~\ref{fig:k2schematic}, and is presented again here to facilitate comparison]. We note that, at late times, the $k^2$ spectrum persists all the way to the box scale (to good approximation). Such scales are not strictly within the domain of validity of our theory (which employs isotropic and homogeneous statistics, only valid at scales much smaller than the box size). Nonetheless, such behaviour should be expected under the statistical-mechanical interpretation of the $k^2$ spectrum [see discussion around~\eqref{k2}], to which our theory is complementary. In principle, it would be desirable to have much larger simulations with increased separation between box and forcing scales in order to rule out any box-scale effects on the development of the $k^2$ spectrum; however, such simulations are unaffordable at present because of the stiff scaling of the simulation cost with size [accounting for the additional time for a larger simulation to reach saturation, which scales as box size squared by~\eqref{kc_passive}, the cost is proportional to the fifth power of the size].

Panel~(b) shows the corresponding development of $\langle \bP^2_V \rangle \propto R^3$, although the split-power-law structure in $\langle \bP^2_V \rangle$, between $\propto R^2$ and $\propto R^3$, is somewhat less pronounced than in the spectrum.

For $b=2$, panel~(c) shows that the saturated spectrum is somewhat steeper than $k^0$, though still shallower than $k^2$. Likewise, the decrease in $k_c$ with time is somewhat faster than~\eqref{kc_passive} predicts, as is shown by the inset to panel~(c). It is plausible that these effects are a consequence of the scale separation between the forcing scale and the scale of the simulation box being insufficient to observe the true $k^0$ large-scale asymptotic: when $b=2$, unlike when $b=4$,~\eqref{general_tgg} only reduces to the asymptotic behaviour $\mcE(k)\propto k^{b-2}$ when
\begin{equation}
    \frac{k}{k_f} \ll \sqrt{\frac{L_{\bF_{\bw}}}{2\pi^2 \beta k_f^2 \mcE_{\bw}(k_f)}}\label{transition_to_asymptotic}.
\end{equation}If the right-hand side of~\eqref{transition_to_asymptotic} happened to be a moderately small number, then the $k^0$ and $k^2$ terms in~\eqref{general_tgg} would be comparable over a range of $k\lesssim k_f$, giving the appearance of a steeper spectrum than $k^0$. To illustrate this possibility, we show in panel~(c) of figure~\ref{fig:k2k4} a numerical fit of the function ${(A k^2/k_p^2 + B)\exp(-C k^2)}$, where $k_p=80$ is the peak forcing wavenumber, and $A$, $B$, $C$ are fitting parameters, to the final data curve. The result reproduces the data well with $A/B \simeq 20$. Alternatively, the discrepancy with the prediction of the passive-velocity model might be a result of the neglected effect of advection by large-scale modes, which do possess a significant proportion of the total energy for a close-to-flat spectrum. As a result, closure schemes such as the popular EDQNM model may be better at capturing this effect than the passive-vector model --- see the discussion at the end of Section~\ref{sec:narrow_band}. Panel~(d) shows that $\langle \bP^2_V \rangle$ follows a scaling reasonably close to~$R^{5}$ at late times, which is consistent with a $k^{0}$ spectrum at large scales, according to~\eqref{P2_asymptotic2}. We note that, for a forcing with $b=2$, running the simulation to later times than those shown in figure~\ref{fig:k2k4} tends to produce a build-up of energy in the largest-scale modes, making the large-scale asymptotic difficult to measure; a similar effect was found in some of the simulations of \cite{AlexakisBrachet19}. Here, we ended our runs before this effect became significant.

\subsection{Narrow-band forcing \label{sec:narrow_band}}

Finally, we discuss the case of forcing in a narrow spectral band. As noted above, the prediction of the passive model for forcing in a finite band may be obtained by setting $C=0$ in \eqref{general_timedepsoln}. This is valid for $k$ much smaller than any other characteristic wavenumber associated with $F_{\bw}(k)$, including the inverse characteristic width of the forcing spectrum; this assumption entered when we used 
\begin{equation}
    k \ll \left[ \frac{1}{\mcE_{\bw}(k)}\frac{\dd \mcE_{\bw}(k)}{\dd k} \right]^{-1} _{k=k_f},\label{k_scale_length}
\end{equation}in order to justify setting $\mcE_{\bw}(k_f + q)\simeq \mcE_{\bw}(k_f)$ in the $q$ integral in \eqref{penultimate}. Of course, \eqref{k_scale_length} is always justified for $k\to 0$. However, if the forcing is concentrated in a narrow band of wavenumbers of width $\Delta k \ll k_f$, which is somewhat artificial compared to the more obviously physically realisable $\Delta k \sim k_f$, but a common choice for numerical simulations~(see, e.g., \citealt{AlexakisBrachet19}), then it is possible that the energy contained in the forced band will greatly exceed the energy contained by the nearby unforced modes. In that case, there will be an extended range of $k$ for which
\begin{equation}
    \left[ \frac{1}{\mcE_{\bw}(k)}\frac{\dd \mcE_{\bw}(k)}{\dd k} \right]^{-1}_{k=k_f} \sim \Delta k \ll k \ll k_f.
\end{equation}For $k$ in this range, \eqref{general_timedepsoln} does not apply. Taking
\begin{equation}
    \mcE_{\bw}(k_f+q) \simeq \begin{dcases*}
                    \frac{C'}{\Delta k}, & if  $|q-k_f| < \Delta k/2$, \\
                     \phantom{-}0, & otherwise,
                 \end{dcases*} \label{narrow_band_forcing}
\end{equation}where $C'$ is a constant, modifies the injection term on the right-hand side of~\eqref{large_scale_equation} to be~$\propto k^3$ rather than~$\propto k^4$, viz.,
\begin{align}
      \p_t \mcE_{\bw}(k) + \beta \kflow^2 k^2 \mcE_{\bw}(k) = \frac{5}{8}\beta C' k^3. \label{narrowband}
\end{align}The saturated spectrum for $\Delta k \ll k \ll k_f$ is then readily obtained from \eqref{narrowband} with $\p_t\to 0$:
\begin{equation}
    \mcE_{\bw}(k) \simeq \frac{5}{8} \frac{C' k}{k_f^2},\label{narrow_saturated}
\end{equation}i.e., $\mcE_{\bw}(k)\propto k$, \emph{not} $k^2$, in this range.

How should we interpret this behaviour? These scalings are reminiscent of two-dimensional turbulence---in 2D, the expansion \eqref{Eexpansion} of $\mcE_{\bw}(k\to 0)$ becomes
\begin{equation}
    \mcE(k) = \frac{L_{\mathrm{2D}}k}{4\pi}+\frac{I_{\mathrm{2D}}k^3}{16\pi}+\dots
\end{equation}where $L_{\mathrm{2D}}$ and $I_{\mathrm{2D}}$ are the two-dimensional analogues of the Saffman and Loitsyansky integrals. In the absence of the inverse cascade, therefore, the two-dimensional forced-turbulence spectrum would consist of a growing $k^3$ part at the largest scales, changing to a growing $k^1$ band at $k_c \propto t^{1/2}$, precisely as we have found for a narrow spectral band in three dimensions. While the latter system is not two-dimensional in real space, the same scalings are obtained because the Fourier modes that dominate the mode-coupling integral in~\eqref{large_scale_equation} are confined to the $k=k_f$ surface in Fourier space.

In practice, saturation precisely according to \eqref{narrow_saturated} is unlikely, because the energy in forced modes usually does not greatly exceed the total energy at the flow scale at saturation. Nonetheless, there can still be some deviation from a precise $k^2$ spectrum, as reported by \cite{AlexakisBrachet19}.

To conclude this section, we note that a relation equivalent to~\eqref{large_scale_equation} was derived by~\cite{Lesieur08} under the EDQNM closure scheme by assuming (\textit{i}) Markovian dynamics --- i.e., neglecting finite-correlation-time effects and (\textit{ii}) Kraichnan's (\citeyear{Kraichnan87b, Kraichnan87a}) distant-interaction algorithm, in which certain types of non-local (in Fourier space) interactions are discarded. We therefore might have derived the corollaries of~\eqref{large_scale_equation} that were discussed in Sections~\ref{sec:passive_theory}, \ref{sec:nonsolenoidal} and~\ref{sec:narrow_band} as consequences of the EDQNM closure scheme (or other similar models). We also note that it might be possible to recover the development of a steeper-than-$k^0$ large-scale spectrum for $k^2$ forcing under the EDQNM scheme [without applying~(\textit{ii})]. We are grateful to an anonymous referee for pointing this out to us. We have not pursued that line of development in the present work, preferring to use the passive-vector model introduced in Section~\ref{sec:passive} as a simple model that clearly illustrates the role of turbulent diffusion of the momentum distribution of initially localised structures as the key process by which the $k^2$ spectrum forms.

\section{Decay of initially forced turbulence\label{sec:decaying_connections}}

\begin{figure}
    \centering
    \includegraphics[width=1.0\textwidth]{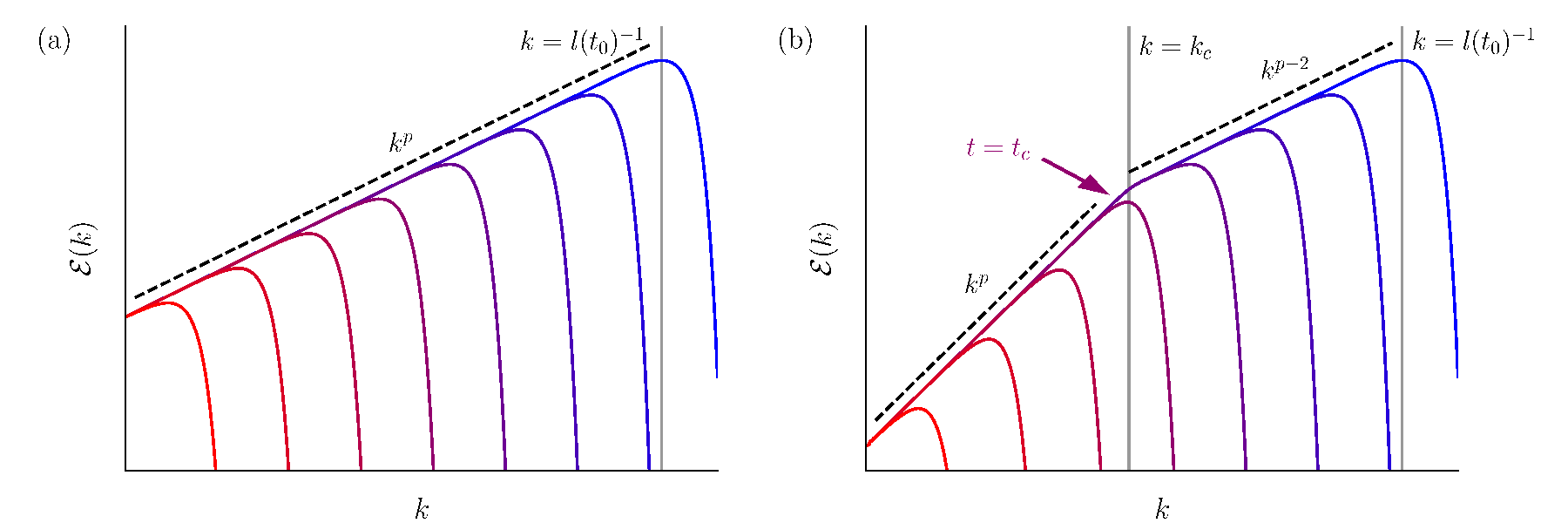}
    \caption{Schematic diagrams of the evolution of the energy spectrum of decaying isotropic turbulence initially forced for a period $t_0$, where (a) $t_0 \lesssim$ the initial eddy-turnover time and (b) $t_0 \gg$ the initial eddy-turnover time. The exponent $p$ is either $4$ or the asymptotic spectral exponent of the forcing as $k\to 0$, whichever is smaller [cf.~\eqref{general_timedepsoln}]. Blue~$\to$~red indicates earlier~$\to$~later times.}
    \label{fig:schematic_decays2}
\end{figure}

In this section, we consider how turbulence that has been forced for a long period decays after the forcing is removed. This situation is somewhat different to the one usually considered in theoretical treatments of decaying turbulence (see~\citealt{Davidson15} for a review and, e.g., \citealt{Panickacheril22} for a recent numerical study), where it is common to consider an initial condition that was generated effectively instantaneously (physically, over a period $\lesssim$ the initial eddy turnover time). For an initial condition generated in this way, the decay is usually believed to be governed by the principle of the `permanence of the large-scale spectrum': while the energy contained at the flow scale cascades to small scales and is dissipated, the large-scale power law is preserved, i.e., ${\mcE(k\ll l(t)^{-1})\simeq \const}$ [see figure~\ref{fig:schematic_decays2}(a)]. In other words, there is no thermalisation of the large scales and no quasi-randomisation of the momentum distribution. In this section, we shall argue that the same is true for previously forced, now decaying turbulence, even if $k_c^{-1}$ is initially much larger than the outer scale. This means that the growing integral scale eventually becomes comparable to $k_c^{-1}$, and, when it exceeds $k_c^{-1}$, the decay laws change---although, as we shall see, this can happen after a long time compared to the duration of the forced stage.

\subsection{Conservation of momentum in decaying turbulence \label{sec:Saffmandecay}}

The large-scale part of the energy spectrum of instantaneously generated turbulence typically follows an unbroken power law, i.e., $\mcE(k\ll l^{-1})\propto k^a$. The ``classical'' exponents $a=2$ and $a=4$ can each arise from initial impulses that do not have long-range spatial correlations [cf.~\eqref{Eexpansion}], although it is also possible to consider other values for $a$. For $a\leq3$, the principle of the permanence of the large-scale spectrum that we outlined above is a consequence of the conservation of linear momentum~\citep{Saffman67, Davidson15}---let us briefly review how this works. According to~\eqref{P2_asymptotic2}, $\langle\bP_V^2\rangle / R^2 \to \infty$ as $R\to\infty$ when $a\leq 3$, indicating that eddies throughout the volume~$V$, not just those at its surface, contribute to~$\langle\bP_V^2\rangle$. On the other hand, $\dd \langle\bP_V^2\rangle/\dd t = O(R^2)$, because~$\langle\bP_V^2\rangle$ only changes as a result of random fluxes through the surface of $V$ [formally, this requires that ${K(r\to\infty)=o(r^{-3})}$; see \eqref{dP2/dt} and \cite{Davidson15}]. Therefore,
\begin{equation}
     \lim_{V\to\infty}\frac{\dd \log \langle\bP_V^2\rangle}{\dd t} = 0,\label{decay_momentum}
\end{equation}so the large-$R$ scaling of $\langle\bP_V^2\rangle$ vs. $R$ is preserved and hence so is the large-scale spectral power law, which imposes the scaling 
\begin{equation}
    u^2 l^{1+a}\sim \const.\label{u2l1+a}
\end{equation}

Assuming that the decay is self-similar and occurs on the turnover timescale $t_{\mathrm{nl}}\sim l/u$ of the largest eddies, so that $\dd u^2/\dd t \propto -u^3/l$, it follows from~\eqref{u2l1+a} that
    \begin{equation}
    u^2(t) \sim  u^2(t_0)\left(\frac{t-t_0}{t_{\mathrm{nl,}0}}\right)^{\!-2(1+a)/(3+a)}, \quad l(t)\sim l(t_0) \left(\frac{t-t_0}{t_{\mathrm{nl,}0}}\right)^{\!2/(3+a)},\label{Generallaws}
\end{equation}where $t_0$ is the time that the forcing ceases and decay commences and ${t_{\mathrm{nl,}0}\equiv t_{\mathrm{nl}}(t_0)\ll t-t_0}$. In the classical case of $a=2$ considered by~\cite{Saffman67}, which can result from a non-solenoidal initial forcing without long-range correlations (see~§\ref{sec:nonsolenoidal}), \eqref{Generallaws} gives $u^2 \propto (t-t_0)^{-6/5}$, $l\propto (t-t_0)^{2/5}$. In the other classical case of $a=4$, \eqref{Generallaws} indicates that $u^2 \propto (t-t_0)^{-10/7}$, $l\propto (t-t_0)^{2/7}$, which are the decay laws predicted by~\cite{Kolmogorov41c}. Note that the $a=4$ laws do not follow from the conservation of $\langle\bP_V^2\rangle$, which constrains the decay only when $a \leq 3$. Instead, they are conventionally justified from the invariance of angular momentum via the Loitsyansky integral~\eqref{Loitsyansky}, although this idea has been challenged---for details, we refer the reader to the footnote on page~\pageref{footnote_Loitsyansky} and references therein.

As we saw in Section~\ref{sec:momentum}, the broken large-scale spectrum developed by forced turbulence is a consequence of changes in the local scaling of $\langle\bP_V^2\rangle$ vs. $R$. This means that any local-power-law energy spectrum must be preserved during decay of previously forced turbulence if the local exponent is $\leq 3$. Therefore, the decaying turbulence must initially follow~\eqref{Generallaws} with $a$ set by the local power-law exponent on the infra-red (small-$k$) side of the wavenumber $k\sim l(t_0)^{-1}$ [see figure~\ref{fig:schematic_decays2}(b)]. This decay will continue until such time when $l(t)^{-1}\sim k_c$.

What happens to $k_c$ during this period? If the $k\to 0$ asymptotic of the spectrum is shallower than $k^3$, it will be preserved by momentum conservation, so $k_c$ must be constant. However, if the $k\to 0$ asymptotic of the spectrum is steeper than $k^3$---and it is $\propto k^4$ when the forcing is solenoidal and local in real space---then it is \textit{a priori} unclear what the evolution of $k_c$ might be. In particular, it does not appear possible to argue for preservation of a $\propto k^4$ asymptotic from the conservation of the Loitsyansky integral~\eqref{Loitsyansky}---as we found in §\ref{sec:passive}, the latter can grow during momentum diffusion, which plausibly could happen during the transient stage of decay, even if it is ruled out for decay with an unbroken power law. We shall therefore employ our passive model of momentum diffusion,~\eqref{passive_velocity_eqn}, to determine the evolution of $k_c$ in decaying turbulence.

\subsection{Passive evolution of the large scales in decaying turbulence\label{sec:mcouplingdecay}}

Let us consider how the small-$k$ part of $\mcE(k)$ evolves during the period of decay when $l(t) \ll k_c^{-1}$, assuming that the largest scales of the velocity field are advected passively by the decaying integral-scale flow, in the sense of~\eqref{passive_velocity_eqn} with $\bF_{\bw}=0$. We again model the energy-containing scales with the Kraichnan ensemble~\eqref{uu_kraichnan_mt}, taking $k_f \to l^{-1}$ and $\kappa_0 \sim u^3 l$ in \eqref{kappa}, which follows from assuming that the energy-containing scales evolve in a self-similar manner. Under these choices,~\eqref{large_scale_equation} becomes
\begin{equation}
    \p_t \mcE(k) + ul k^2 \mcE = ul^{3-a} A k^4\label{dtE},
\end{equation}where we have used~\eqref{u2l1+a} to obtain $\mcE(l^{-1}) \sim A l^{-a}$, where $A$ is a constant. We note that a relation similar to \eqref{u2l1+a} may also be derived from the EDQNM closure approximation under certain assumptions --- see the discussion at the end of §\ref{sec:narrow_band}. Let us define $s = \int_{t_0}^t \dd t'\, u(t')l(t')$, where $t_0$ is the time at which forcing ceases. Because $l\sim u(t-t_0)$ by~\eqref{Generallaws}, $s\sim l^2$ for $t \gg t_{\mathrm{nl,0}}$ independently of $a$. Changing variables from $t$ to $s$, we find that~\eqref{dtE} can be rewritten as
\begin{equation}
    \p_s \mcE(k) + k^2 \mcE(k) = l^{2-a} A k^4.\label{dsE}
\end{equation}The characteristic scale in $s$ that is associated with the diffusion term on the left-hand side of~\eqref{dsE}, $k^2 \mcE(k)$, is constant, $\sim 1/k^2$. However, this is also the value of $s$ at which the growing outer scale $l$ reaches $\sim1/k$, at which point~\eqref{dsE} ceases to be a valid description of $\mcE(k)$ at the particular $k$ under consideration. We conclude that the diffusion term can be neglected when the turbulence is decaying. It is straightforward to show that the solution for $t\gg t_{\mathrm{nl},0}$ of~\eqref{dtE} without the diffusion term $ul k^2 \mcE$ is\footnote{When $a=4$, there is an additional factor of $\log t$ in the second term in~\eqref{dtE_soln}.}
\begin{equation}
    \mcE(k, t)\simeq \mcE(k, t_0) + \mathrm{const} \times (kl)^4 \mcE(l^{-1}, t)  \label{dtE_soln}.
\end{equation}In words, $\mcE(k, t)$ is given by the larger of the initial condition or the Batchelor ($\propto k^4$) spectrum that would correspond to the instantaneous outer scale of the turbulence.

Thus, the energies of large-scale modes are preserved as the turbulence decays, provided that these exceed the energies that would correspond to a Batchelor spectrum. For unbroken spectra, this is simply a recovery of the permanence of the large-scale spectrum that was derived from the conservation of $\langle\bP_V^2\rangle$ in Section~\ref{sec:Saffmandecay}. However, for broken spectra (i.e., for $k_c \ll l^{-1}$),~\eqref{dtE_soln} also shows that there is no evolution of $k_c^{-1}$ before the integral scale reaches it [see figure~\ref{fig:schematic_decays2}(b)].

\subsection{Decay laws for initially forced turbulence}

We now report the analogues of the classical decay laws described above but for turbulence that was forced for a long period by a body force without long-range correlations in real space. If the forcing was solenoidal, so that the spectrum at $t=t_0$ is $\propto k^4$ for $k \ll k_c$ and $\propto k^2$ for $k_c \ll k \ll l(t_0)^{-1}$, then the decay is according to the Saffman laws, viz.,~\eqref{Generallaws} with $a=2$, until $l\sim k_c^{-1}\sim l(t_0)(t_0/t_{\mathrm{nl,0}})^{1/2}$, which occurs at $t = t_c \sim t_0 + t_0 (t_0/t_{\mathrm{nl,0}})^{1/4} \gg t_0$. After this, the decay follows the Batchelor laws, viz.,~\eqref{Generallaws} with $a=4$ and $t_0$ replaced by $t_c$. An interesting feature of these results is that memory of the forcing is retained for a long time---the Saffman laws are followed for a period that is asymptotically longer (when $t_0/t_{\mathrm{nl,0}} \gg 1$) than the duration of the forcing stage.

The other interesting case is one with non-solenoidal forcing, for which the spectrum at $t=t_0$ is generically $\propto k^2$ for $k \ll k_c$ and, as per  \eqref{general_timedepsoln}\footnote{In~§\ref{sec:nonsolenoidal}, we already saw numerical evidence that the $k^0$ law may not actually be developed by real turbulence. This is likely due to the importance of advection by large-scale modes, which is neglected in our model---see discussion in §\ref{sec:nonsolenoidal}. However, a power law reasonably close to $k^0$ does appear to be reached (see figure~\ref{fig:k2k4}), so we assume the $k^0$ scaling here, for simplicity and lack of a better theory.}, $\propto k^0$ for $k_c \ll k \ll l(t_0)^{-1}$. In this case, the decay follows~\eqref{Generallaws} with $a=0$, viz., $u^2\propto (t-t_0)^{-2/3}$, $l\propto (t-t_0)^{2/3}$, until $l\sim k_c^{-1}$ at $t = t_c \sim t_0 + t_0 (t_0/t_{\mathrm{nl,0}})^{-1/4} \ll t_0$. Thus, unlike solenoidal forcing, non-solenoidal forcing tends to generate turbulence that has a \textit{short} memory---the $a=0$ laws are followed for a period that is $\ll t_0$ when $t_0/t_{\mathrm{nl,0}} \gg 1$. After $t=t_c$, the decay follows the Saffman laws, viz., \eqref{Generallaws} with $a=2$ and $t_0$ replaced by $t_c$.

These results indicate that both types of forcing encourage turbulence to decay in the Saffman regime---for solenoidal forcing, this is because the transient period of Saffman-like decay is asymptotically long compared with the forcing period, while for non-solenoidal forcing, this is because the transient period of non-Saffman decay is asymptotically short compared with the same. A numerical study to test these predictions would be extremely valuable, although it would also incur significant numerical cost---it would be necessary to resolve two scale separations, first between the box size and $k_c(t_0)^{-1}$, and second between $k_c(t_0)^{-1}$ and $l(t_0)$, and to run the forced part of the simulation for long enough for the latter scale separation to be reached while furthermore also ensuring that the Reynolds number at the integral scale were always large enough for the Saffman decay laws~[\eqref{Generallaws} with $a=2$] to be valid. We therefore defer such a numerical study to future work (or invite the reader to undertake it).

\section{Conclusion\label{sec:conclusion}}

In this work, we have addressed the apparent disconnect between the ``decaying-turbulence view'' of the large-scale structure of turbulence, i.e., the notion that it is determined kinematically by the values of certain invariants that describe statistical properties of the flow field, and the increasingly popular idea that the large-scale spectral tail might (in some cases) constitute an isolated subsystem in thermal equilibrium. If the latter were true, equipartition of energy between Fourier modes would imply $\mcE(k\to 0)\propto k^2$, which corresponds to a non-zero Saffman integral, $L$ [see~\eqref{Eexpansion}]. However, as we found in §\ref{sec:correlations}, $L$ is an invariant not only of decaying turbulence, but also of forced turbulence, provided that the forcing is solenoidal and sufficiently localised in real space. This invariance is a manifestation of the conservation of linear momentum: solenoidal, localised forcing can only generate eddies with vanishing total momentum, so the total momentum contained within a sufficiently large volume of the turbulence must always vanish (in the sense that only surface contributions matter to the total), in which case $L=0$ [see~\eqref{Saffman_momentum}]. 

Nonetheless, the total momentum contained within a \emph{finite} volume of turbulence need not vanish, provided that one waits long enough for the momentum distribution to become stochastic on the relevant scale. As explained in §\ref{sec:momentum}, this scale-dependent stochasticisation is the net result of the momentum-conserving interactions between eddies, and leads to a broken-power-law spectrum at large scales. The ``knee'' wavenumber that separates the developing $k^2$ spectrum from the ``Batchelor'' asymptotic $\mcE(k\to 0) \propto k^4$ is a decreasing function of time: numerically, we have found it to decrease like $k_c \propto t^{-1/2}$, as would be expected for a turbulently diffusing field (of which a good solvable model can be constructed using the Kraichnan flow: see §\ref{sec:passive}). In a finite system, such as a numerically simulated turbulence in a periodic box, the spectral knee eventually reaches the box scale, at which point the turbulence is essentially identical in character to a finite-system-size approximation of ``Saffman turbulence'', i.e., turbulence with $L\neq 0$. In particular, $\mcE(k)\propto k^2$ at all resolved scales larger than the forcing scale. However, in an infinite domain, the spectral knee grows indefinitely, and increasingly slowly, as increasingly distant points become correlated. Interestingly, these conclusions may be modified somewhat if the forcing is non-solenoidal and thus momentum-injecting, as explained in §\ref{sec:nonsolenoidal} (see also below), and also if it is in a narrow spectral band, as explained in §\ref{sec:narrow_band}.

We anticipate that there may be a number of applications of the ideas developed in this work to more complex variants of fluid turbulence---in particular, to naturally occurring astrophysical turbulence. One such astrophysical application might be to the evolution of primordial magnetic fields in the early universe. Typically, these magnetic fields are assumed to have a magnetic-energy spectrum $\propto k^4$ at large scales, because it is usually thought that long-range correlations in the magnetic field, of the sort required for a shallower spectrum, are excluded by causality constraints imposed by cosmological inflation models \citep{DurrerCaprini03, Brandenburg15, Brandenburg17,ReppinBanerjee17}. In the statistically isotropic case, their decay then proceeds via reconnection of magnetic-field lines, conserving magnetic helicity, either in a ``net'' \citep{Hatori84, BiskampMuller99, Brandenburg17, HoskingSchekochihin20decay} or ``fluctuating'' sense \citep{HoskingSchekochihin20decay}. However, intuitively, it seems likely that a process akin to the one described in this work may be able to induce a magnetic-energy spectrum $\propto k^2$ over a finite range of scales, if the magnetic energy is maintained by an effective ``forcing'' from the velocity field (i.e., magnetic dynamo). In this case, the process of momentum diffusion would be replaced by ``flux diffusion'', which presumably can occur due to magnetic reconnection. While a dedicated study would be necessary for a complete understanding of this effect, we note that numerical simulations of the MHD fluctuation dynamo that have scale separation between the box size and forcing scale do indeed appear to saturate with a magnetic-energy spectrum $\propto k^2$ at large scales \citep{MaronBlackman02,Brandenburg22}. This would motivate consideration of a $k^2$ large-scale spectrum in the primordial magnetic field. In the process of  decay of such a field, the ``Saffman flux invariant'',
\begin{equation}
    L_{\bB} = \int \dd^3 \br \,\langle \bB(\bx) \bcdot \bB (\bx + \br)\rangle,\label{SaffmanB}
\end{equation}which is the analogue of the Saffman integral~\eqref{saffman_integral} but for magnetic flux, should be conserved---the interested reader will find discussion of the effect of $L_{\bB}$ on MHD decay laws in the Supplementary Information of~\cite{HoskingSchekochihin21primordial}. We also note that momentum and flux diffusion, in the sense described here, provides a mechanism for the transfer of some energy to large spatial scales. It might therefore be useful to consider this effect in the context of the fluctuation (non-helical) dynamo, whose saturated energy spectrum is an outstanding theoretical problem (see, e.g.,~\citealt{Galishnikova22},~\citealt{Rincon19}, and references therein). 

Another application of the ideas presented here to MHD turbulence concerns the 2D spectra of the latter in the presence of a strong mean field, in which case the idea of thermalisation turns out to have some traction in treating the scales perpendicular to the mean field that lie in the inertial range but are larger than the ``critical-balance'' scale---an intrigued reader will find the details in Appendix~B of \cite{Schekochihin20}. Such 2D spectral scalings turn out to be of some consequence also in the theory of phase-space turbulence in kinetic plasmas~\citep{Schekochihin16}.

Returning to hydrodynamics, one intriguing finding of this study is that turbulence does \emph{not} develop a thermal large-scale spectrum when the forcing generates eddies with non-zero linear momentum directly. This can occur even with a spatially localised forcing, provided it is non-solenoidal. In that case, long-range correlations can be generated when the non-solenoidal part is removed, an idea that underpins the realisability of decaying Saffman turbulence \citep{Saffman67}. The measurement of a flatter-than-$k^2$ large-scale spectrum in a forced-turbulence experiment, such as the one under construction at ENS Paris, would therefore represent a neat direct demonstration of the physics underlying Saffman's theory of decaying turbulence, as would direct measurement of the Saffman decay laws (see~§\ref{sec:Saffmandecay}) in turbulence forced locally and solenoidally until a large-scale thermal spectrum developed, and then allowed to decay. From the statistical-mechanics perspective, the thermodynamical motivation for an equilibrium spectrum (see §\ref{sec:introduction}) relies on weak interaction between the forcing scales and the much larger ones. This condition is satisfied when the forcing is solenoidal, because then the fluid response is essentially local in real space (see §\ref{sec:batchelor_argument}), but it is violated when the forcing is non-solenoidal, because then long-range interactions between distant points via exchange of pressure waves become an important feature of the dynamics, even though the forcing itself might be local in real space. 

In light of this observation, we suggest that an interesting topic for further study would be the large-scale structure of compressible turbulence, in which pressure (sound) waves propagate with finite velocity. In the large-Mach-number limit of highly supersonic motions, turbulent diffusion should be the dominant mechanism of momentum transport, so equilibration of the large scales should be possible. At finite Mach number, however, sound waves may correlate distant points before the turbulent diffusion of momentum can, owing to the fact that sound waves propagate ballistically, rather than diffusively. Whether this precludes thermalisation of the large scales in forced compressible turbulence is an intriguing question.

\acknowledgements 
D.N.H. was supported by a UK STFC studentship. The work of A.A.S. was supported in part by the UK EPSRC grant EP/R034737/1. This work used the ARCHER2 UK National Supercomputing Service (https://www.archer2.ac.uk).
\newline
\newline
\textbf{Declaration of interests.} The authors report no conflict of interest.

\appendix

\section{Invariance of the Saffman integral in forced turbulence \label{BatchelorProof}}

In this appendix, we formalise the arguments presented in §\ref{sec:batchelor_argument} for the invariance of the Saffman integral in forced turbulence. Our key assumption is that \textit{the value of the forcing function at any given point in space and time is statistically independent of the value it has at all other finite times at arbitrarily distant points in space}. More precisely, we assume
\begin{quotation}
\noindent A-I: \textit{All cumulants of $f(\bx,t)$}
\begin{equation}
    \langle f_i(\bx, t) f_j(\bx', t')f_k(\bx'', t'') \dots \rangle_c
\end{equation}
\textit{decay sufficiently quickly with distance for their integral moments (with respect to $\bx'-\bx$, $\bx''-\bx$, etc.) to converge.}\footnote{For $n>2$, the cumulant $\langle \dots \rangle_c$ of the $n$th-order correlation function $\langle \dots \rangle$ is the difference between $\langle \dots \rangle$ and the value in terms of second-order correlators that it would have if the underlying statistics were Gaussian. For example, $\langle f_i f_j'f_k'' f_l''' \rangle_c \equiv \langle f_i f_j'f_k'' f_l''' \rangle - \langle f_if_j'\rangle \langle f_k''f_l'''\rangle - \langle f_if_k''\rangle \langle f_j'f_l'''\rangle - \langle f_if_l'''\rangle \langle f_j'f_k''\rangle$, where $f_i \equiv f_i(\bx, t)$, $f_i' \equiv f_i(\bx', t')$, etc. For $n\leq 2$, $\langle \dots \rangle_c = \langle \dots \rangle$.}
\end{quotation}

A-I is a natural generalisation to forced turbulence of the assumption adopted by \cite{BatchelorProudman56} for decaying turbulence, namely that integral moments of cumulants of $\bu$ at the initial time converge\footnote{Our results reduce to theirs for $\bF(\bx, t)\to \bF(\bx)\delta(t)$.} [also see \cite{Saffman67} for a similar analysis of the decaying-turbulence problem in Fourier space]. Using methods analogous to theirs, we shall use A-I to determine from the \textit{forced} Navier-Stokes equation all of the initial ($t=0$) time derivatives of the velocity correlation functions in the large-separation limit. We further follow~\cite{BatchelorProudman56} in assuming that
\begin{quotation}
    A-II: \textit{The large-separation asymptotics of correlators at $t>0$ can be written as convergent Taylor series in $t$, with derivatives evaluated at $t=0$. }
\end{quotation}
The time derivative in such a Taylor series that decays slowest with the separation $r$ gives the large-$r$ asymptotic of the correlator at $t>0$.

We note that A-II is neither trivial nor unquestionable. Though he uses a similar assumption in his theory of decaying turbulence, \cite{Saffman67} points out that there exist initial conditions for which the resulting velocity field is not an analytic function of time, which suggests that correlation functions need not be either. Furthermore, correlation functions are not, in general, analytic functions of time in diffusing systems. For example, with~$u$ a statistically homogeneous and isotropic solution of the 1D diffusion equation ${\p u/\p t =  D\p^2 u/\p x^2}$, the correlation function ${C(r,t)=\langle u(x,t)u(x+r,t)\rangle}$ also solves a diffusion equation: $\p C/\p t =  2D\p^2 C/\p r^2$. Therefore, if $C(r,t=0)$ has compact support $r<R$, then $C(r>R,t)$ is non-analytic in $t$ --- we see from the diffusion equation for $C$ that all of its $t$ derivatives are zero at $t=0$ for $r>R$, but $C(r>R,t)$ is not zero at all times: the support of $C(r,t)$ is only compact at $t=0$. We are grateful to an anonymous referee for pointing this example out to us. It should be noted, however, that such examples do not necessarily contradict A-II, as the latter is an assumption only about the behaviour of the large-separation asymptotics of correlators, rather than about the correlators (or solutions) in general. For example, even in the case of diffusion of a quantity with initially compact support, A-II does give the correct asymptotic behaviour at large separation: at arbitrary finite $t$, the large-$r$ limit of $C(r,t)$ is super-exponentially small. \cite{BatchelorProudman56} similarly note that A-II is an assumption about asymptotics of correlation functions of solutions to the Navier-Stokes equation, which, being averaged properties, are likely to be better behaved than the solutions themselves. For these reasons, we shall take A-II as acceptable in what follows, but we do flag to the reader that it is far from trivial, and remains an assumption, rather than a theorem.

Equation~\eqref{dL/dt} contains two terms that can cause growth of the Saffman integral~$L$: one that involves the triple correlator $K(r)$ and the other involving $\uf$. From the isotropy and solenodiality of $\bu$ and $\bF$, we have that [cf.~\eqref{uiuj(f)}]
\begin{equation}
    \Psi_{ij}\equiv \langle u_i f_j'\rangle = \frac{1}{2r}\left[\frac{\p}{\p r}(r^2 \psi) \delta_{ij} - \frac{\p\psi}{\p r} r_i r_j\right] \implies \uf = \frac{1}{r^2}\frac{\p}{\p r} (r^3\psi). 
\end{equation}Therefore,~\eqref{dL/dt} may be rewritten as
\begin{equation}
    \frac{\dd L}{\dd t} = 4\pi \lim_{r\to\infty}\left[\frac{1}{r} \frac{\p}{\p r} (r^4 u^3 K)\right] +8\pi \lim_{r\to\infty}(r^3 \psi),\label{dL/dt_APP}
\end{equation}which demands the examination of the large-$r$ asymptotics of $K(r)$ and $\psi(r)$. We consider each in turn, with the goal of showing that both of the limits appearing in~\eqref{dL/dt_APP} are zero.

\subsection{Asymptotic of \texorpdfstring{$K(r)$}{K(r)}}

Consider the triple correlator $S_{ijk}\equiv \langle u_i u_j u'_k\rangle$. We wish to estimate all of its time derivatives at $t=0$. Let us assume that $\bu = 0$ at $t=0$. It follows immediately that $S_{ijk}= \p_t S_{ijk} = \p_t^2 S_{ijk}=0$ at $t=0$, while we have from the Navier-Stokes equation~\eqref{NavierStokes} that $\p_t^3 S_{ijk}=\langle f_i f_j f_k'\rangle$, which, by A-I, decays rapidly in space (i.e., faster than any power law). Similarly, 
\begin{multline}
    \frac{\p^4S_{ijk}}{\p t^4} = \bigg\langle \left(\nu \nabla^2 f_i + \frac{\p f_i}{\p t}\right)f_j f_k'\bigg\rangle \\+ \bigg\langle f_i\left(\nu \nabla^2 f_j + \frac{\p f_j}{\p t}\right) f_k'\bigg\rangle + \bigg\langle f_i f_j\left(\nu \nabla'^2 f'_k + \frac{\p f'_k}{\p t}\right)\bigg\rangle,
\end{multline}which also must decay rapidly with increasing $r$. 

In contrast, the fifth derivative in time of $S_{ijk}$ is not guaranteed by A-I to decay rapidly with $r$ at $t=0$, because it contains terms of the sort $\langle f_i f_j \p_t^2 \p_k' p'\rangle$. The pressure field $p'$ is determined non-locally by~\eqref{p_integral} [note that pressure did not appear in the first four time derivatives of $S_{ijk}$ because, according to \eqref{p_integral}, both $\bnabla p$ and its first derivative in time are zero at $t=0$]. Differentiating~\eqref{p_integral} twice with respect to time, we have
\begin{align}
    \frac{\p^2 p(\bx)}{\p t^2} = \frac{1}{4\pi} \int \frac{\dd^3 \bx'}{|\bx' -\bx|} \frac{\p}{\p x_i'}\frac{\p}{ \p x_j'} \frac{\p u_i'}{\p t} \frac{\p u_j'}{\p t}=\frac{1}{4\pi} \int \frac{\dd^3 \bx'}{|\bx' -\bx|} \frac{\p}{\p x_i'}\frac{\p}{ \p x_j'} f_i' f_j' \label{d2t_pAPP},
\end{align}where the second equality is valid only at $t=0$. It follows that
\begin{equation}
    \frac{\p^5 S_{ijk}}{\p t^5} \sim \bigg\langle f_i f_j \frac{\p}{\p x'_k}\frac{\p^2 p'}{\p t^2} \bigg\rangle = \frac{1}{4\pi} \frac{\p}{\p r_k} \int \frac{\dd^3 \bx''}{|\bx'' -\bx'|} \frac{\p}{\p x_l''}\frac{\p}{ \p x_m''} \langle f_l'' f_m'' f_i f_j \rangle.\label{d5S}
\end{equation}Here, and in what follows, we use ``$\sim$'' to relate expressions that have a common asymptotic scaling with ${r=|\bxp - \bx|}$ at large $r$. To determine the large-$r$ asymptotic of~\eqref{d5S}, we make use of the fact that $\langle f_l'' f_m'' f_i f_j \rangle - \langle f_l'' f_m''\rangle \langle f_i f_j \rangle=\langle f_l'' f_m'' f_i f_j \rangle_c $ + $\langle f_l''f_i \rangle \langle f_m'' f_j \rangle $ + $\langle f_l'' f_j \rangle \langle f_m'' f_i \rangle$, and, therefore, has convergent integral moments, by A-I. This means that we can make use of the large-$r$ expansion
\begin{equation}
    \frac{1}{|\bx''-\bx|} = \frac{1}{|\br + \bs|}= \frac{1}{r}+s_i \frac{\p}{\p r_i}\frac{1}{r}+s_i s_j \frac{\p}{\p r_i}\frac{\p}{\p r_j}\frac{1}{r} +  O\left(\frac{1}{r^4}\right),\label{r_expansion}
\end{equation}where $\bs \equiv \bx''-\bx'$, to write~\eqref{d5S} as
\begin{equation}
    \frac{\p^5 S_{ijk}}{\p t^5} \sim  \frac{\p}{\p r_k}\frac{\p}{\p r_l}\frac{\p}{ \p r_m} \frac{1}{r} \int \dd^3 \bs\, \big(\langle f_l'' f_m'' f_i f_j \rangle - \langle f_l'' f_m''\rangle \langle f_i f_j \rangle\big) + O\left(\frac{1}{r^5}\right).\label{Sr-4}
\end{equation}The expansion is justified because A-I guarantees convergence of the higher-order (in~$1/r$) terms.

Equation~\eqref{Sr-4} shows that the large-$r$ asymptotic of $\p_t^5 S_{ijk}$ decays like $1/r^4$. Thus, we expect the large-$r$ asymptotic of $S_{ijk}$ to decay like $1/r^4$ for $t>0$, unless even slower-decaying terms are present in higher time-derivatives of $S_{ijk}$. Let us check that they are not. Using Leibnitz' theorem, the derivatives in question can be written as
\begin{align}
    \frac{\p^n S_{ijk}}{\p t^n} & = \sum_{m_1+m_2+m_3 = n} \frac{n!}{m_1!\,m_2!\,m_3!} \bigg\langle \frac{\p^{m_1} u_i}{\p t^{m_1}} \frac{\p^{m_2} u_j}{\p t^{m_2}} \frac{\p^{m_3} u_k'}{\p t^{m_3}}\bigg\rangle \nonumber \\ & \sim  \sum_{m_1+m_2+m_3=n}\frac{n!}{m_1!\,m_2!\,m_3!}\bigg\langle \frac{\p^{m_1-1} f_i}{\p t^{m_1-1}} \frac{\p^{m_2-1} f_j}{\p t^{m_2-1}} \frac{\p^{m_3-1}}{\p t^{m_3-1}}\frac{\p p'}{\p x'_k}\bigg\rangle,\label{d^nS}
\end{align}In writing the final expression in~\eqref{d^nS}, i.e., in identifying that terms of this form are the ones that decay slowest with $r$ at large $r$, we used the facts that (i) correlators involving multiple instances of pressure will contain more powers of $1/r$ after the expansion~\eqref{r_expansion} is taken than those with a single instance, (ii) correlators of $f'_k$ (or its time derivatives) with a product of $f_i$ and $\p_j \p_t^2 p$ (or their time derivatives) are vanishingly small\footnote{This is because, after~\eqref{d2t_pAPP} is used to eliminate pressure, the correlation functions to be integrated in $\bx''$ contain both $f$ and $f'$. They are thus vanishingly small in the $r\to \infty$ limit, for any value of $\bx''$.}, and (iii) the terms in $\p_t^n S_{ijk}$ that can be formed by eliminating time-derivatives of $\bu$ in favour of spatial ones via $\p_t u_i = \p_j(u_i u_j)+\dots$ (thus producing a higher-order correlator in the sense of number of correlated fields) may be seen straightforwardly to be the same or even higher order in the expansion in $1/r$. Substituting for pressure using~\eqref{p_integral} and for $|\bx''-\bx'|^{-1}$ using~\eqref{r_expansion}, we have
\begin{align}
    \frac{\p^n S_{ijk}}{\p t^n} &\sim \sum_{m_1+m_2+m_3 = n} \frac{n!}{m_1!m_2!m_3!} \sum_{m_4+m_5=m_3-3}\frac{(m_3-3)!}{m_4!m_5!} \nonumber\\ & \quad\quad\times\frac{\p}{\p r_k} \int \frac{\dd^3 \bx''}{|\bx'' -\bx'|} \frac{\p}{\p x_l''}\frac{\p}{ \p x_m''}\bigg\langle \frac{\p^{m_1-1} f_i}{\p t^{m_1-1}} \frac{\p^{m_2-1} f_j}{\p t^{m_2-1}} \frac{\p^{m_4}f_l''}{\p t^{m_4}}  \frac{\p^{m_5}f_m''}{\p t^{m_5}}\bigg\rangle\ \nonumber\\
    & \sim \sum_{\{m_{\alpha}\}}\,\frac{\p}{\p r_k} \frac{\p}{\p r_l}\frac{\p}{ \p r_m} \frac{1}{r} \int \dd^3 \boldsymbol{s}\, \bigg( \bigg\langle \frac{\p^{m_1-1} f_i}{\p t^{m_1-1}} \frac{\p^{m_2-1} f_j}{\p t^{m_2-1}} \frac{\p^{m_4}f_l''}{\p t^{m_4}}  \frac{\p^{m_5}f_m''}{\p t^{m_5}}\bigg\rangle \nonumber \\&\phantom{\sum_{\{m_{\alpha}\}}\,\frac{\p}{\p r_k} \frac{\p}{\p r_l}\frac{\p}{ \p r_m}\left(\frac{1}{r}\right)}- \bigg\langle \frac{\p^{m_1-1} f_i}{\p t^{m_1-1}} \frac{\p^{m_2-1} f_j}{\p t^{m_2-1}} \bigg\rangle \bigg\langle\frac{\p^{m_4}f_l''}{\p t^{m_4}}  \frac{\p^{m_5}f_m''}{\p t^{m_5}}\bigg\rangle \bigg) + O\left(\frac{1}{r^5}\right)\nonumber\\&=O\left(\frac{1}{r^4}\right),\label{S_n_int}
\end{align}where, after the second equality, we have for brevity suppressed the combinatoric factors and the explicit forms of the sums over $\{m_{\alpha}\}$ that appear in the first line. In moving from the final expression in~\eqref{d^nS} to the first equality of~\eqref{S_n_int}, we used the fact that the slowest-decaying correlators that are obtained by eliminating $\p u''_i/ \p t$ using the Navier-Stokes equation are the ones containing $f''_i$, rather than the inertial terms or the pressure gradient. This is because the latter correlators are higher order in $1/r$, for the same reasons as explained in points (ii) and (iii) above. We conclude from~\eqref{S_n_int} that $S_{ijk}=O(r^{-4})$ for $t>0$, from which it follows that ${u^3 K(r) = S_{xxx} = O(r^{-4})}$. The term containing $K(r)$ in \eqref{dL/dt_APP} is therefore zero. 

\subsection{Asymptotic of \texorpdfstring{$\psi(r)$}{psi(r)}}

To establish the large-$r$ scaling of $\psi$, we evaluate time derivatives at $t=0$ of $\langle u_i f_j'\rangle$ in the same manner as we have done for the triple correlator $S_{ijk}$. Long-range pressure-induced correlations appear first in the third derivative of $\Psi_{ij}$ with respect to time, in terms such as
\begin{multline}
    \frac{\p^3 \Psi_{ij}}{\p t^3}\sim \bigg\langle f_i \frac{\p}{\p x'_j}\frac{\p^2 p'}{\p t^2} \bigg\rangle = \frac{1}{4\pi} \frac{\p}{\p r_j} \int \frac{\dd^3 \bx''}{|\bx'' -\bx'|} \frac{\p}{\p x_l''}\frac{\p}{ \p x_m''} \langle f_l'' f_m'' f_i \rangle \\ \sim  \frac{\p}{\p r_j}\frac{\p}{\p r_l}\frac{\p}{ \p r_m} \frac{1}{r} \int \dd^3 \bs\, \langle f_l'' f_m'' f_i \rangle + O\left(\frac{1}{r^5}\right).\label{d3Psi}
\end{multline}The final integral in~\eqref{d3Psi} vanishes because $\p\langle f_l'' f_m'' f_i \rangle/\p r_i=0$. Therefore, $\p_t^3 \Psi_{ij} \sim O(r^{-5})$. 

As we did in the previous section, we now check that there are no slower-decaying higher-order time derivatives of $\Psi_{ij}$. The $n^{\mathrm{th}}$ derivative of $\Psi_{ij}$ with respect to time satisfies:
\begin{equation}
     \frac{\p^n\Psi_{ij}}{\p t^n} = \sum^{n}_{m=0}\frac{n!}{m!(n-m)!} \bigg\langle \frac{\p^{m} f_i}{\p t^{m}} \frac{\p^{n-m} u_j'}{\p t^{n-m}}\bigg\rangle \\ \sim \sum^{n}_{m=0}\frac{n!}{m!(n-m)!} \bigg\langle \frac{\p^{m} f_i}{\p t^{m}} \frac{\p^{n-m}}{\p t^{n-m}}\frac{\p p'}{\p x'}\bigg\rangle.
\end{equation}The final expression follows because eliminating time derivatives of $u_i$ in favour of spatial derivatives by means of the Navier-Stokes equation introduces terms that are higher-order in $1/r$. An expansion analogous to~\eqref{d3Psi} then shows that $\p_t^n\Psi_{ij}= O(r^{-5})$, whence $\Psi_{ij}= O(r^{-5})$. Thus, $\psi=O(r^{-5})$, and hence the term containing $\psi$ in \eqref{dL/dt_APP} is zero. 

We conclude that $L$ is constant under the assumptions A-I and A-II stated at the start of this appendix.

\section{Long-range correlations induced by finite-band forcing\label{app:finite_band_forcing}}

In §\ref{sec:correlations}, we approached the problem of the conservation of the Saffman integral [i.e., the coefficient of $k^2$ in the expansion \eqref{Eexpansion} of $\mcE(k)$ as $k\to 0$] in terms of the strength of real-space correlations possessed by the forcing function. This is natural for turbulence occurring in nature or in a laboratory, where the spatial profile of the forcing may be measured, or prescribed. However, in numerical studies, forcing is commonly implemented in spectral space---in particular, it is common to force in a finite band of wavenumbers. Here, we examine the strength of correlations in real space induced by such a forcing.

As we have seen, the strongest long-range correlations in $\uf$ are typically induced by the last term on the right-hand side of \eqref{uf_soln}, so that, according to~\eqref{ff'},
\begin{equation}
    \lim_{r\to\infty}\uf = \lim_{r\to\infty}\int^t_0 \dd s \langle \bF(s) \bcdot \bF'(t) \rangle = \lim_{r\to\infty} \frac{1}{r^2}\frac{\p}{\p r} \big[r^3 H(t,r)\big].\label{forcing_correlator}
\end{equation}The function $H(t,r)$ is related to the spectral function $F(k)$ by a Fourier transform: dropping explicit dependence on time,
\begin{equation}
    \frac{1}{r^2}\frac{\p}{\p r} \big[r^3 H(r)\big] = 2 \int^{\infty}_0 \dd k F(k) \frac{\sin (kr)}{kr},
\end{equation}whence it follows that
\begin{equation}
    H(r) = 2 \int^{\infty}_0\dd k\,  F(k)\, \frac{\sin(kr)-kr\cos(kr)}{(kr)^3}.\label{h(E_f)_maintext}
\end{equation}

Let us now suppose that $F(k)$ is non-zero only for $k_1 \leq k \leq k_2 $. Under the assumption that $F$ is smooth, it is possible to show that $H(r)$ decays arbitrarily quickly (i.e., more quickly than any power law) at large $r$. This is intuitive, given the absence of any non-zero large-scale Fourier modes of $\bF$. A proof is as follows.

Integrating \eqref{h(E_f)_maintext} by parts, we have
\begin{equation}
    H(r) = \frac{2}{r^3} \int_0^{\infty}\dd k\frac{1}{k} \left(\frac{\p}{\p k}\frac{F}{k}\right)\sin(kr) = \frac{2}{r^3} \mathrm{Im} \left[ \int_0^{\infty}\dd k \frac{1}{k}\left(\frac{\p}{\p k}\frac{F}{k}\right)e^{ikr}\right],
\end{equation}where the boundary terms vanish because $F$ and all its derivatives vanish at $0$ and $\infty$. If we integrate by parts a further $n$ times, boundary terms continue to vanish, leaving
\begin{align}
    H(r) = \frac{2}{r^{3+n}} \mathrm{Im} \left[i^{n} \int_0^{\infty}\dd k \left(\frac{\p^n}{\p k^n}\frac{1}{k}\frac{\p}{\p k}\frac{F}{k}\right)e^{ikr}\right],
\end{align}so that
\begin{align}
    | H(r) | & = \frac{2}{r^{3+n}} \Bigg| \int_0^{\infty}\dd k \left(\frac{\p^n}{\p k^n}\frac{1}{k}\frac{\p}{\p k}\frac{F}{k}\right)e^{ikr} \Bigg| < \frac{2}{r^{3+n}} \int_0^{\infty}\dd k \Bigg| \frac{\p^n}{\p k^n}\frac{1}{k}\frac{\p}{\p k}\frac{F}{k}\Bigg| = \frac{C_n}{r^{3+n}},
\end{align}where $C_n$ is a finite constant. Thus, as $r\to \infty$, $| H(r) | \leq \const \times r^{-m}$, for any $m$. \textit{Q.E.D.}

However, $F(k)$ is not typically chosen to be smooth in numerical studies. Instead, $F(k)$ is often discontinuous at $k_1$ and $k_2$, with ${F(k<k_1)=F(k>k_2)=0}$. In this case, \eqref{h(E_f)_maintext} becomes, after integrating by parts,
\begin{equation}
    H(r) = \frac{2}{r^3}\left[\frac{F(k_1)\sin(k_1 r)}{k_1^2} -\frac{F(k_2)\sin(k_2r)}{k_2^2}\right] + \frac{2}{r^3} \int_{k_1}^{k_2}\dd k\frac{1}{k} \left(\frac{\p}{\p k}\frac{F}{k}\right)\sin(kr).\label{h_limitband}
\end{equation}The boundary term that has arisen in \eqref{h_limitband} is dominant over the integral, in the limit $r \gg 1/k_1$, $1/k_2$, by the Riemann-Lebesgue lemma. Subdominant terms in the expansion may be computed by continuing to integrate by parts: at every order, the resulting terms will be of the form (oscillating part at frequency $k_1$ or $k_2$)~$\times \,r^{-n}$. Thus, the effect of the discontinuity in $F(k)$ at $k_1$ and $k_2$ is to induce correlations that, at large distances, oscillate about zero with component wavenumbers $k_1$ and $k_2$, and an amplitude decreasing as a power law in $r$.

While a forcing of this sort will not inject energy into large-scale modes directly [i.e., via \eqref{forcing_correlator}], oscillatory behaviour will inevitably propagate into other correlators via the von K\'{a}rm\'{a}n-Howarth equation~\eqref{KH} and its higher-order analogues. It may then be the case that the oscillatory correlations induced in $\chi(r)$ decay more slowly with distance than the monotonically decaying ones implied by \eqref{h_limitband}. However, this should not affect the properties of the small-$k$ part of the energy spectrum, as it turns out that oscillatory behaviour in $\chi(r\to\infty)$, with the amplitude of oscillations decaying as a power law, always has negligible effect on $\mcE(k\to 0)$, independently of the power-law exponent.

To see why, let us suppose that
\begin{equation}
    \uu(r\geq r_0)= A\,\bigg[\frac{\sin(k_f r)}{r^n} + \frac{1}{r_0^n}\left(\frac{r_0}{r}\right)^m \bigg],
\end{equation}i.e., that the behaviour of $\uu$ at $r$ larger than some $r_0$ is the sum of an oscillation with frequency $k_f\gg 1/r_0$, with an amplitude that decays as $r^{-n}$, and a monotonically decaying part that decays as $r^{-m}$, where $m$ and $n$ are both larger than $1$ for finite $\mcE(k)$ as $k \to 0$. According to \eqref{E(k)exact}, the contribution of the oscillating part of $\uu$ to the energy spectrum at small $k$ is 
\begin{equation}
    \frac{Ak}{\pi}\int_{r_0}^{\infty}\dd r \frac{1}{r^{n-1}} \sin(k_f r) \sin (kr) = \frac{Ak^2}{\pi k_f r_0^{n-2}} \cos(k_f r_0)\left[ 1 + O\left(\frac{1}{k_f r_0}\right) \right],\label{osc1}
\end{equation}where we have integrated $\sin(k_f r)$ by parts and applied the Riemann-Lebesgue lemma. Meanwhile, the contribution of the monotonically decaying part is
\begin{equation}
    \frac{A k r_0^{m-n}}{\pi}\int_{r_0}^{\infty}\dd r \frac{1}{r^{m-1}} \sin (kr) = \frac{A F_m }{\pi k r_0^{n}} (k r_0)^{\min[m,\,3]} \left[ 1 + O\left(k r_0\right) \right], \label{osc2}
\end{equation}where $F_m$ is a positive number that depends on $m$. The equality in \eqref{osc2} is non-trivial: it is proven by direct application of asympototic methods to the integral, with a case distinction for $m<3$ and $m \geq 3$, or else via evalutation of the integral directly (resulting in a hypergeometric function), and taking the limit of small $k r_0$.

The right-hand side of \eqref{osc2} is always large compared to that of \eqref{osc1}; for $m > 3$, which is likely the only case of physical significance, the difference is $k_f r_0\gg 1$. This confirms our assertion that decaying oscillatory behaviour in $\chi(r\to\infty)$ has negligible effect on $\mcE(k\to 0)$: the latter is always dominated by any part of $\chi(r\to\infty)$ that decays monotonically. In conclusion, therefore, while forcing with a discontinuous spectrum does induce long-range correlations (as it must, owing to the sharp features induced in the spectrum of $\bu$), these are likely of little significance to the dynamics of the large scales.

\section{Asymptotic form of \texorpdfstring{$\chi(r)$}{X(r)} for turbulence with a power-law energy spectrum\label{app:asymptotics}}

In this Appendix, we present a derivation of \eqref{asymptotic_result}, which gives the longitudinal correlation function, $\chi(r)$, corresponding to turbulence with spectrum
\begin{equation}
    \mcE(k) = C k^a \label{E=Ck^a}
\end{equation}at scales $k_1 \leq k \leq k_2$, for values of $r$ satisfying $k_2^{-1}\ll r \ll k_1^{-1}$. The constant $C$ may be expressed in terms of the total energy contained at scales $k_1 \leq k \leq k_2$: since
\begin{equation}
\int^{k_2}_{k_1} \dd k \,\mcE(k) \simeq \begin{cases*}
                    C k_1^{a+1}/|1+a| & if  $a<-1$, \\
                    C \ln{(k_2/k_1)} & if  $a=-1$, \\
                    C k_2^{a+1}/(1+a) & if  $a>-1$,
                 \end{cases*}
\end{equation}we have
\begin{equation}
    C = \int^{k_2}_{k_1} \dd k \,\mcE(k) \times \begin{cases*}
                    |1+a| k_1^{-1-a} & if  $a<-1$, \\
                    1/\ln{(k_2/k_1)} & if  $a=-1$, \\
                    (1+a) k_2^{-1-a} & if  $a>-1$.
                 \end{cases*}\label{C}
\end{equation}

Let us start from~\eqref{f(E)_maintext},
\begin{equation}
    u^2 \chi(r) = 2 \int^{\infty}_0\dd k\,  \mcE(k)\, \frac{\sin(kr)-kr\cos(kr)}{(kr)^3},\label{f(E)}
\end{equation}and consider first the part of the integral with $k<k_1$, which we denote $J_1$. A straightforward Taylor expansion in $kr \ll 1 $ yields
\begin{align}
    J_1 & \equiv 2 \int^{k_1}_0 \dd k  \mcE(k) \frac{\sin(kr)-kr\cos(kr)}{(kr)^3} = \frac{2}{3} \int_0^{k_1} \dd k\, \mcE(k) +  O\left[(k_1 r)^3\right].
\end{align}Then~\eqref{f(E)} becomes
\begin{equation}
    u^2 \chi(r) = \frac{2}{3} \int_0^{k_1} \dd k\, \mcE(k) + J_2 + J_3 + O\left[(k_1 r)^3\right], \label{app_intermediate_step}
\end{equation}where $J_2$ and $J_3$ are integrals that correspond to the parts of \eqref{f(E)} with $k_1<k<k_2$ and $k>k_2$, respectively, so that
\begin{equation}
    J_2 + J_3 = 2 \int^{\infty}_{k_1} \dd k  \mcE(k) \frac{\sin(kr)-kr\cos(kr)}{(kr)^3}.\label{J2+J3}
\end{equation}Depending on the particular value of $a$, it can be convenient to integrate \eqref{J2+J3} by parts before choosing precise definitions for $J_2$ and $J_3$. This allows the boundary term at $k = k_2$ to be redistributed between $J_2$ and $J_3$. For some values of $a$ in~\eqref{E=Ck^a}, it will then be possible to show that $J_2$ is the dominant contribution to \eqref{app_intermediate_step}, allowing the dominant part of $\chi(r)$ to be computed despite our incomplete knowledge of $\mcE(k)$.  We proceed by considering different ranges of $a$ in turn, defining and computing $J_2$ and $J_3$ for each case. The final asymptotic expressions for $\chi(r)$ will be assembled using \eqref{app_intermediate_step} at the end of this Appendix.

\subsection{Case of \texorpdfstring{$a<2$}{a<2}}

\subsubsection{Calculation of \texorpdfstring{$J_2$}{J2}}

In this case, it is unnecessary to integrate by parts. We define
\begin{align}
    J_2 & \equiv 2 \int^{k_2}_{k_1} \dd k \mcE(k) \frac{\sin(kr)-kr\cos(kr)}{(k r)^3} =   2 C r^{-a-1}\int^{k_2 r}_{k_1 r} \dd x x^{a} \frac{\sin(x)-x\cos(x)}{x^3}.\label{J2}
\end{align}

For $-1<a<2$, \eqref{J2} is convergent for $k_1 r \to 0$, $k_2 r \to \infty$: to leading order in $k_1 r$ and $1/ k_2 r$,
\begin{align}
    J_2 = 
    - 2 C r^{-a -1}(a-1)\Gamma(a-2)\sin \frac{a \pi}{2},\quad\quad -1<a<2. \label{J2_-1<a<2}
\end{align}Therefore, $J_2$ is large compared to $J_1\sim Ck_1^{1+a}$ by a factor of $(k_1 r)^{-1-a}\gg 1$. 

For $a<-1$, the integral in \eqref{J2} is divergent at the lower limit. Its leading-order asymptotic as $k_1 r \to 0$ is
\begin{equation}
    J_2 = \frac{2Ck_1^{1+a}}{3|1+a|}\left\{1+O\left[(k_1 r)^2\right]\right\},\quad\quad a<-1,\label{J2_a<-1}
\end{equation}which is the same size as $J_1\sim Ck_1^{1+a}$, and is independent of $r$.

In the particular case of $a=-1$, taking the leading-order asymptotic of~\eqref{J2}, we have
\begin{equation}
    J_2 = \frac{2C}{3}\left[\ln{(k_1 r)}+O\left(1\right)\right],\quad\quad a=-1.\label{J2_a-1}
\end{equation}Despite appearances, $J_2$ does not blow up as $k_1 r \to 0$, because $C\to 0$ as $k_1 r\to 0$ in order for the total energy to be finite, as \eqref{C} shows. Nonetheless, $J_2$ dominates over $J_1\sim C$ in \eqref{app_intermediate_step}, by a factor of $\ln (k_1 r)$.

\subsubsection{Calculation of \texorpdfstring{$J_3$}{J3}}

In $J_3$, we define $x = k/k_2$ and integrate by parts:
\begin{align}
    J_3 & \equiv 2 \int^{\infty}_{k_2} \dd k \mcE(k) \frac{\sin(kr)-kr\cos(kr)}{(k r)^3} \nonumber \\ 
    & = - 2 k_2 \left[ \mcE(k_2 x) \frac{\cos(k_2 r x)+k_2 r x\sin(k_2 r x)}{(k_2 r x)^4} \right]^{\infty}_{1} \nonumber \\ & \phantom{=}\,\,+ 2 k_2 \int^{\infty}_{1} \dd x  \left[\frac{\p}{\p x} \frac{\mcE(k_2 x)}{(k_2 r x)^3} \right] \frac{\cos(k_2 r x)+k_2 r x\sin(k_2 r x)}{k_2 r x}.\label{RL}
\end{align}The boundary term at $\infty$ is exponentially small, by assumption, while the remaining integral is small compared to the boundary term by a factor of $(k_2 r)^{-1} \ll 1$, by the Riemann-Lebesgue lemma. This leaves
\begin{align}
    J_3 & = 2 k_2 \mcE(k_2 ) \frac{\cos(k_2 r )+k_2 r\sin(k_2 r)}{(k_2 r)^4} + O\left[ (k_2 r)^{-1} \right]\\
    & = 2 C r^{-a -1 } (k_2 r)^{a-3} \left[\cos(k_2 r )+k_2 r\sin(k_2 r)\right] + O\left[ (k_2 r)^{-1} \right],
\end{align}which is manifestly small compared to $J_2$ for all $a<2$.

\subsection{Case of \texorpdfstring{$2 \leq a \leq 3$}{2<a<3} \label{sec:2<a<3}}

In this range of $a$, it is more convenient to use integration by parts in \eqref{J2+J3} before splitting the integration domain. We have
\begin{equation}
    2 \int^{\infty}_{k_1}\dd k  \mcE(k) \frac{\sin(kr)-kr\cos(kr)}{(kr)^3}
    = 2 C (k_1 r)^{a-1} r^{-1 - a} + 2\int^{\infty}_{k_1}\dd k  \left[\frac{\p}{\p k}\frac{\mcE(k)}{kr}\right]\frac{\sin (kr)}{kr^2}, \label{asy_intermediate2}
\end{equation}where the boundary term at $\infty$ vanishes by assumption.

\subsubsection{Calculation of \texorpdfstring{$J_2$}{J2}}

We take
\begin{align}
    J_2 & \equiv 2 C (k_1 r)^{a-1} r^{-1 - a} + 2\int^{k_2}_{k_1} \dd k \left[ \frac{\p}{\p k}\frac{\mcE(k)}{kr}\right]\frac{\sin (kr)}{kr^2} \nonumber \\ & = 2 C (k_1 r)^{a-1} r^{-1 - a} + 2 Cr^{-1-a}(a-1)\int^{k_2r}_{k_1r} dx x^{a-3} \sin x.\label{J2_a>2}
\end{align}When $2\leq a<3$, the integral is convergent for $k_1 r \to 0$, $k_2 r \to \infty$, so
\begin{equation}
    J_2 = - 2 C r^{-a -1}(a-1)\Gamma(a-2)\sin \frac{a \pi}{2}\label{J2_2a3}
\end{equation} to leading order, the boundary term being small by a factor of $(k_1 r)^{a-1}\ll 1$. As before, $J_2$ is large compared to $J_1\sim Ck_1^{1+a}$ by a factor of $(k_1 r)^{-1-a}\gg 1$. 

If $a=3$, then \eqref{J2_a>2} does not converge as $k_2 r \to \infty$. Instead, we have
\begin{equation}
    J_2 = 4 Cr^{-4}\left[1-\cos(k_2 r)\right],\label{J2_a3}
\end{equation}to leading order in $k_1 r\ll 1$.

\subsubsection{Calculation of \texorpdfstring{$J_3$}{J3}}

Taking $J_3$ to be the part of \eqref{asy_intermediate2} not included in $J_2$ as defined in~\eqref{J2_a>2}, integrating by parts and applying the Riemann-Lebesgue lemma as in \eqref{RL} gives
\begin{align}
    J_3 & \equiv 2\int^{\infty}_{k_2} \dd k \left[\frac{\p}{\p k}\frac{\mcE(k)}{kr}\right]\frac{\sin (kr)}{kr^2} \nonumber \\
    & = 2 (a-1) C r^{-1-a}(k_2 r)^{a-3} \cos (k_2r) + O\left[ (k_2 r)^{-1} \right].
\end{align}Comparison with \eqref{J2_2a3} shows that $J_3$ is small compared to $J_2$ for $a<3$. If $a=3$, then the leading-order part of $J_3$ cancels with the term proportional to $\cos(k_2 r)$ in \eqref{J2_a3}. The remaining term in \eqref{J2_a3} is precisely \eqref{J2_2a3} with $a=3$, therefore we conclude that \eqref{J2_2a3} is valid for $2 \leq a \leq 3$, and provides the leading order part of \eqref{app_intermediate_step} in this range.

\subsection{Case of \texorpdfstring{$a > 3$}{a3}}

The procedure for $a>3$ is similar to the one followed in Appendix~\ref{sec:2<a<3}: we continue to integrate~\eqref{asy_intermediate2} by parts, split the resulting integral into pieces with $k<k_2$ and $k>k_2$, and use the Riemann-Lebesgue lemma to infer that the $k>k_2$ piece is subdominant. While a different number of integrations by parts will be required depending on the particular value of $a$, let us treat them all simultaneously, and integrate the integral appearing in \eqref{asy_intermediate2} by parts $n$ times. Using the complex representation of trigonometric terms for convenience, we have
\begin{align}
    2\int^{\infty}_{k_1}\dd k  \left[\frac{\p}{\p k}\frac{\mcE(k)}{kr}\right]\frac{\sin (kr)}{kr^2} & = \frac{2}{r^3}\mathrm{Im} \Bigg\{ \sum_{m=1}^n (-1)^{m-1}\left[\frac{e^{ikr}}{(ir)^m}\frac{\p^{m-1}}{\p k^{m-1}}\frac{1}{k}\frac{\p}{\p k}\frac{\mcE(k)}{k}\right]^{\infty}_{k_1} \nonumber \\ & \phantom{=}+ (-1)^n\int^{\infty}_{k_1} \dd k \frac{e^{ikr}}{(ir^n)} \frac{\p^{n}}{\p k^{n}}\frac{1}{k}\frac{\p}{\p k}\frac{\mcE(k)}{k}  \Bigg\}\\
    & = 2\mathrm{Im} \Bigg[ \sum_{m=1}^n i^{m}e^{i k_1 r} \frac{\Gamma (a-2)}{\Gamma (a-m-1)} (k_1 r)^{a-m-2} C r^{-a-1}  \nonumber \\ & \phantom{=}+  \frac{(-1)^n}{r^3}\int^{\infty}_{k_1} \dd k \frac{e^{ikr}}{(ir)^n} \frac{\p^{n}}{\p k^{n}}\frac{1}{k}\frac{\p}{\p k}\frac{\mcE(k)}{k}  \Bigg].\label{J2+J3_a>3}
\end{align}The last expression was obtained by simplifying the boundary term using the fact that $\mcE(k)= Ck^a$ around $k=k_1$.

\subsubsection{Calculation of \texorpdfstring{$J_2$}{J2}}

As before, we split \eqref{asy_intermediate2} into two components, $J_2$ and $J_3$. Formally, $J_2$ should be defined to include the $k=k_1$ boundary terms in \eqref{asy_intermediate2} and \eqref{J2+J3_a>3}. However, these are all small compared to $Cr^{-a-1}$ as long as we choose $n<a-2$. Doing so, we have
\begin{align}
    J_2 & = 2 \mathrm{Im}\left[ \frac{(-1)^n}{r^3}\int^{k_2}_{k_1} \dd k \frac{e^{ikr}}{(ir)^n} \frac{\p^{n}}{\p k^{n}}\frac{1}{k}\frac{\p}{\p k}\frac{\mcE(k)}{k}\right] \left\{ 1+O\left[(k_1 r)^{a-n-2}\right]\right\}\nonumber\\ & \simeq 2\frac{\Gamma(a-2)}{\Gamma(a-n-2)}(a-1)C r^{-1-a} \mathrm{Im} \left( i^n \int^{k_2r}_{k_1r}\dd x e^{ix} x^{a-n-3} \right).\label{J2_a>3}
\end{align}Now, if $-1<a-n-3<0$ (which is consistent with our earlier choice of $n<a-2$), the integral in \eqref{J2_a>3} is convergent as we take $k_1 r \to 0$, $k_2 r \to \infty$. In that case, \eqref{J2_a>3} becomes the now-familiar
\begin{equation}
    J_2 = - 2 C r^{-a -1}(a-1)\Gamma(a-2)\sin \frac{a \pi}{2}.\label{J2_anonint}
\end{equation}It follows  that \eqref{J2_anonint} is valid for all non-integer $a>3$, since then it is always possible to choose $n$ such that $-1<a-n-3<0$. 

If $a$ is an integer, we choose $n = a -3$ instead. In this case, \eqref{J2_a>3} becomes
\begin{equation}
    J_2 = 2(a-1)\Gamma(a-2)Cr^{-1-a}  \,\mathrm{Im} \big\{ i^{a} \left[e^{ik_2r}-1+O(k_1 r)\right] \big\}.\label{J2_aint>3}
\end{equation}

\subsubsection{Calculation of \texorpdfstring{$J_3$}{J3}}

Taking $J_3$ to be the part of \eqref{asy_intermediate2} not included in $J_2$ and using \eqref{J2+J3_a>3}, we have
\begin{align}
    J_3 & \equiv 2 \mathrm{Im}\left[ \frac{(-1)^n}{r^3}\int^{\infty}_{k_2} \dd k \frac{e^{ikr}}{(ir)^n} \frac{\p^{n}}{\p k^{n}}\frac{1}{k}\frac{\p}{\p k}\frac{\mcE(k)}{k}\right] \nonumber\\ & = 2 \mathrm{Im} \left(i^{n+1}e^{ik_2r}\right)\frac{\Gamma(a-2)}{\Gamma(a-n-2)}(a-1)Cr^{-1-a}(k_2 r)^{a-n-3}\nonumber 
    \\ & \phantom{=}+ 2 \mathrm{Im}\left[ \frac{i^{n+1}}{r^3}\int^{\infty}_{k_2} \dd k \frac{e^{ikr}}{r^{n+1}} \frac{\p^{n+1}}{\p k^{n+1}}\frac{1}{k}\frac{\p}{\p k}\frac{\mcE(k)}{k}\right].\label{J3_a>3}
\end{align}In the case of non-integer $a$, our choice of $a-n-3<0$ ensures the boundary term in~\eqref{J3_a>3} is small compared to $J_2$, while the integral is lower order still, by the Riemann-Lebesgue lemma. For integer $a$, our choice of $n=a-3$ means that the boundary term in \eqref{J3_a>3} cancels with the term that is proportional to $\mathrm{Im} (i^{a}e^{ik_2 r})$ in \eqref{J2_aint>3}. The remaining term in \eqref{J2_aint>3} is precisely the right-hand side of \eqref{J2_anonint}, which dominates over the remaining integral in \eqref{J3_a>3} (by the Riemann-Lebesgue lemma), provided that $a$ is odd.

However, special care must be taken when $a$ is an even integer greater than $3$. After the term proportional to $\mathrm{Im} (i^{a}e^{ik_2 r})$ in \eqref{J2_aint>3} cancels with its partner in \eqref{J3_a>3}, the other term in \eqref{J2_aint>3} is proportional to $\mathrm{Im}(i^a)$, which vanishes if $a$ is even. This means that $J_2 = Cr^{-1-a}\times O(k_1 r)$. The leading-order non-vanishing term in the asymptotic expansion in $k_2 r\gg 1$ is contained within the integral in \eqref{J3_a>3}. However, unlike for other values of $a$, this term cannot be extracted from~\eqref{J3_a>3} by any number of integrations by parts, as the boundary terms so generated always vanish. This is because the desired leading-order contribution to $\chi(r)$ comes from the part of $\mcE(k)$ with $k>k_2$, where the form of $\mcE(k)$ is unknown. At best, we can constrain the dependence on $r$ by noting that, for~$n=a-3$,
\begin{align}
     \Bigg| 2 \mathrm{Im}\left[ \frac{i^{n+1}}{r^3}\int^{\infty}_{k_2} \dd k \frac{e^{ikr}}{r^{n+1}} \frac{\p^{n+1}}{\p k^{n+1}}\frac{1}{k}\frac{\p}{\p k}\frac{\mcE(k)}{k}\right] \Bigg| & < 2r^{-1-a}   \int^{\infty}_{k_2} \dd k \Bigg| \frac{\p^{a-2}}{\p k^{a-2}}\frac{1}{k}\frac{\p}{\p k}\frac{\mcE(k)}{k} \Bigg|. \label{upperbound}
\end{align}The integral on the right-hand side of \eqref{upperbound} is independent of $r$. It has the same dimensions as $C$, and typically will be $\sim C$, in which case the integral in \eqref{J3_a>3} can be large compared to the $O(k_1 r)$ part of $J_2$, and therefore it provides the dominant contribution to $\chi(r)$ in \eqref{app_intermediate_step}. Thus, for even integer $a>3$, we are only able to conclude that $\chi(r)$ decays as $Cr^{-1-a}$ or faster as $r\to\infty$.

\subsection{The leading-order correction in \texorpdfstring{$k_1 r\ll 1$}{k1 r<<1} \label{sec:app-correction}}

The conclusion that $\chi(r\to\infty)\leq O(Cr^{-1-a})$ may fail if the contribution to $\chi(r)$ from $\mcE(k>k_2)$ is much smaller than the upper limit enforced by \eqref{upperbound}, with the result that the strongest long-range correlations are instead determined by the $k<k_1$ part of the spectrum. An example of this is the superposition of a small-scale velocity field  with exponentially decaying correlations and a second velocity field at much larger scales. On scales longer than those of the former field, correlations are dominated by the latter, so $\chi(r)\sim \const$ if $r$ is small compared to the characteristic scale of the large-scale field.

A convenient way to obtain this result formally, without the need to keep track of higher-order terms in the expansion in $k_1 r$ above, is to define an auxiliary spectrum
\begin{equation}
    \mathscr{E}(k) = \begin{dcases*}
                    C k^a & if  $k<k_1 $,\\
                    \mcE(k) & otherwise,
                 \end{dcases*}
\end{equation}i.e., $\mathscr{E}(k)$ is the spectrum obtained by extending the power-law behaviour of $\mcE(k)$ at $k_1<k<k_2$ to $k<k_1$. Let us denote by $\chi_{\mathscr{E}}(r)$ the function obtained by replacing $\mcE(k)$ by $\mathscr{E}(k)$ in \eqref{f(E)} while retaining the definition of $u^2$.  In this notation, $\chi(r)=\chi_{\mcE}(k)$ is the correlation function that we have been concerned with thus far. Then,~$\chi_{\mathscr{E}}$ may be computed by simply setting $k_1 = 0$ in the calculation presented above:
\begin{equation}
    \chi_{\mathscr{E}}(r) = \chi_{\mcE}(r)\big|_{k_1=0}.
\end{equation}Now we define $\Delta(k) = \mcE(k) - \mathscr{E}(k)$ to be the spectrum of the ``excess energy'' contained at large scales (which can be negative). Owing to the linearity of \eqref{f(E)} in $\mcE$, we have
\begin{equation}
    \chi_{\mcE}(r) = \chi_{\mathscr{E} + \Delta}(r) =  \chi_{\mathscr{E}}(r) + \chi_{\Delta}(r) = \chi_{\mcE}(r)\big|_{k_1=0} + \chi_{\Delta}(r).
\end{equation}
Thus, the finite-$k_1 r$ correction to the correlation function corresponding to $\mcE(k)$ is exactly the function obtained by replacing $\mcE(k)$ by $\Delta(k)$ in \eqref{f(E)} (again, while retaining the definition of $u^2$). This is easily computed to leading order in $k_1 r$: since $\Delta(k>k_1) = 0$,
\begin{equation}
    u^2 \chi_{\Delta}(r) \equiv 2 \int^{k_1}_0\dd k\,  \Delta(k)\, \frac{\sin(kr)-kr\cos(kr)}{(kr)^3} = \frac{2}{3} \int_0^{k_1} \dd k\, \Delta(k) +  O\left[(k_1 r)^3\right].\label{k1r_correction}
\end{equation}As anticipated above, the leading-order correction is $\chi_{\Delta}(r)\sim \const$. The size of the correction depends on the amount of ``excess energy'' contained at $k<k_1$, i.e., on the difference between the actual energy and the energy that would be present if the power law $\mcE(k)\propto Ck^a$ extended to $k<k_1$. We note that this excess energy can be negative, in which case $\chi_{\Delta}(r)$ will also be negative, resulting in a tendency to introduce anti-correlations.

It should, however, be noted that this discussion is largely academic: while it is formally possible that the leading-order term in the $k_1 r$ expansion, \eqref{k1r_correction}, will dominate~$\chi(r)$, this is an artificial situation, because correlations do not fall off exponentially quickly in real turbulence~\citep{BatchelorProudman56}. Instead, $\chi(r)$ typically decays as $r^{-6}$ in isotropic Batchelor turbulence (see \citealt{Davidson13} for a discussion). In the general case, it should therefore be expected that the integral in \eqref{J3_a>3} provides the dominant part of~$\chi(r)$.

\subsection{Final expression}

Assembling all the results derived in this appendix, and using \eqref{C} to eliminate~$C$, \eqref{app_intermediate_step} becomes~\eqref{asymptotic_result} [in the case of $a = 2,4,6,\dots$, we assume that the dominant correlations come from the part of $\mcE(k)$ with $k>k_2$; see the discussion in appendix~\ref{sec:app-correction}].

\section{Evolution of mean square momentum in forced turbulence \label{momentum_evolution_appendix}}

In this Appendix, we examine the evolution of $\blangle \bP_V^2 \brangle$ under the forced Navier-Stokes equation, as \cite{Davidson15} did for decaying turbulence. 

Ignoring viscous forces, the evolution of the total momentum contained within a volume~$V$ is given by
\begin{equation}
    \frac{\dd \bP_V}{\dd t} = -\int_{\p V} \bu (\bu \bcdot \dd \boldsymbol{S})-\int_{\p V} p\dd \boldsymbol{S} + \int_V \dd^3\bx \bF.\label{dP/dt}
\end{equation}The three terms appearing on the right-hand side are identified straightforwardly as the advection of momentum out of $V$, the net pressure force on $V$, and the net external forcing. Therefore,
\begin{equation}
    \frac{\dd \bP_V^2}{\dd t} = 2\int_V \dd^3 \bx' \bu' \bcdot \bigg[-\int_{\p V} \bu (\bu \bcdot \dd \boldsymbol{S})-\int_{\p V} p\dd \boldsymbol{S}+\int_V \dd^3\bx \bF\bigg].\label{dP^2dt_intermediate}
\end{equation}For simplicity, we shall assume that the correlation time of the forcing is short compared to the eddy-turnover time. In that case, \eqref{uf_soln} and \eqref{ff'} give
\begin{equation}
    \uf = \frac{1}{r^2}\frac{\p}{\p r}\left[r^3 H(t,r)\right],\label{uf_shorttc}
\end{equation}while isotropy also demands [cf. \eqref{uiuj(f)}]
\begin{equation}
    \langle u_i f_j'\rangle = \frac{1}{2r}\left[(r^2 H)' \delta_{ij} - H'(r) r_i r_j\right].\label{fifj_shorttc}
\end{equation}Taking an ensemble average of \eqref{dP^2dt_intermediate}, using \eqref{uf_shorttc} and \eqref{fifj_shorttc}, and restricting attention to spherical $V$ with radius $R$, one can show that
\begin{multline}
    \frac{\dd \langle \bP_V^2 \rangle}{\dd t}
     = 4\pi^2 R^2u^3 \int_0^{2R} \dd r \left[1-\left(\frac{r}{2R}\right)^2\right]\frac{1}{r}\frac{\p}{\p r}(r^4 K) \\ + 8\pi^2 R^2 \int_0^{2R} \dd r \left[1-\left(\frac{r}{2R}\right)^2\right]r^3 H\label{dP2/dt}.
\end{multline}The derivation of this equation is closely analogous to the one presented in \cite{Davidson15} for $\dd \langle \bP_V^2 \rangle / \dd t$ in decaying turbulence, to which we refer the reader for details. 

Equation \eqref{dP2/dt} shows that there are two relevant processes that can change the expectation value of the squared linear momentum contained in a volume of size $V$. The first, represented by the first term on the right-hand side, encodes the effect of the advection of momentum by the flow, the second the injection of momentum by the forcing. Both terms are at most $O(R^2)$ as $R\to\infty$, as long as $K(r\to\infty)=o(r^{-3})$ and $H(r\to\infty)=o(r^{-3})$. This makes sense, given that both effects are surface processes (as long as long-range correlations in the forcing are weak). This means that neither term can spontaneously generate $\blangle \bP_V^2 \brangle \propto R^3$ for arbitrarily large $R$, as is consistent with the conclusion of §\ref{sec:correlations}.

In fact, the net effect of the two terms on the right-hand side of~\eqref{dP2/dt} may be a scaling of $\dd \blangle \bP_V^2 \brangle/\dd t$ vs. $R$ even weaker than $R^2$, because there can be partial cancellation between them, or else because $\p (r^4 K)/\p r$ may change sign at some $r$. Of course, some cancellation between the two terms is inevitable: the forcing cannot perpetually increase $\blangle \bP_V^2 \brangle$ unchecked, because, even if turbulence could maintain very short-range correlations and so not grow a $k^2$ spectrum, there would still be a cascade of energy to smaller scales (encoded in the first term) resulting in the destruction of the local structures. Nonetheless, it appears robustly the case that the net size of the right-hand side of~\eqref{dP2/dt} must scale at most as $R^2$ for large $R$.

Let us then consider some $R\gg l$, and suppose that, indeed, the aggregate of the terms on the right-hand side of~\eqref{dP2/dt} scales as $R^2$. In that case, on dimensional grounds,
\begin{equation}
    \frac{\dd \langle \bP_V^2 \rangle}{\dd t} \sim u^3 R^2 l^3. \label{dP2/dt_rough}
\end{equation}
Suppose, roughly, that $\langle \bP_V^2 \rangle$ increases according to \eqref{dP2/dt_rough} until it saturates. Saturation occurs when $\langle \bP_V^2 \rangle \sim u^2 R^3 l^3$---this must be the case on dimensional grounds, because the $R$ dependence is fixed by \eqref{P2_asymptotic2} with $a=2$. We then find that the time taken for saturation at scale $R$ is $t_{\mathrm{sat}}\sim R/u$. Equivalently, the scale $R_c$ at which the growth of $\langle \bP_V^2 \rangle$ has just saturated at time $t$ is given by 
\begin{equation}
    R_c \sim u t. \label{Rc_sim_t}
\end{equation}That this should be the strongest-allowed scaling of $R_c$ vs. $t$ is intuitive: \eqref{Rc_sim_t} simply represents the limit on the growth of $R_c$ imposed by causality. At distances greater than $R_c \sim u t$, two points in the flow cannot have exchanged momentum, because there has not been enough time for local processes, namely, advection at speed $u$ (or else the cumulative effect of local forces acting on scales $\sim l$ with timescales $\sim l/u$), to act.

\section{Derivation of the passive-momentum equations \label{app:passive}}

In this Appendix, we show how the mode-coupling equation \eqref{mode_coupling_equation_kraichnan_mt} is obtained under the assumptions explained in §\ref{sec:passive_theory}. Our goal is to compute the evolution of the spectrum $\mcE_{\bw}(t, k)$ of the passive vector field $\bw$ satisfying~\eqref{passiveNS}. To do this, we shall derive an evolution equation for the correlation function $\blangle w_i (t, \bk) w_j (t, \bk')\brangle$.

\subsection{Homogeneous and isotropic forms of relevant correlators}

Let us first note the restrictions imposed by symmetries on the various correlators involved.
Due to statistical homogeneity of $\bw$, this function is restricted to satisfy
\begin{align}
    \blangle w_i (t, \bk) w_j (t, \bk')\brangle = (2\pi)^3 \delta ( \bk + \bk') \Psi_{ij}(t, \bk). \label{ww}
\end{align}The form of the tensor $\Psi_{ij}(t, \bk)$ is further restricted by isotropy and incompressibility:
\begin{align}
    \Psi_{ij}(t, \bk) = \frac{1}{2}\Psi(t, k) \mcP_{ij} (\bk),
\end{align}where
\begin{equation}
    \mcP_{ij}(\bk) = \delta_{ij} - \frac{k_i k_j}{k^2}
\end{equation}is the usual projection operator onto the plane perpendicular to $\bk$, and the isotropic function $\Psi(t, k)$ is related to $\mcE_{\bw}(t, k)$ via
\begin{equation}
    \Psi(t, k) = 2\pi^2 \frac{\mcE_{\bw}(t, k)}{k^2}.
\end{equation}

In order to compute~\eqref{ww}, we require similar correlation functions for $\bu$ and $\bF_{\bw}$. In particular, we shall need $\blangle u_i(t, \bk) u_j (t', \bk')\brangle$, whose general form is also restricted by homogeneity and statistical invariance of $\bu$ in time:
\begin{align}
    \blangle u_i(t, \bk) u_j (t', \bk')\brangle & = (2\pi)^3 \delta (\bk + \bk') \kappa_{ij}(\bk, t-t'). \label{uu_kraichnan}
\end{align}As explained in §\ref{sec:passive_theory}, we shall take $\bu$ to have zero correlation time:
\begin{equation}
    \kappa_{ij}(\bk, t-t') = \kappa_{ij}(\bk)\delta(t-t').
\end{equation}Together, incompressibility and isotropy further imply
\begin{align}
    \kappa_{ij}(\bk) & = \kappa (k) \mcP_{ij}(\bk).
\end{align}As explained in §\ref{sec:passive_theory}, we shall assume $\bu$ to have a single scale, so that the isotropic function $\kappa(k)$ is
\begin{equation}
    \kappa (k) = \kappa_0 \delta (k-k_f),
\end{equation}

Finally, we shall require $\blangle f_i(t, \bk) f_j (t', \bk')\brangle$, which satisfies
\begin{align}
    \blangle f_i(t, \bk) f_j (t', \bk')\brangle & = \frac{1}{2}(2\pi)^3 \delta (\bk + \bk') \delta (t-t')\Phi(k)\mcP_{ij}(\bk)\label{fkfk}
\end{align}for the same reasons as the other two correlation functions.

\subsection{Derivation of the mode-coupling equation}

Equation~\eqref{passiveNS} reads, after dropping the subscript $\bw$ from $p_{\bw}$ and $\bF_{\bw}$ (there can be no confusion with the corresponding fields for $\bu$, as $\bu$ is now prescribed artificially),
\begin{align}
    \p_t w_i = - \p_j (u_j w_i) - \p_i p + \nu \nabla^2 w_i + f_i. 
\end{align}In Fourier space, this is
\begin{align}
    \p_t w_i(\bk) + \nu k^2 w_i(\bk) = - i k_l \mcP_{ip}(\bk)\intpikp u_l(\bk') w_p(\bk - \bk') + f_i(\bk),  \label{dtw}
\end{align}where we have taken $\bF$ to be solenoidal---there is no loss of generality here, as if one is interested in a non-solenoidal forcing, one can interpret $\bF$ to be its solenoidal part. Then,
\begin{multline}
     \p_t \blangle w_i(\bk) w_j(\bk') \brangle + \nu \left( k^2+k'^2 \right) \blangle w_i(\bk) w_j(\bk') \brangle  \\ = - i \intpikpp \bigg[ k_l \mcP_{ip}(\bk) \blangle u_l(\bk'') w_p(\bk - \bk'') w_j(\bk') \brangle + k'_l \mcP_{jp}(\bk') \blangle u_l(\bk'') w_p(\bk' - \bk'') w_i(\bk) \brangle\bigg]\\
    +\langle  f_i(\bk)w_j(\bk')\rangle+\langle f_j(\bk')w_i(\bk)\rangle.\label{dt<ww>}
\end{multline}

In order to simplify the correlators appearing on the right-hand side of~\eqref{dt<ww>}, we can make use of the zero-correlation time assumption for $\bu$ and $\bF$. In the latter case, we note that integrating \eqref{dtw} over time, multiplying by $f_i(\bk)$ and ensemble-averaging yields
\begin{align}\langle w_i(\bk)f_j(\bk')\rangle & = 
        \int^t\dd t' \Bigg[ -\nu k^2 \langle w_i(\bk, t')f_j(\bk',t) \rangle \nonumber\\ 
        & \phantom{=}- i k_l \mcP_{ip}(\bk)\intpikp \langle u_l(\bk',t') w_p(\bk - \bk',t')f_j(\bk',t)\rangle + \langle f_i(\bk,t')f_j(\bk',t) \rangle\Bigg] \nonumber
       \\ 
       & = \frac{1}{2}(2\pi)^3 \delta (\bk + \bk') \Phi(k)\mcP_{ij}(\bk),\label{wf}
\end{align}where, in the second equality, we have taken the  contributions of the first two correlators inside the $t'$ integral to vanish, as demanded by causality, and used~\eqref{fkfk} to express the final correlator in terms of $\Phi(k)$.

A similar strategy may be employed to treat the triple correlations appearing in~\eqref{dt<ww>}. Multiplying by $u_l(t,\boldsymbol{k}_3)$ the formal solution of the unaveraged version of~\eqref{dt<ww>}, which may be written as
\begin{multline}
    w_i(\bk_1) w_j(\bk_2) = \int^t \dd t' \Bigg\{- \nu \left( k_1^2+k_2^2 \right) w_i(\bk_1, t') w_j(\bk_2, t') \\ - i \intpikpp \bigg[ k_{1n} \mcP_{iq}(\bk_1) u_n(\bk'', t') w_q(\bk_1 - \bk'', t') w_j(\bk_2, t')  \\+ k_{2n} \mcP_{jq}(\bk_2) u_n(\bk'', t') w_q(\bk_2 - \bk'', t') w_i(\bk_1, t') \bigg] \\+f_i(\bk_1,t')w_j(\bk_2,t')+f_j(\bk_2,t')w_i(\bk_1,t')\Bigg\}, \label{correlation_fn_new}
\end{multline}ensemble averaging, and, finally, using the short-correlation-time assumption to split correlators by invoking causality, yields
\begin{multline}
    \blangle w_i(t, \bk_1) w_j(t, \bk_2) u_l(t, \bk_3) \brangle = - \frac{i}{2} (2 \pi)^3 \kappa_{ln}(\bk_3) \delta (\bk_1 + \bk_2 + \bk_3) \\\times\big[k_{1n}\mcP_{iq}(\bk_1) \Psi_{qj} ( \bk_1 + \bk_3 ) + k_{2n}\mcP_{jq}(\bk_2) \Psi_{iq} ( \bk_2 + \bk_3 )  \big].\label{tripleaverages_new}
\end{multline}

Using \eqref{wf} and \eqref{tripleaverages_new},~\eqref{dt<ww>} becomes
\begin{align}
     \p_t \blangle w_i(\bk) w_j(\bk') &\brangle + \nu \left( k^2+k'^2 \right) \blangle w_i(\bk) w_j(\bk') \brangle = \nonumber -\frac{1}{2} \intkpp \kappa_{ln}(\bk'') \delta (\bk + \bk') \times\\ \bigg\{ k_l \mcP_{ip}(\bk) & \bigg[ (k_n-k''_n)\mcP_{pq}(\bk - \bk'') \Psi_{qj} ( \bk ) +  k'_{n}\mcP_{jq}(\bk') \Psi_{pq} ( \bk' + \bk'' ) \bigg] \nonumber \\  +\, k'_l \mcP_{jp}(\bk') & \bigg[ k_n \mcP_{iq}(\bk) \Psi_{qp} ( \bk + \bk'' ) + (k'_{n}-k''_{n})\mcP_{pq}(\bk'-\bk'') \Psi_{iq} ( \bk') \bigg]\bigg\}\nonumber\\
     & \quad \quad \quad + (2\pi)^3 \delta (\bk + \bk') \Phi(k)\mcP_{ij}(\bk). \label{intermediate}
\end{align}Integrating \eqref{intermediate} over $\bk'$, and taking the trace (i.e., setting $i=j$ and summing over $i$), we obtain, after a small amount of algebra,
\begin{multline}
      \p_t \Psi(k) + 2 \nu k^2 \Psi(k) \\= - \frac{1}{2}\intpikpp \kappa_{ln}(\bk'') k_l k_n \mcP_{pq} (\bk) \mcP_{pq} (\bk + \bk'') \left[\Psi(k) - \Psi(|\bk +\bk''|)\right] + 2 \Phi(k),\label{intermediate2}
\end{multline}where a useful step in the simplification of the integrand was noting that $\kappa_{ln}(\bk'')k''_n \propto \mcP_{ln} (\bk'')k''_n=0$.

Now, $\Psi(k)$ may be brought outside the first integral in \eqref{intermediate2}, giving rise to a turbulent-viscosity term, $-\nu_T(k)k^2 \Psi(k)$, where
\begin{align}
    2 \nu_T(k) & \equiv \frac{1}{2}\intpikpp \kappa_{ln}(\bk'') \frac{k_l k_n}{k^2} \mcP_{pq} (\bk) \mcP_{pq} (\bk + \bk'') \nonumber\\ 
    & = \frac{1}{(2\pi)^3}  \int_0^{\infty} \dd k'' k''^2 \kappa (k'')  \int \dd^2 \Omega''  \left[1-\frac{(\bk\bcdot\bk'')^2}{k^2 k''^2}\right] \left[1 - \frac{(\bk \times \bk'')^2}{2 k^2 (\bk + \bk'')^2}\right] \nonumber\\   & \equiv \frac{1}{(2\pi)^2}  \int_0^\infty \dd k''\, k''^2 \kappa (k'')  G\left(\frac{k''}{k}\right).\label{turbulent_viscosity}
\end{align}The function $G(x)$ may be obtained by choosing a spherical coordinate system about $\bk$. Computing the angle integral then yields
\begin{equation}
    G(x) = \frac{1}{96 x^3} \bigg[ 2 x (3+ 53x^2 - 11x^4 +3x^6) - 3 (1-x^2)^4 \,\log\left|\frac{1+x}{1-x}\right| \bigg]. \label{G}
\end{equation}Finally, substituting $\kappa(k'')=\kappa_0 \delta(k''-k_f)$, we get
\begin{equation}
    2 \nu_T(k) = \frac{1}{(2\pi)^2} \kappa_0 k_f^2\,  G\left(\frac{k_f}{k}\right).
\end{equation}

The other term inside the integral in \eqref{intermediate2}, containing $\Psi(|\bk+\bk''|)$, is handled in the following manner:
\begin{align}
    &\frac{1}{2}\intpikpp \kappa_{ln}(\bk'') k_l  k_n \mcP_{pq} (\bk) \mcP_{pq} (\bk + \bk'') \Psi ( \big| \bk + \bk'' \big|) \nonumber \\
    &= \frac{1}{2}\frac{1}{(2\pi)^3}\int_0^{\infty} \dd k'' k''^2 \kappa (k'')  \int \dd^2 \Omega''  k_l  k_n\mcP_{ln}(\bk'')\mcP_{pq} (\bk) \mcP_{pq} (\bk + \bk'')  \Psi ( \big| \bk + \bk'' \big|) \nonumber\\
    &= \frac{1}{2}\frac{1}{(2\pi)^3} \int_0^{\infty} \dd k'' k''\kappa (k'')\int^{2\pi}_0 \dd \varphi'' \int^{k+k''}_{|k-k''|} \dd k' \frac{k'}{k} k_l  k_n \mcP_{ln}(\bk'') \mcP_{pq} (\bk) \mcP_{pq} (\bk')\Psi ( k')\nonumber\\
    &= \frac{1}{(2\pi)^2} \int_0^{\infty} \dd k'' k''\kappa (k'') \int^{k+k''}_{|k-k''|} \dd k' \frac{k'}{k} K(k'', k', k) \Psi ( k'),
\end{align}where, in the second equality, we set $\bk' = \bk + \bk''$, chose the polar axis to be along $\bk$, and changed variables from the polar angle $\theta ''$ to $k'=|\bk'|$, noting that $k' \dd k' = -k k'' \sin\theta''\dd\theta''$. In the final equality, we have used $\bk\bcdot \bk''=(k'^2-k^2-k''^2)/2$ and $\bk\bcdot\bk' = (k'^2+k^2 - k''^2)/2$ to write the intergration kernel $K(k'', k',k)$ in terms of the magnitudes of $k$, $k'$ and $k''$ only. Explicitly, it is
\begin{multline}
    K(k'', k',k)= -\frac{(k^2-k''^2)^4}{32 k^2 k''^2}\frac{1}{k'^2}-\frac{(k^2-k''^2)^3}{8k^2 k''^2}+\frac{(5k^4+6k^2k''^2 - 3k''^4)}{16k^2 k''^2}k'^2\\- \frac{k^2 - k''^2}{8k^2k''^2}k'^4 - \frac{1}{32 k^2 k''^2}k'^6.
\end{multline}Finally, taking $\kappa(k'')=\kappa_0 (k''-k_f)$ and integrating over $k''$, we find
\begin{multline}
    \frac{1}{2}\intpikpp \kappa_{ln}(\bk'') k_l  k_n \mcP_{pq} (\bk) \mcP_{pq} (\bk + \bk'') \Psi ( \big| \bk + \bk'' \big|) \\= \frac{\kappa_0}{(2\pi)^2} k_f \int^{k+k_f}_{|k-k_f|} \dd k' \frac{k'}{k}   K(k_f, k', k) \Psi ( k').\label{modecoupling}
\end{multline}

Using \eqref{turbulent_viscosity} and \eqref{modecoupling}, \eqref{intermediate2} becomes
\begin{multline}
      \p_t \Psi(k) + 2 \left[\nu+\nu_T(k)\right] k^2 \Psi(k) = \frac{\kappa_0}{(2\pi)^2} k_f \int^{k+k_f}_{|k-k_f|} \dd k' \frac{k'}{k}    K(k', k) \Psi ( k') + 2 \Phi(k),
\end{multline}where we have suppressed the explicit dependence of $K(k_f, k', k)$ on $k_f$.
Using ${\mcE(k)= k^2 \Psi(k) / 2\pi^2}$, and identifying ${F=k^2\Phi(k)/\pi^2}$ as the spectrum of the energy injection, the evolution equation for the spectrum of the passive vector becomes
\begin{align}
      & \p_t \mcE(k) + 2 \left[\nu + \nu_T(k)\right] k^2 \mcE(k) = \frac{\kappa_0 \kflow}{ (2\pi)^2} k \int^{k+\kflow}_{|k-\kflow|} \frac{\dd k' }{k'}   K(k',k) \mcE ( k') + F(k),\label{mode_coupling_equation_kraichnan}
\end{align}which is~\eqref{mode_coupling_equation_kraichnan_mt}, as promised. This is the analogue for the passive vector field $\bw$ of the similar mode-coupling equations for the magnetic field~\citep{KulsrudAnderson92} and the passive scalar~\citep{Schekochihin04_passivescalar}.

\subsection{Small-\texorpdfstring{$k$}{k} limit of the mode-coupling equation}

In the present study, we require only the limit of~\eqref{mode_coupling_equation_kraichnan} with $k\ll k_f$. Noting that $\lim_{x\to\infty}G(x)=4/5$, and
\begin{equation}
    \lim_{k,\,q\,\to 0}\left[\frac{k_f k}{k_f + q}K(k'=k_f+q,k)\right]=\frac{k^4 - q^4 }{2 k},
\end{equation}we find that~\eqref{mode_coupling_equation_kraichnan} becomes
\begin{align}
      & \p_t \mcE(k) + \beta k_f^2 k^2 \mcE(k) = \frac{5}{8} \frac{\beta}{k} \int^{k}_{-k} \dd q (k^4 - q^4) \mcE ( k_f+q) + F(k),\label{smallk}
\end{align}where $\beta = \kappa_0 / 5\pi^2$ and we have assumed that $\nu_T\gg \nu$. This is~\eqref{penultimate}.

\section{Alternative derivation of the scaling of \texorpdfstring{$\langle\bP_V^2\rangle$}{the mean square momentum} vs. \texorpdfstring{$R$}{R} in turbulence forced with long-range correlations\label{app:anomalousP2}}

In this Appendix, we show how the scalings~\eqref{anomalousP2}, viz.,
\begin{equation}
    \langle \bP^2_V \rangle \propto \begin{dcases*}
            R^{7-b} & if  $l\ll R\ll R_c$, \\
            R^{5-b} & if  $R \gg R_c$ \& $b<3$, \\
            R^{2} & if  $R \gg R_c$ \& $3<b<4$, \\
            \end{dcases*}  
\end{equation}which were derived in~§\ref{sec:passive_theory} from the spectral evolution equation~\eqref{large_scale_equation} for the passive vector field,
can be obtained by instead considering momentum diffusion as the sum of many instances of a decaying passive vector field. 

Between the times $t_{\mathrm{inj}}$ and $t_{\mathrm{inj}}+\dd t_{\mathrm{inj}}$, the forcing causes the spectral energy density to increase by~${\dd\mcE_{\bw} = Ck^b \dd t_{\mathrm{inj}}}$. As~\eqref{large_scale_equation} is linear, we may consider the evolution of the spectral-energy-density increment~$\dd\mcE_{\bw}$ in isolation from the rest of $\mcE_{\bw}$. Assuming that $k\ll k_f$ and $b<4$,~\eqref{large_scale_equation} implies that $\dd\mcE_{\bw}$ decays with time $t$ according~to
\begin{equation}
    \dd\mcE_{\bw}(k, t) = C k^b \,\dd t_{\mathrm{inj}}\, \exp(-\beta k_f^2 k^2 \Delta t ),\label{decaying}
\end{equation}where $\Delta t = t-t_{\mathrm{inj}}>0$. The energy-containing scale and total energy associated with $\dd \mcE_{\bw}$ are respectively ${\lambda \sim (\beta k_f^2 \Delta t)^{1/2}}$ (note that $\lambda$ is not the same as $l$, the latter being the energy-containing scale associated with the full field $\bw$, rather than solely the increment generated at $t_{\mathrm{inj}}$) and
\begin{equation}
    \dd E = C \,\dd t_{\mathrm{inj}}\, \int_0^{\infty}\dd k \,k^b \exp(-\beta k_f^2 k^2 \Delta t ) = \frac{1}{2} \Gamma \left( \frac{1+b}{2}\right)C \dd t_{\mathrm{inj}}\, (\beta k_f^2 \Delta t)^{-(1+b)/2}.
\end{equation}

Let us now determine the contribution of $\dd\mcE_{\bw}(k, t)$ to the mean square momentum $\dd \langle \bP^2_V \rangle$ contained within a spherical control volume $V$ with radius $R$. There are three different cases to consider:

\begin{enumerate}
    \item If $R\ll \lambda$, then \eqref{P2_asymptotic2} gives $\dd \langle \bP^2_V \rangle \propto R^{6}\,\dd E \propto R^{6} (\Delta t)^{-(1+b)/2} \dd t_{\mathrm{inj}}$, as the decaying field has reached scales much larger than $R$, so there is little variation of $\bw$ within $V$.
    \item If $R\gg \lambda$ and $b<3$, then \eqref{P2_asymptotic2} gives $\dd \langle \bP^2_V \rangle \propto R^{5-b} \dd t_{\mathrm{inj}}$, with no time dependence, owing to the fact that the momentum scaling is set by the power-law part of \eqref{decaying} [see~\eqref{P2_asymptotic} and surrounding discussion], which is constant in time, a consequence of momentum conservation.
    \item If $R\gg \lambda$ and $3<b<4$, then \eqref{P2_asymptotic2} gives $\dd \langle \bP^2_V \rangle \propto R^2 \, c(\Delta t) \dd t_{\mathrm{inj}}$. An as-yet undetermined function $c(\Delta t)$ appears here because the integral~\eqref{P2_asymptotic} is dominated by the evolving contribution from the energy-containing scales, rather than from the invariant $k^b$ tail of \eqref{decaying}. In principle, we can determine $c(\Delta t)$ by substituting~\eqref{decaying} into our equation for $\chi(r)$ in terms of $\mcE(k)$, \eqref{f(E)_maintext}, and then evaluating the integral \eqref{P2_bp} exactly. However, some effort can be spared by using the self-similar form of \eqref{decaying}. Substituting $\mcE(k) = k^b g(k^2 \Delta t)$ in \eqref{f(E)_maintext}, and changing the integration variable to $x=kr$, we have
\begin{equation}
    u^2 \chi(r) = \frac{2}{r^{1+b}}\int_0^{\infty} \dd x x^b g \left(\frac{x^2 t}{r^2}\right)\frac{\sin x - x \cos x}{x} \equiv \frac{2}{r^{1+b}} G\left(\frac{\Delta t}{r^2}\right).
\end{equation}Then, with a change of variables to $y = r'/\sqrt{\Delta t}$, \eqref{P2_bp} becomes
\begin{equation}
    \langle \bP^2_V \rangle = 4\pi^2 u^2 (\Delta t)^{(3-b)/2}\int^{2R}_0 \dd r\,r \int^{r/\sqrt{\Delta t}}_0 \dd y\,y^{2-b}\,G\left(\frac{1}{y}\right).\label{E5}
\end{equation}For $b>3$, \eqref{P2_asymptotic2} demands that $\langle \bP_{V}^2\rangle \propto R^2$ at any fixed time, so the integral over $y$ in~\eqref{E5} must be independent of its upper limit. In this case, the result of the double integration is time independent, so we deduce $c(\Delta t) = (\Delta t)^{(3-b)/2}$.
\end{enumerate}
 
To summarise these results, we have established that:
\begin{equation}
    \dd \langle \bP^2_V \rangle \propto \dd t_{\mathrm{inj}}\, \begin{dcases*}
            R^{6}\, (\Delta t)^{-(1+b)/2} & if  $R\ll \lambda\sim (\beta k_f^2 \Delta t)^{1/2}$, \\
            R^{5-b} & if  $R \gg \lambda \sim (\beta k_f^2 \Delta t)^{1/2}$ \& $b<3$, \\
            R^{2} \, (\Delta t) ^{(3-b)/2}& if  $R \gg \lambda\sim (\beta k_f^2 \Delta t)^{1/2}$ \& $3<b<4$. \\
            \end{dcases*}\label{P2_decaying}
\end{equation}

Under the passive-field assumption, and owing to the fact that the forcing at any given time is uncorrelated with the forcing at any other time, the total squared momentum contained within a volume of continually forced turbulence may be obtained as the sum of contributions from passive decays initialised continuously and uniformly in time. Let us consider a fixed sphere of radius $R$ and let $t_c = R^2/\beta k_f^2$ be the time at which the energy-containing scale of the decaying field initialised at $t_{\mathrm{inj}}=0$ reaches the scale $R$. Then, for $t<t_c$, each part of the sum comes from the $R \gg \lambda$ part of \eqref{P2_decaying}, so
\begin{equation}
    \langle \bP^2_V \rangle = \int \dd \langle \bP^2_V \rangle = \int^{t}_0 \dd t_{\mathrm{inj}} \frac{\dd \langle \bP^2_V \rangle}{\dd t_{\mathrm{inj}}}\propto \begin{dcases*}
            R^{5-b} t & if $b<3$, \\
            R^{2} t^{(5-b)/2} & if  $3<b<4$. \\
            \end{dcases*} \label{t<tc}
\end{equation}If, instead, $t>t_c$, then
\begin{equation}
    \langle \bP^2_V \rangle = \int \dd \langle \bP^2_V \rangle = \int^{t}_{t-t_c} \dd t_{\mathrm{inj}} \frac{\dd \langle \bP^2_V \rangle}{\dd t_{\mathrm{inj}}} + \int^{t-t_c}_{0} \dd t_{\mathrm{inj}} \frac{\dd \langle \bP^2_V \rangle}{\dd t_{\mathrm{inj}}}.\label{Pdecaysintegral}
\end{equation}The first integral encodes all the decays that have not yet reached the scale $R$, as ${t-t_{\mathrm{inj}}<t_c}$ for them. Therefore, we may again substitute for $\dd \langle \bP^2_V \rangle$ using the $R \gg \lambda$ part of~\eqref{P2_decaying}, giving
\begin{align}
    \int^{t}_{t-t_c} \dd t_{\mathrm{inj}} \frac{\dd \langle \bP^2_V \rangle}{\dd t_{\mathrm{inj}}} & \propto \begin{dcases*}
            R^{5-b} t_c & if $b<3$, \\
            R^{2} t_c^{(5-b)/2} & if  $3<b<4$, \\
            \end{dcases*} \nonumber \\[5pt]
            & \propto R^{7-b}.
\end{align}The second integral in \eqref{Pdecaysintegral} encodes all the decays that were initialised at $t_{\mathrm{inj}}< t-t_c$, so have reached a scale larger than $R$ at time $t$. It is
\begin{align}
    \int^{t}_{t_c} \dd t_{\mathrm{inj}} \frac{\dd \langle \bP^2_V \rangle}{\dd t_{\mathrm{inj}}} \propto R^6 \int^{t-t_c}_{0} \dd t_{\mathrm{inj}} (t-t_{\mathrm{inj}})^{-(1+b)/2}\propto R^6\, t_c^{(1-b)/2} \propto R^{7-b}.
\end{align}Thus, decays that have reached $\lambda>R$ and ones with $\lambda<R$ both contribute a term $\propto R^{7-b}$ to $\langle \bP_V^2\rangle$, explaining the scaling found in~\eqref{anomalousP2}. In particular, this calculation explains why forcing with different values of $3<b<4$ results in saturation with different scalings for $\langle \bP_V^2\rangle$, despite both having $\langle \bP_V^2\rangle \propto R^2$ at $R>R_c$: the total energy contained within an initally more diffuse blob of momentum decays more slowly than an initally less diffuse one (i.e., one with smaller $b$), even though the energy-containing scales of both grow at the same rate.

As an aside, we note that the saturated spectrum is also derivable directly from the ensemble of decaying states, \eqref{decaying}:
\begin{align}
    \mcE_{\bw}(k, t) = \int \dd \mcE_{\bw}(k,t) = \int^t_0 \dd t_{\mathrm{inj}} C k^b \exp(-\beta k_f^2 k^2 (t-t_{\mathrm{inj}}) ) = \left[1- e^{-\beta k^2 k_f^2 t }\right]\frac{C k^{b-2}}{\beta k_f^2},
\end{align}which is~\eqref{general_timedepsoln} [with the $\beta \mcE_{\bw}(k_f)k^4$ term neglected].

\section{Non-solenoidal forcing \label{app:nonsolenoidal}}

In this Appendix, we formalise the discussion of non-solendoidal forcing in~§\ref{sec:nonsolenoidal}. The essential result is due to \cite{Saffman67}, which may be stated in the notation of present study as follows:\vspace{3mm}

\textbf{Theorem} (Saffman)\textbf{.} \textit{Let $\bF(\bx)$ be a random function of $\bx$ that is statistically isotropic and homogeneous, and that has an analytic spectral tensor,}
\begin{equation}
    M_{ij}(\bk, t, t') \equiv \int \dd^3 \br \langle f_{\alpha}(t, \bx) f_{\beta}(t', \bx + \br)\rangle e^{-i\bk \bcdot \br}.
\end{equation}\textit{Let $\bF^{\mathrm{(s)}}$ be the solenoidal part of $\bF$, with spectral tensor $M^{\mathrm{(s)}}_{ij}(\bk, t, t')$. Then, the ``Saffman integrals'' associated with $\bF$ and $\bF^{\mathrm{(s)}}$, given by ${L_{\bF}= \int^t \dd t' \,M_{ii}(\boldsymbol{0}, t, t')}$ and ${L_{\bF^{\mathrm{(s)}}}= \int^t \dd t' \, M^{\mathrm{(s)}}_{ii}(\boldsymbol{0}, t, t')}$, respectively [cf.~\eqref{Lf}], satisfy}
\begin{equation}
    L_{\bF^{\mathrm{(s)}}}=\frac{2}{3}L_{\bF}. \label{Saffman_result}
\end{equation}

\subsection{Proof of Saffman's theorem \label{app:saffman_proof}}

A proof of Saffman's theorem, adapted from \cite{Saffman67}, is as follows (we are grateful to an anonymous referee for suggesting the following proof, which is more direct than the proof that we had originally). Because $\bF$ is statistically isotropic and homogeneous, we may write
\begin{align}
    M_{ij}(\bk,t,t') & = a(k,t,t')P_{ij}(\bk) + b(k,t,t')\frac{k_i k_j}{k^2}  \nonumber\\ & = a(k,t,t')\delta_{ij} +[b(k,t,t') - a(k,t,t')]\frac{k_i k_j}{k^2},\label{Saffman_proof1}
\end{align}from which it follows that
\begin{equation}
    M^{(s)}_{ij}(\bk,t,t') =  a(k,t,t')P_{ij}(\bk).\label{Saffman_proof2}
\end{equation}Since $M_{ij}(\bk,t,t')$ is analytic in $\bk$ by assumption, it must be the case that ${a(0,t,t')=b(0,t,t')}$. It follows immediately that
\begin{equation}
    M^{(s)}_{ii}(\boldsymbol{0},t,t') = 2 a(0,t,t') = \frac{2}{3}M_{ii}(\boldsymbol{0},t,t').\label{Saffman23}
\end{equation}Integrating~\eqref{Saffman23} over $t'$, we recover~\eqref{Saffman_result}, \textit{q.e.d.}

\subsection{Consequences of Saffman's theorem \label{app:saffman_consequences}}

We make the following remarks regarding~\eqref{Saffman_result}.
\begin{enumerate}
    \item The utility of Saffman's theorem for forced turbulence is as follows. Equations~\eqref{dL/dt} and~\eqref{uf_soln} together imply that
\begin{equation}
    \frac{\dd L}{\dd t} = 4\pi \lim_{r\to\infty}\left(\frac{1}{r} \frac{\p}{\p r} r^4 u^3 K\right) +L_{\bF^{\mathrm{(s)}}} \label{dL/dt_compressible1}
\end{equation}for a turbulence forced by a series of realisations of $\bF$ that are delta-correlated in time. According to the argument presented in §\ref{sec:batchelor_argument}, the term involving $K(r)$ vanishes.\footnote{Note that, if $L_{\bF^{\mathrm{(s)}}}\neq 0$, $K(r)=O(r^{-3})$ as $r\to\infty$, not $O(r^{-4})$ [as in~\eqref{Ksim-4}]. This is because long-range correlations in $\bF$ can propagate into $K$ via the correlators of~$\bu$ and~$\bF$ that are present in~\eqref{KH_for_K}. Nonetheless, such a decay of $K(r\to\infty)$ is still sufficiently rapid for the term containing $K(r)$ in~\eqref{dL/dt_compressible1} to be negligible.} Thus, according to~\eqref{Saffman_result},
\begin{equation}
    \frac{\dd L}{\dd t} = \frac{2}{3}L_{\bF} ,\label{dL/dt_compressible2}
\end{equation}so the rate of growth of $L$ in forced turbulence is finite, and proportional to $L_{\bF}$.
    \item Saffman's theorem requires that $M_{ij}$ be analytic. This condition is guaranteed by choosing $\bF$ so that $\langle f_i(\bx) f_j(\bx +\br)\rangle$ decays rapidly with distance -- for example, we might take $\bF$ to consist of an ensemble of local patches of uniformly directed force, whose magnitude decays exponentially away from their centre. However, it should be emphasised that Saffman's result does fail if $\langle f_i(\bx) f_j(\bx +\br)\rangle$ decays slowly with $r$. An immediate example of this is the case where $\bF$ is itself solenoidal, so $\bF = \bF^{\mathrm{(s)}}$, and therefore naive application of \eqref{Saffman_result} would suggest $L_{\bF}=0$. However, this does not mean that $L_{\bF}=0$ in general for solenoidal forcing, as the argument applies only to functions $\bF(\bx)$ with an analytic spectral tensor. Long-range correlations in $\bF^{(s)}(\bx)$ are necessary for $L_{\bF^{(s)}}\neq 0$, in which case its spectral tensor is not analytic~\citep{Saffman67}. 
    \item Saffman's theorem does not imply that \emph{any} non-solenoidal forcing with an analytic spectral tensor will have $L_{\bF^{\mathrm{(s)}}}\neq 0$, and so induce a flow with $\mcE(k\to 0)\propto k^2$. This is because it remains possible that $L_{\bF} = 0$ [i.e., $a(0,t,t')=0$ in~\eqref{Saffman23}]. An example of this would be a ensemble of \emph{pairs} of oppositely directed, local instantaneous impulses, separated by a small distance~[cf.~§\ref{sec:randomisation_mechanism}]. Another example, pertinent to numerical studies of forced turbulence, is forcing in a finite spectral band. As shown in Appendix~\ref{app:finite_band_forcing}, long-range correlations decay arbitrarily rapidly for such a forcing, meaning that $M_{ij}$ is analytic---nonetheless, $L_{\bF}$ is manifestly zero for such a $\bF$, and, therefore, so is $L_{\bF^{\mathrm{(s)}}}$.
\end{enumerate}

\bibliographystyle{jfm}
\bibliography{bibliography.bib}

\end{document}